\documentclass[floatfix,aps,amsmath,prd,twocolumn,notitlepage,eqsecnum,showpacs,nofootinbib,aas_macros]{revtex4-1}

\usepackage{hyperref,url}
\usepackage[hyphenbreaks]{breakurl}
\usepackage{color,graphicx,graphics,amsfonts,amsmath,mathrsfs,amssymb,bm,psfrag,natbib,dcolumn,textcase}
\usepackage{array}
\usepackage{enumerate}
\usepackage{grffile}
\usepackage[utf8]{inputenc}
\usepackage[normalem]{ulem}
\usepackage{verbatim}
\allowdisplaybreaks
\newcommand{\be}{\begin{equation}}
\newcommand{\ee}{\end{equation}}
\newcommand{\bea}{\begin{eqnarray}}
\newcommand{\eea}{\end{eqnarray}}

\newcommand{\bs}{\begin{subequations}}
\newcommand{\es}{\end{subequations}}

\newcommand{\arun}[1]{}

\begin{document}
\title{Constraining the orbital eccentricity of inspiralling compact binary systems\\ with Advanced LIGO}
\author{Marc Favata}
\email{marc.favata@montclair.edu}
\affiliation{Department of Physics \& Astronomy, Montclair State University, 1 Normal Avenue, Montclair, New Jersey 07043, USA}
\author{Chunglee Kim}
\email{chunglee.kim@ewha.ac.kr}
\affiliation{Department of Physics, Ewha Womans University, \\
52, Ewhayeodae-gil, Seodaemun-gu,
Seoul 03760, Korea}
\author{K.~G.~Arun}
\email{kgarun@cmi.ac.in}
\affiliation{Chennai Mathematical Institute, Siruseri 603103, India}
\author{JeongCho Kim}
\email{jeongcho.kim@gmail.com}
\author{Hyung Won Lee}
\email{hwlee@inje.ac.kr}
\affiliation{Institute of Basic Science and Department of Computer Simulation, Inje University, 197 Inje-ro, Gimhae 50834, Korea}
\date{Submitted 12 August 2021; accepted 3 November 2021}
\begin{abstract}
The detection of $\sim 50$ coalescing compact binaries with the Advanced LIGO and Virgo detectors has allowed us to test general relativity, constrain merger rates, and look for evidence of tidal effects, compact object spins, higher waveform modes, and black hole ringdowns. An effect that has not yet been confidently detected is binary eccentricity, which might be present in a small fraction of binaries formed dynamically. Here we discuss general limits on eccentricity that can, in-principle, be placed on all types of compact object binaries by a detector operating at the design sensitivity of Advanced LIGO. Using a post-Newtonian model for gravitational-wave phasing valid in the small eccentricity regime, we assess the relative measurement error for eccentricity for a variety of spinning and nonspinning binaries. Errors and correlations involving the mass and spin parameters are also investigated. We find that decreasing the low frequency limit of a detector's observational frequency band is one of the key design factors for increasing the odds of measuring binary eccentricity. We also introduce and analytically explore the \emph{eccentric chirp mass} parameter, which replaces the chirp mass as the key measurable parameter combination in eccentric gravitational waveform models. The eccentric chirp mass parameter explains a degeneracy between the chirp mass and the eccentricity. This degeneracy leads to a bias in the standard chirp mass parameter. We also investigate the systematic parameter bias that arises when eccentric systems are recovered using circular waveform templates. We use both Fisher matrix and Bayesian-inference-based Markov Chain Monte Carlo (MCMC) methods to investigate these parameter estimation issues, and we find good agreement between the two approaches (for both statistical and systematic errors) in the appropriate signal-to-noise ratio regime. This study helps to quantify how effectively one can use eccentricity measurements as a probe of binary formation channels.\\
\end{abstract}
\maketitle

\section{Introduction and Motivation}
The first three observing runs of the Advanced LIGO \cite{LIGO-stdrd-ref1} (hereafter LIGO) and Advanced Virgo \cite{aVirgo-stdref} detectors have led to the discovery of $48$ binary black holes (BBHs)~\cite{detectionPRL2016,gw151226-PRL2016,GW170104,GW170608,GW170814,GWTC1-LVC2018,GW190412,GW190814,GW190521-PRL2021,GWTC-2-PRX2021}, two binary neutron star (BNS) mergers~\cite{GW170817,GW190425}, and two neutron-star/black-hole (NS/BH) mergers \cite{NSBH-ApJL2021}. Several known features in the gravitational waveforms of compact binaries were observed or constrained via these systems. These include the tidal deformability in binary neutron star mergers GW170817 \cite{GW170817} and GW190425~\cite{GW190425}, the detection of subdominant harmonics in GW190412~\cite{GW190412} and GW190814~\cite{GW190814}, and the detection of a weak signature of spin-induced precession of the orbital plane in the case of GW190412~\cite{GW190412}. Another class of compact binaries that is yet to be confirmed by advanced ground-based detectors are those in elliptical orbits (but see our discussion regarding GW190521 and other recent work in Sec.~\ref{subsec:eccPEworks} below).

This paper addresses the parameter estimation problem of eccentric compact binaries in detail, focusing on limits that an Advanced-LIGO like detector could (in principle) place in the near future (once design sensitivity is reached). We also pay significant attention to the role of systematic parameter bias if eccentricity in waveform models is ignored.

An important open problem in GW astrophysics is identifying the formation
channels of compact binaries~\cite{GWTC1-LVC2018}. Two competing models are field binary evolution~\cite{LIGOScientific:2018jsj} and dynamical formation in dense stellar clusters~\cite{Benacquista:2011kv,bae-etal-MNRAS2013,Rodriguez:2017pec,Samsing:2017xmd}. The latter predicts $\sim 5\%$ to $10\%$ of BBHs merging in globular clusters will have eccentricities $e_0>0.1$ in the LIGO frequency band.\footnote{Throughout this paper, we use the following notation when discussing eccentricity. The symbol $e_0$ denotes a constant parameter representing the value of the binary eccentricity at a particular reference frequency $f_0$; this is taken to be $f_0=10$ Hz (unless stated otherwise) when discussing our results and the results of references that we discuss below.  The value $f_0=10$ Hz is the reference frequency chosen in most studies, corresponding to the low-frequency limit of the Advanced LIGO frequency band. The notation $e$ (no subscript) is used when quoting results in other papers where the reference frequency is not specified or easily determined. The notation $e_t$ denotes a time-evolving eccentricity. It is equivalent to the ``time-eccentricity'' parameter that is introduced in the quasi-Keplerian formalism \cite{DGI} (see Sec.~III of \cite{moore-etal-PRD2016} for a discussion and additional references).} It is well known that gravitational-wave (GW) emission in an inspiralling compact binary leads to decreasing orbital eccentricity \cite{petersmathews,peters}. Measurement of nonzero eccentricity has been proposed as a potential smoking gun for the dynamical formation channel (see references in Sec.~\ref{subsec:eccexpect} below). In this case, a tight eccentric binary forms via multi-body interactions and is not able to shed its eccentricity before coalescence. Hence, accurate measurement of orbital eccentricity could play an important role in understanding the formation and evolution of compact binaries.

Previous work (e.g., \cite{favata-PRL2014} and references therein) has shown that even relatively small eccentricities ($e_0 \sim 10^{-3} \mbox{--} 10^{-2}$) can produce parameter biases. Here, we revisit and extend this work with the aim to ask (and answer) the following questions: (i) how well can LIGO measure orbital eccentricity and (ii) what bias does eccentricity induce if neglected. We explore these questions for a variety of binary systems: binary neutron-stars (NSs), binary black holes (BHs), and NS/BH binaries. In addition to selected ``generic'' examples of such systems, we also examine systems with parameters similar to the first detected LIGO signals, GW150914 \cite{detectionPRL2016} and GW151226 \cite{gw151226-PRL2016}. These are chosen as representatives of the two classes (high and low masses) of binary black holes (BBHs) that LIGO/Virgo has seen during the first two observing runs~\cite{GWTC1-LVC2018}. We focus on the case when the eccentricity $e_0$ at $10$ Hz is small, typically $e_0 \sim 0.001 \mbox{--} 0.1$ or less. The rapid circularization effect of GW emission makes this the more astrophysically likely regime.  

\begin{table*}[t]
\caption{\label{tab:eccevol} Eccentricity decay time for compact-object binaries as a function of measured eccentricity in the LIGO band. The first column lists the eccentricity $e_0$ detected in the Advanced LIGO band at 10 Hz. The remaining columns show the time $\Delta T(e_t)$ prior to the binary emitting at 10 Hz, at which the eccentricity had a value of $e_t$. This is shown for three systems discussed in the text: BBHs with masses similar to GW150914 and GW151226, as well as a BNS system with masses 1.4 and 1.25 $M_{\odot}$. Times are listed in days (d) or years (yr).}
\begin{tabular}{|l|lll|lll|lll|}
	\hline 
 & \multicolumn{3}{c|}{GW150914} & \multicolumn{3}{c|}{GW151226} & \multicolumn{3}{c|}{BNS} \\
\hline
\multicolumn{1}{|c|}{$e_0$ (10 Hz)} & \multicolumn{1}{c}{$\Delta T (0.9)$} & \multicolumn{1}{c}{$\Delta T (0.99)$} & \multicolumn{1}{c|}{$\Delta T (0.999)$} & \multicolumn{1}{c}{$\Delta T (0.9)$} & \multicolumn{1}{c}{$\Delta T (0.99)$} & \multicolumn{1}{c|}{$\Delta T (0.999)$}  & \multicolumn{1}{c}{$\Delta T (0.9)$} & \multicolumn{1}{c}{$\Delta T (0.99)$} & \multicolumn{1}{c|}{$\Delta T (0.999)$}\\
	\hline
	$10^{-1}$ & 0.0525 d & 0.293 d & 1.08 d & 0.356 d & 1.99 d & 7.31 d & 12.5 d & 69.9 d & 257 d \\ 
    $10^{-2}$ & 18 d & 0.282 yr & 1.04 yr & 0.343 yr & 1.91 yr & 7.04 yr & 12.1 yr  & 67.3 yr & 248 yr \\ 
    $10^{-3}$ & 17.0 yr & 94.9 yr & 349 yr & 115 yr & 643 yr & 2370 yr & 4050 yr & 22\,600 yr  & 83\,200 yr \\
    $10^{-4}$ & 5710 yr & 31\,900 yr & 117\,000 yr & 38\,700 yr  & 216\,000 yr  & 795\,000 yr & 1.36 Myr & 7.60 Myr & 28.0 Myr \\
    $10^{-5}$ & 1.92 Myr & 10.7 Myr & 39.4 Myr & 13.0 Myr & 72.6 Myr & 267 Myr & 0.457 Gyr & 2.55 Gyr & 9.39 Gyr \\
    \hline 
\end{tabular}
\end{table*}

\subsection{\label{subsec:eccexpect}Expected eccentricities of compact-object binaries}
A variety of studies have examined the eccentricity of binaries when they enter the frequency band of ground or space-based detectors. (A brief summary of these studies prior to 2016 is provided in Sec.~IA of \cite{moore-etal-PRD2016}.) In particular, we note that currently observed galactic BNSs will have very small eccentricities ($\lesssim 7 \times 10^{-6}$) when they enter the LIGO frequency band (10 Hz; see Table II of \cite{moore-etal-PRD2016}). Reference \cite{kowalska-etal-eccentricity-distribution-AA2011} used a population synthesis code to estimate that $0.3\%$, $0.7\%$, and $2\%$ of BBH, NS/BH, and NS/NS binaries (respectively) will have eccentricities exceeding $0.01$ at $30$ Hz. A study of isolated triple systems \cite{silsbee-tremaine2016} (containing a BBH inner binary) indicated that a few percent of these systems could also produce very large eccentricities above 10 Hz.

Simulations of binary formation in globular clusters predict a wide range of eccentricities in the LIGO band. Reference \cite{antonini-etalApJ2015} predicts that $\sim 20\%$ of BBHs formed via dynamical interactions will have eccentricity $e>0.1$  at 10 Hz. In Ref.~\cite{rodriguez-chatterjee-rasio2016} $\sim 1\%$ of globular cluster (GC) BBHs will have $e_0>10^{-3}$. (See also earlier predictions in Ref.~\cite{wen-eccentricity-ApJ2003,antonini-murray-mikkola-ApJ2014,antognini-etalMNRAS2014,gultekin-miller-hamilton2004ApJ,oleary-etal-BHmergersGC-ApJ2006}.) In Ref.~\cite{haster-etal2016} the capture and inspiral of a single stellar-mass BH around a cluster intermediate-mass BH was simulated and found to have a very small eccentricity in the LIGO band.   Reference \cite{dorazio-samsingMNRAS2018} found that BBHs formed via 3-body interactions within GCs may make up to $\sim 5\%$ of the dynamically-formed population and could have $e_0\sim 0.1$ in the LIGO band. A study focusing on NS/BH binaries \cite{fragione-loebMNRAS2019} found that those merging in isolation are nearly circular in the LIGO band, while a large fraction of NS/BH formed in triples have a high eccentricity ($e_0 \gtrsim 0.1$). A recent study \cite{zevin-etal2021} attempts to constrain the role of clusters in producing detectable eccentric BBHs. They find that around $\sim 7\%$ of potentially detectable cluster BBH sources will have measurable eccentricity, and that one detection of eccentricity in the GWTC-2 catalog would suggest that dense star clusters produce $>14\%$ of the detectable BBH population. Observation of eccentricity via space-based interferometers can also discriminate between cluster and field origins for BBHs \cite{breivik-etal2016}.

Along with globular clusters, the dense environments near galactic nuclei can also produce eccentric BBHs \cite{oleary-kocsis-loeb-MNRAS2009,antonini-peretsApJ2012,hong-lee_MNRAS2015}, with Ref.~\cite{oleary-kocsis-loeb-MNRAS2009} predicting quite high eccentricities in the LIGO band ($e>0.9$) and Ref.~\cite{antonini-peretsApJ2012} predicting that $10\%$ of BBHs formed near a supermassive BH  will have $e>0.1$ when entering the LIGO band.  Studies of BBHs formed via GW capture in galactic nuclei \cite{laszlo-etalApJ2018,gondan-bence-MNRAS2021} find that a substantial fraction of stellar mass BBHs formed via this route will have $e_0>0.1$ (see especially Tables 1 through 4 in \cite{laszlo-etalApJ2018}). Another study of the BBH eccentricity distribution in galactic nuclei \cite{takatsy-etalMNRAS2019} found that $\sim 75\%$ of BBHs formed in galactic nuclei via gravitational capture will have $e_0>0.1$.  The scattering of BBHs with singles in AGN disks was found to efficiently produce eccentric mergers with $e_0>0.1$ for LIGO/Virgo \cite{samsing-etal2020}. A more recent study \cite{tagawa-etalApJL2021} found that binary-single interactions in AGN accretion disks could yield $e_0\gtrsim 0.03$ to $0.3$ in the LIGO band.

\subsection{Eccentricity decay timescales}
 The eccentricity constraints discussed in this paper are typically phrased in terms of $e_0$, taken to be the eccentricity when the binary radiates GWs at 10 Hz. To put this number into context, it is useful to understand how quickly the eccentricity increases ``backwards in time'' from 10 Hz (e.g., how large was the eccentricity some number of days or years before it entered the LIGO band at 10 Hz). To compute this we numerically integrated the equations for the eccentricity and periastron separation as a function of time at leading-order in a post-Newtonian (PN) expansion (using equations in \cite{favata-eccentricmemory}, equivalent to the so-called Peters-Mathews approximation \cite{peters}; see Appendix \ref{app:eccevolve} for details). Table \ref{tab:eccevol} quantifies the results, showing how long it took a particular value of eccentricity $e_0$ (measured when the emitted primary GW harmonic frequency is at $10$ Hz) to decay from a larger (earlier) value of $e_t = 0.9 \mbox{--} 0.999$. For example, a BBH with parameters similar to GW151226 that is observed to have an eccentricity $e_0=0.01$ at $10$ Hz had an eccentricity of $e_t=0.99$ only $\Delta T(0.99) \approx 1.9$ years earlier. We note that the timescales discussed here assume that no dynamical interactions with a third body occur between the high and low eccentricity states.
 
 Table \ref{tab:eccevol} indicates that BBH or BNS systems that are observed with $e_0\approx 0.1$ were highly eccentric only hours to weeks earlier. As our analysis will show that only values of $e_0\gtrsim 0.01$ are plausibly detectable with Advanced LIGO detectors, this suggests that any eccentric binaries observed by 2nd-generation ground-based detectors must have been formed almost immediately prior to detection. Even if eccentricities as low as $e_0 \sim 0.001$ could be detected (which is plausible for some sources with third-generation detectors), such binaries would likewise have had eccentricities near $\approx 1$ on a timescale of order $\sim 10$ to $10^4$ years earlier---extremely short in comparison to typical astrophysical timescales associated with stellar systems (e.g., the relaxation time or crossing time for a cluster). Hence, the observation of eccentricity with ground-based detectors is a powerful indicator that the compact object binary was recently formed via a large-eccentricity capture in a high-density stellar environment such as a globular cluster or nuclear star cluster. 

\subsection{\label{subsec:eccPEworks}Searches and parameter estimation constraints for eccentric binaries}
In the case of ground-based detectors, several studies have considered the impact of ignoring eccentricity on detection \cite{mandel-brown,martel-poisson-eccentric-PRD1999,tessmer-gopu-PRD2008,cokelaer-pathak-detecteccentric-CQG2009,brown-zimmerman-eccentric-PRD2010,huerta-brown2013PRD}. They find that (depending on the binary mass) circular templates are sufficient for eccentricities $e \lesssim 0.02 \mbox{--} 0.15$. (The analogous problem for supermassive BH binaries in the eLISA band is treated in  Ref.~\cite{porter-sesana2010}.) Other studies~\cite{kocsis-levinPRD2012,east-etalPRD2013,tai-mcwilliams-pretoriusPRD2014,coughlin-etalPRD2015,tiwari-etal-eBBHsearch2016PhRvD} considered the problem of detecting LIGO-band binaries with moderate to large eccentricities. Methods such as {\tt Coherent Wave Burst}~\cite{Klimenko:2008fu}, an algorithm to search for GW transients, were also employed to search for eccentric binary black holes in the O1 and O2 data \cite{Salemi:2019owp} (with none found in that search). A recent analysis of the O3 data \cite{LVKallskyBurst2021} also found no detections but improved on the O2 rate estimate of highly eccentric mergers by a factor $\lesssim 2$. A comparison of such morphology-independent analyses with those using theoretical waveforms from eccentric binaries can be found in Ref.~\cite{Ramos-Buades:2020eju}.

Some works have directly considered parameter estimation for eccentric binaries.  One of the present authors \cite{favata-PRL2014} considered the systematic bias induced on the mass or tidal parameters if eccentricity, spin, tides, or high PN effects are neglected (focusing only on BNS). Reference \cite{kyutoku-seto2014MNRAS} considered the effect of eccentricity on the sky-localization by a second-generation GW detector network, while Ref.~\cite{sun-cao-etal2015PhRvD} examined the ability of Advanced LIGO and ET to measure eccentricity, as well as the effect of eccentricity on other signal parameters. Another set of studies \cite{Gondan:2017hbp,Gondan:2018khr} assessed the accuracy with which LIGO and Virgo could estimate the parameters of eccentric binaries. They found that, compared to circular binaries, important parameters of the binary (masses, distance, source location) will improve if $e_0 \geq 0.1$. This improvement may be attributed to the eccentricity-induced higher harmonics of the orbital phase, which plays an important role in parameter estimation. Using the Bayesian parameter estimation package {\tt Bilby} \cite{bilby2019ApJS} and nonspinning inspiral-only templates, Ref.~\cite{Lower:2018seu} found that 2nd generation GW observatories can measure BBHs with $e_0\geq 0.05$.

Several works have examined the LIGO/Virgo data from the first two observing runs for signatures of eccentric binaries. In Refs.~\cite{detection-PEpaper2016,detection-Astropaper2016ApJL} it was initially reported that $e_0 \lesssim 0.1$ would not produce any measurable deviation from the parameters of GW150914 that were determined with circular waveforms. More recent work by the LIGO-Virgo Collaborations looked at a range of possible systematic errors~\cite{Abbott:2016wiq}.  In the case of eccentricity, they found that an eccentricity $e >0.05 - \mbox{0.1}$ at $25$ Hz will begin to show a bias in the chirp mass for GW150914 [with the mass ratio and effective spin parameter unaffected by eccentricities as large as the maximum value they considered, $e (25\, {\rm Hz})=0.13$]. Using a spin-aligned eccentric effective-one-body (EOB) model \cite{cao-han-SEOBeccPRD2017} and {\tt Bilby} \cite{bilby2019ApJS}, Ref.~\cite{romero-shaw2019MNRAS} analyzed the GWTC-1 \cite{GWTC1-LVC2018} BBH events and found that all are consistent with zero eccentricity; they also provide 90\% upper limits on the events, ranging from $e_0 \leq 0.024 \mbox{--} 0.054$. (For GW151226, which we consider in detail below, they find an upper limit of $e_0\leq 0.029$.) Using a frequency-domain nonspinning eccentric waveform \cite{huerta-etal-PRD2014} and {\tt Bilby}, Ref.~\cite{wu-cao-zhu2020} also performed parameter estimation on the GWTC-1 BBHs. They find 90\% upper limits ranging from $0.033$ to $0.181$ (with the latter for GW151226).

Reference \cite{nitz-lenon-brown2020ApJ} performed a search for eccentric BNS in the LIGO data from 2015 to 2017, finding no evidence for eccentric sources. Those same authors \cite{lenon-nitz-brown2020MNRAS} also performed Bayesian parameter estimation via the {\tt PyCBC Inference} package \cite{PyCBCInfereceA2019PASP}
on the BNS events GW170817 and GW190425, making use of the 3PN inspiral-only low-eccentricity waveform developed and implemented by the present authors  \cite{moore-etal-PRD2016,TaylorF2EccURL}. They find upper limits (at 90\% confidence) of $e_0 \leq 0.024$ for GW170817 and $e_0 \leq 0.048$ for GW190425, but with significant dependence on the priors.

As this manuscript was being finalized, we became aware of recent work in Ref.~\cite{OSheaKumar2021}, who also investigate the bias when recovering eccentric signals with circular templates. Those authors further analyzed the public LIGO/Virgo data for GW151226 and GW170608. They find clear evidence for eccentricity if nonspinning templates are used, but both events are consistent with zero eccentricity if aligned-spin and eccentric waveform templates are used. Upper limits (90\% confidence) of $e_0<0.15$ and $e_0<0.12$ are set on GW151226 and GW170608 (respectively). More recently, Ref.~\cite{romero-shaw-eeccGWTC2} analyzed 36 of the BBHs in GWTC-2 \cite{GWTC-2-PRX2021} using {\tt Bilby} and assuming aligned spins. Making use of the waveform model in \cite{cao-han-SEOBeccPRD2017}, they find twelve events with some support for eccentricity $e_0\geq 0.05$. Two events have more than $50\%$ of their posterior probability distribution above $e_0\geq 0.05$: GW190521 and GW190620A.

Following the detection of the first intermediate mass BH system GW190521 by LIGO/Virgo \cite{GW190521-PRL2021}, several works have considered the possibility that this system is an eccentric binary. The LIGO/Virgo Collaboration analysis  \cite{GW190521-astroApJL} raised the possibility that spin precession or very-high eccentricity (or a head-on collision \cite{Bustillo-headon190521PRL2021}) could not be distinguished in this source, given the small number of cycles from such a large mass binary. A subsequent analysis in Ref.~\cite{romero-shaw-etal2020ApJ} suggests that the data prefers a signal with $e_0\geq 0.1$ over a precessing circular signal; however they do not confidently determine if the source is eccentric or precessing/circular. Work by~\cite{gayathri-etal2020} also suggests evidence for a high-eccentricity merger over a precessing/circular one. Given that GW190521 has only $\sim 4$ GW cycles detected in the LIGO/Virgo frequency band, with the inspiral representing $\lesssim 2$ of those cycles (see Fig.~1 of \cite{GW190521-PRL2021}), a firm measurement of eccentricity from that source seems unlikely. Similarly,  GW190620A was identified (along with GW190521) in Ref.~\cite{romero-shaw-eeccGWTC2} as having support for eccentricity, using spin-aligned waveforms. This is also a high-mass BBH (total mass $M \approx 92 M_{\odot}$)  with few inspiral cycles. It remains to be seen if indications of eccentricity in these systems remain when analyzed with waveforms that also include spin precession. Our work will focus on inferring eccentricity from lower-mass binaries ($M\lesssim 65 M_{\odot}$), where the effect of eccentricity manifests itself over several inspiral cycles.

\begin{figure*}[th]
$
\begin{array}{cc}
\includegraphics[angle=0, width=0.48\textwidth]{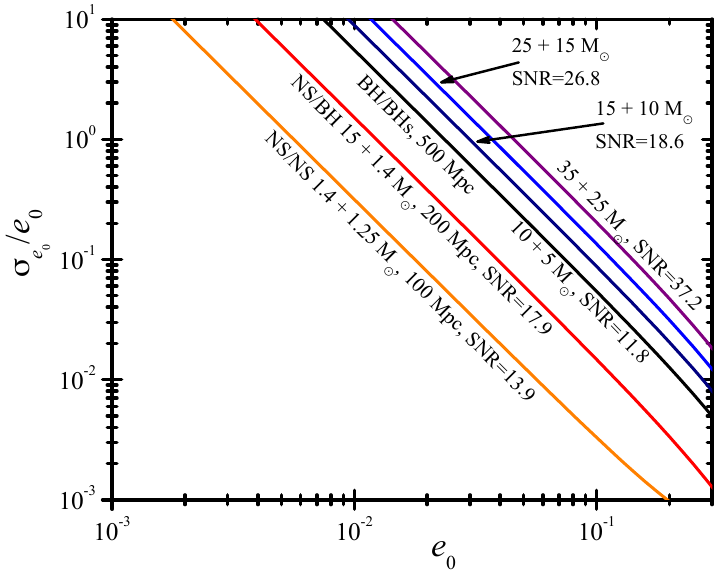} &
\includegraphics[angle=0, width=0.49\textwidth]{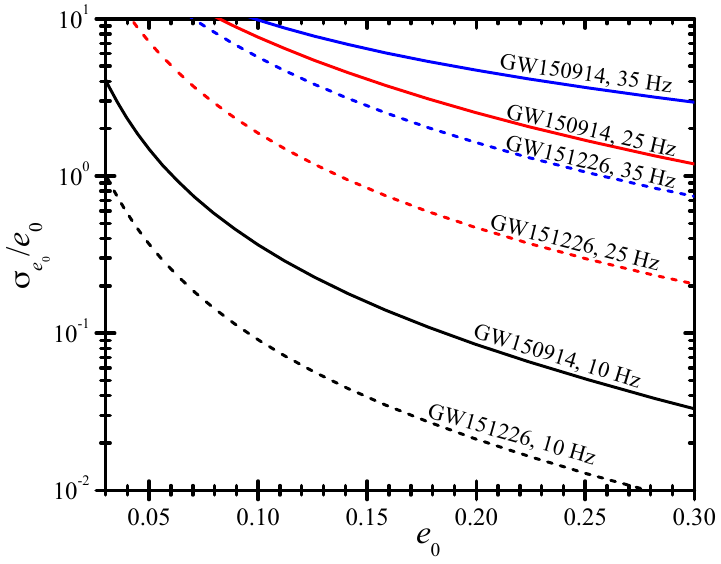}
\end{array}
$
\caption{\label{fig:PEe0results}Fractional ($1$-sigma) statistical error in eccentricity $e_0$ at 10 Hz as a function of $e_0$. The left panel shows the fractional errors for a variety of fiducial systems discussed in the main text. From left to right, the four BBH curves correspond to the systems labeled BBH1, BBH2, BBH3, and BBH4 defined in Sec.~\ref{sec:statistical} and with masses and signal-to-noise ratios as indicated in the figure. Results for a NS/BH and a BNS system are also shown. The right panel shows these same errors for systems with parameters similar to GW150914 (solid curves) or GW151226 (dashed curves). The SNRs are chosen to be the same as the actual observed events, but the calculations were performed using the LIGO design sensitivity. The different colors indicate different low-frequency limits used in evaluating the Fisher matrix integrals ($35$ Hz, $25$ Hz, and $10$ Hz). Reducing the detector's low frequency sensitivity dramatically improves the eccentricity measurement precision.}
\end{figure*}
\begin{figure*}[t]
$
\begin{array}{ccc}
\includegraphics[angle=0, width=0.3\textwidth]{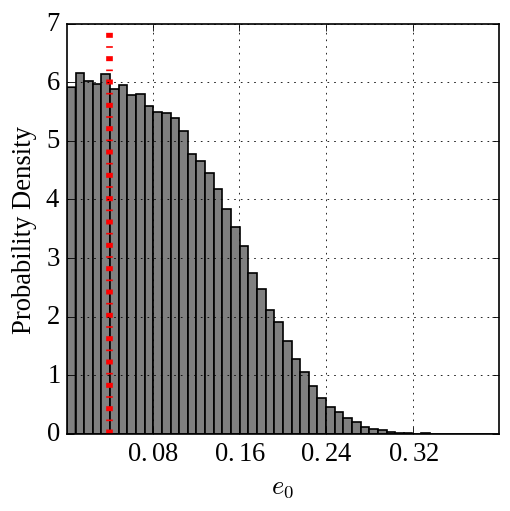} &
\includegraphics[angle=0, width=0.3\textwidth]{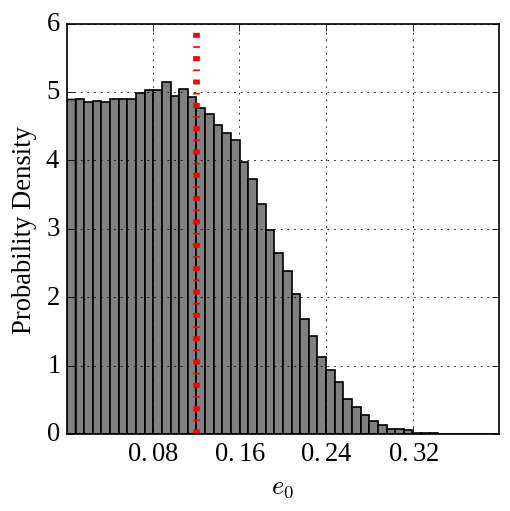} &
\includegraphics[angle=0, width=0.3\textwidth]{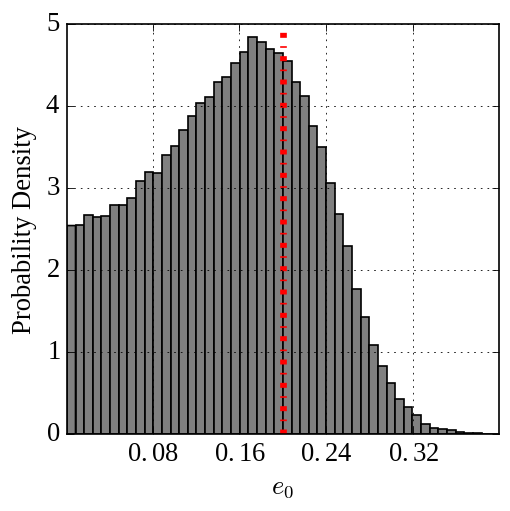} \\
\includegraphics[angle=0, width=0.3\textwidth]{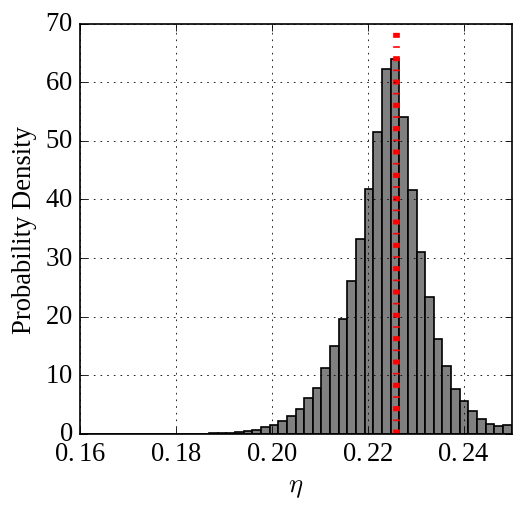} &
\includegraphics[angle=0, width=0.3\textwidth]{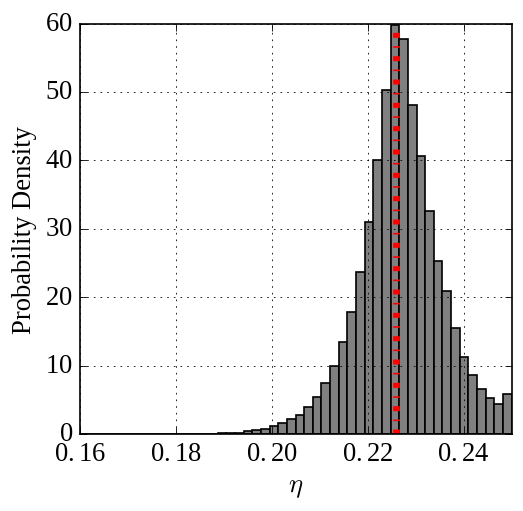} &
\includegraphics[angle=0, width=0.3\textwidth]{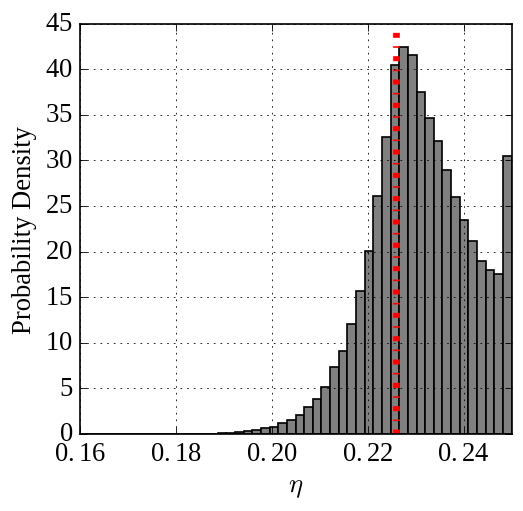} \\
\includegraphics[angle=0, width=0.3\textwidth]{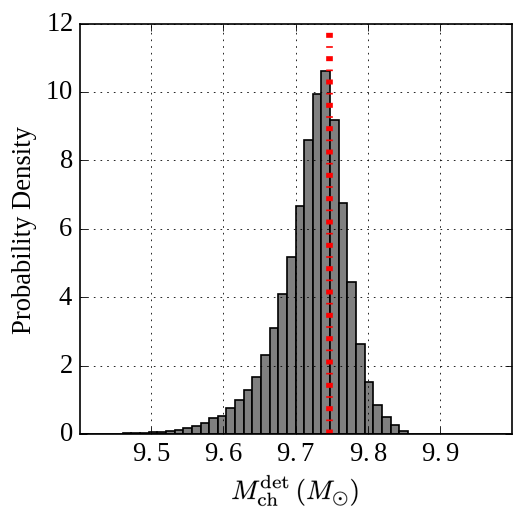} &
\includegraphics[angle=0, width=0.3\textwidth]{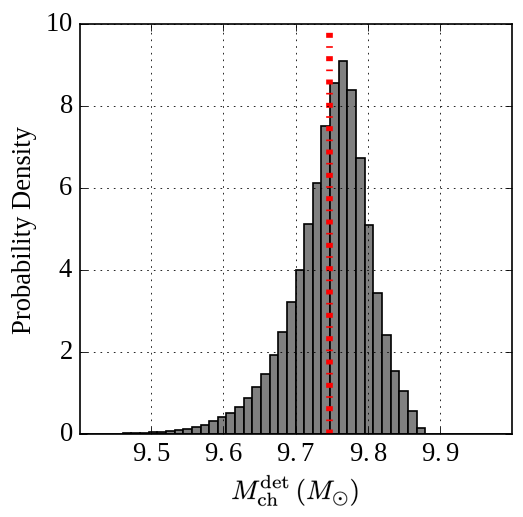} &
\includegraphics[angle=0, width=0.3\textwidth]{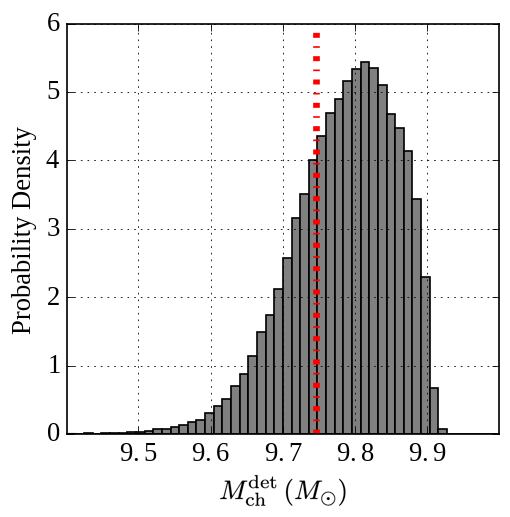} 
\end{array}
$
\caption{\label{fig:MCMCpdf-stat} Posterior probability distributions from a Bayesian inference MCMC calculation of a binary black hole system with parameters similar to GW151226, showing three intrinsic parameters as a function of injected eccentricity. The top row shows the eccentricity parameter $e_0$, the middle shows the reduced mass ratio $\eta$, and the bottom row shows the detector-frame chirp mass $M_{\rm ch}^{\rm det}$. The injected eccentricity varies by column over $e_0^{\rm inj}= [0.04, 0.12, 0.2]$ from left to right. The vertical dotted lines show the injected value of the corresponding parameter. The posterior distributions shown here were marginalized over the other parameters. Relevant detector-frame mass parameters are $m_1^{\rm det}=15.6 M_{\odot}$, $m_2^{\rm det} = 8.2 M_{\odot}$, $M_{\rm ch}^{\rm det}= 9.746 M_{\odot}$, and $\eta = 0.226$. The binary components are assumed to be nonspinning and the SNR is set to $20.12$. Parameter estimation is performed using the {\tt TaylorF2Ecc} waveform as described in the main text. See also Table \ref{tab:MCMCcompare_e0}. Note the growing bias in the recovered chirp mass as the injected value of  $e_0$ is increased (bottom row). This arises from the degeneracy between $e_0$ and $M_{\rm ch}^{\rm det}$ (see Sec.~\ref{sec:degen}).
}  
\end{figure*}
\subsection{Present work}
In this paper we take an agnostic approach as to the likelihood of binary eccentricity in the LIGO band. We view binary eccentricity as another signal parameter which should be measured, with the resulting constraints providing feedback to astrophysical models. Our goals are simply to estimate how well LIGO can---in principle---constrain binary eccentricity, and to assess the impact of ignoring eccentricity in GW signal templates. To do this, we first apply the Fisher matrix formalism to investigate a range of compact-binary types. We also provide some limited comparisons using a Bayesian inference Markov-chain Monte Carlo (MCMC) approach.

We expand on previous work in several ways. Our focus is on the capability of LIGO to constrain eccentricity once it reaches design sensitivity. We focus on two sets of binary systems: (i) a set of ``fiducial'' BNS, NS/BH, and BBH binaries, and (ii) systems with parameters similar to GW150914 and GW151226. (Our analysis uses only the published parameters for these systems and does not directly analyze the events' strain data.) Unlike some previous studies, our Fisher-matrix analysis includes both spin and eccentricity effects in our waveform model. (Our MCMC analysis does not include spins.) Focusing on the low-eccentricity limit, $[e_0 \equiv e_t(f_0) \lesssim 0.2$], our eccentric waveform consistently incorporates all secular, eccentric phase corrections to 3PN order and to order $e_0^2$. A complete description of our eccentric waveform is given in \cite{moore-etal-PRD2016}. Here we also incorporate spin corrections to the circular phasing, but work in the regime where coupled eccentric/spin phase corrections are negligible.    

For the two sets of systems mentioned above, we consider the following issues: (a) with what error can LIGO (at design sensitivity) measure $e_0$, and (b) what error in the mass and spin parameters is induced by ignoring eccentricity entirely. These issues are addressed via a Fisher matrix analysis, supplemented with the systematic parameter bias formalism developed by Cutler and Vallisneri \cite{cutler-vallisneri-systematicerrors-PRD2007}. For the case of GW151226, we also performed a separate parameter estimation analysis using {\tt LALInferenceMCMC} \cite{rover-meyer-christensen-PRD2007, vandersluys-etal-ApJ2008, veitch-etalPEpaper-PRD2015}, a Bayesian inference MCMC code implemented in the LIGO Algorithm Library (\texttt{LALSuite}) \cite{LALSuite}. We compare those calculations with our Fisher matrix estimates.

\begin{figure*}[t]
$
\begin{array}{cc}
\includegraphics[angle=0, width=0.48\textwidth]{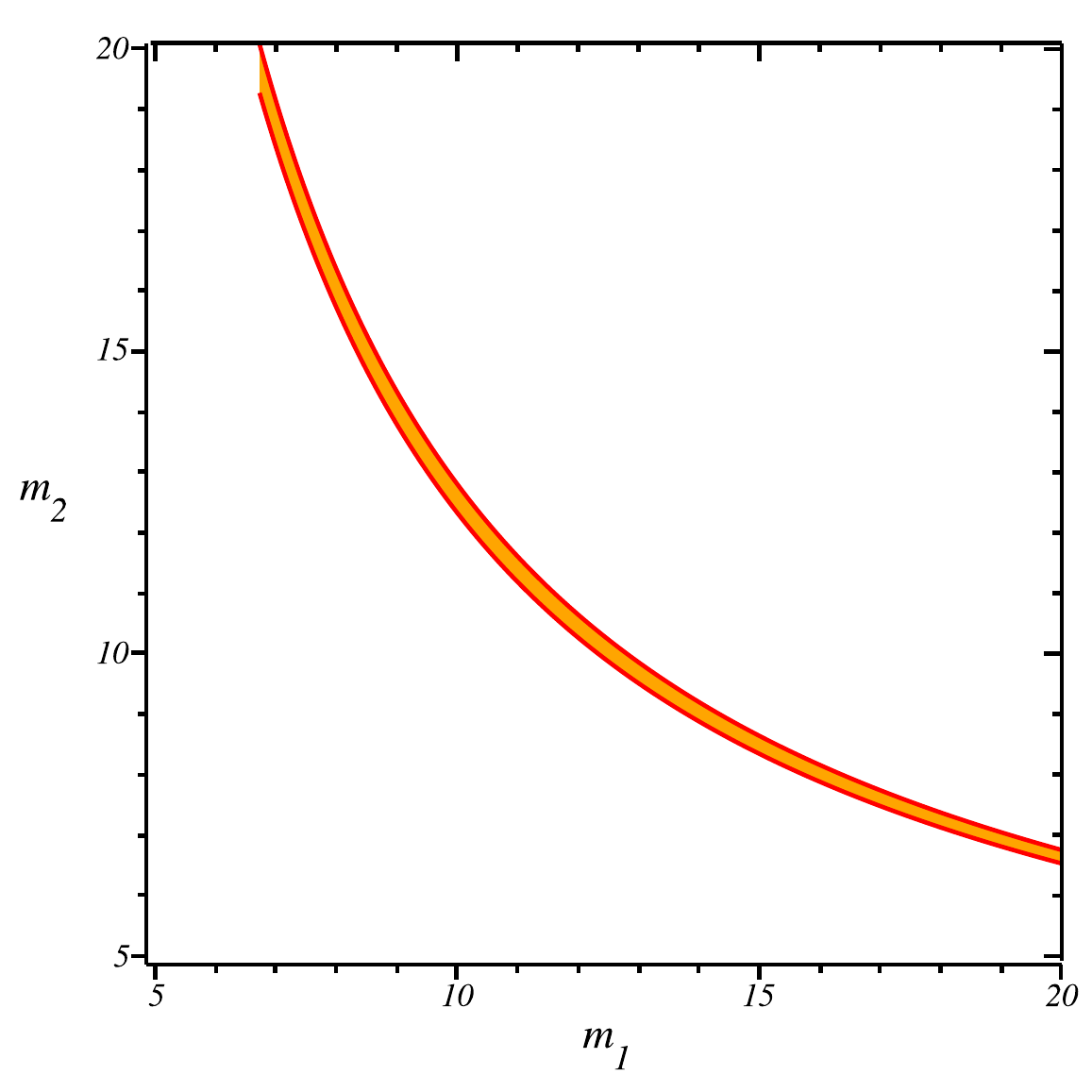} &
\includegraphics[angle=0, width=0.48\textwidth]{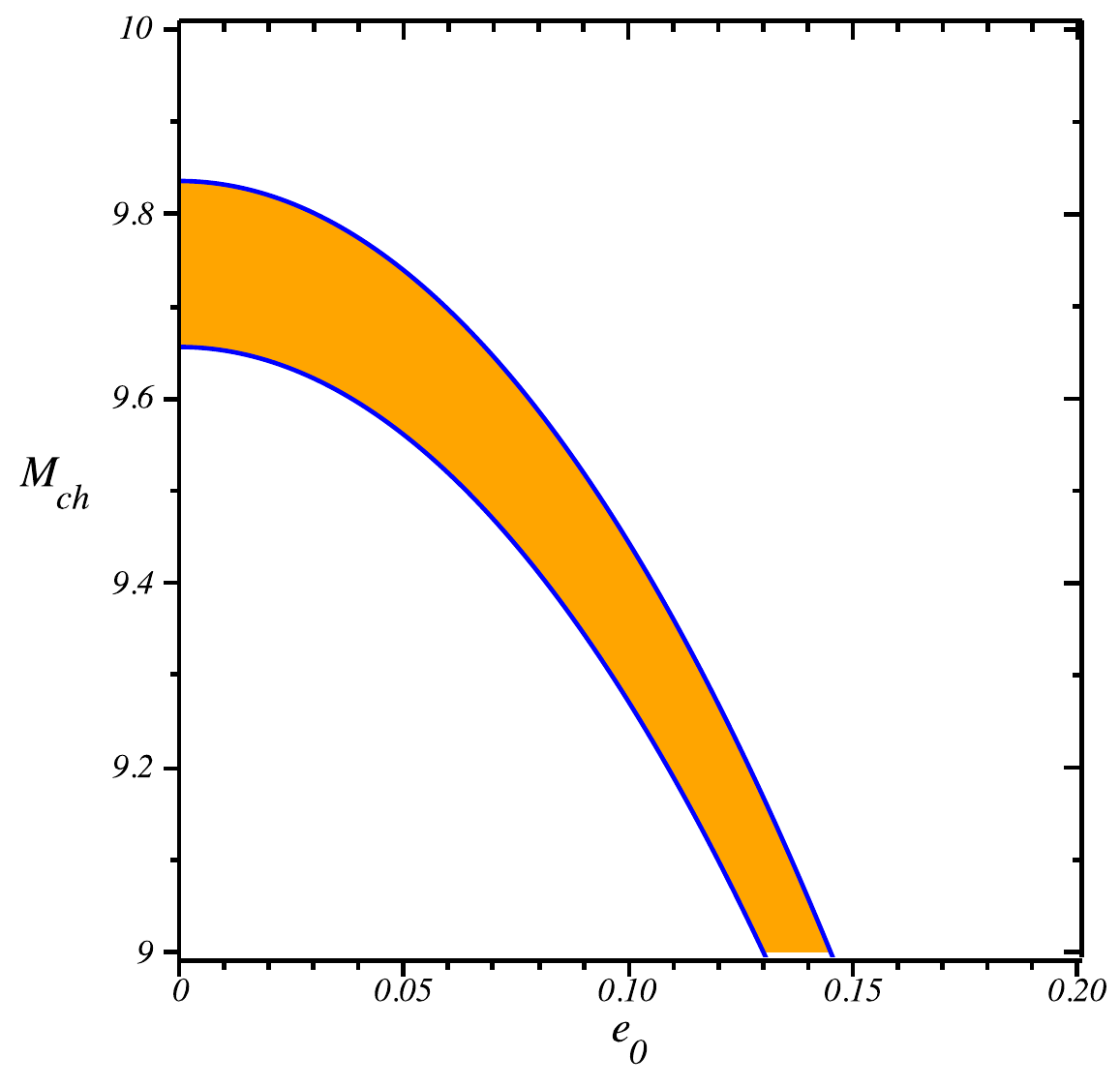}
\end{array}
$
\caption{\label{fig:m1m2chirp}Comparison of contours of constant chirp mass and the eccentric chirp mass. The left plot shows the standard curve of constant chirp mass $M_{\rm ch}=\eta^{3/5}M = (m_1 m_2)^{3/5} (m_1+m_2)^{-1/5}$ in the $m_1$-$m_2$ plane, with the colored band delimiting the region with $M_{\rm ch}= 9.746 \pm 0.09\, M_{\odot}$. Analogously, the right plot shows a band of constant eccentric chirp mass $M_{\rm ch}^{\rm ecc} = M_{\rm ch} (1-\frac{157}{24} e_0^2)^{-3/5}$ in the $e_0$-$M_{\rm ch}$ plane, with $M_{\rm ch}^{\rm ecc}= 9.746 \pm 0.09\, M_{\odot}$. Masses in both plots are in units of $M_{\odot}$.}
\end{figure*}
\subsection{Summary of results}
This paper contains a variety of results that we summarize here. 
\begin{enumerate}
\item Measurements of eccentricity better than $\approx 30\%$ (1-sigma fractional error) require a binary's eccentricity to exceed $e_0 \approx 0.01$ for BNS, $\approx 0.03$ for NS/BH binaries, and $\approx 0.05 \mbox{--} 0.1$ for BBHs. Generally, larger mass binaries require a larger $e_0$ to attain a specific fractional error in $e_0$. This is quantified in the left panel of Fig.~\ref{fig:PEe0results}. Additional Fisher-matrix parameter error estimates and parameter correlations are shown in Figs.~\ref{fig:PEotherresults} and \ref{fig:corr}. 

\item Efforts to improve the low-frequency sensitivity of ground-based detectors are of paramount importance for improving eccentricity measurements. We investigated the eccentricity measurement precision for binaries with parameters similar to GW150914 and GW151226, as well as the measurement's dependence on the detector's low-frequency sensitivity. This is shown in the right panel of Fig.~\ref{fig:PEe0results}. In the case of a GW150914-like binary, no strong constraint on $e_0$ can be placed for any $e_0<0.3$ {\it unless} the low-frequency limit ($f_{\rm low}$) of the detector goes to 10 Hz. In the $f_{\rm low}=10$ Hz case, a better than $\sim 30\%$ constraint on $e_0$ can be set if $e_0\gtrsim 0.1$. For a GW151226-like system a $\sim 30\%$ constraint can be set if the detector low-frequency limit is 25 Hz and $e_0 \gtrsim 0.25$. If the limit is 10 Hz, this same constraint is relaxed to $e_0 \gtrsim 0.06$. The improvement in the eccentricity measurement as the detector's low-frequency limit is reduced simply follows from the fact that binary circularization implies larger eccentricity at lower frequencies. 

\item A Bayesian MCMC parameter estimation analysis was performed for a BBH system with masses similar to GW151226. Overall, the MCMC parameter estimation results are consistent with those obtained from our Fisher matrix calculations. Marginalized posterior probability distributions for the recovered parameters for increasing binary eccentricity are shown in Fig.~\ref{fig:MCMCpdf-stat}. There we see that a clear measurement of eccentricity in a GW151226-like system with a signal-to-noise ratio $\approx 20$ requires $e_0 \gtrsim 0.1 \mbox{--} 0.2$. See Table \ref{tab:MCMCcompare_e0} for details of the comparison between the MCMC and Fisher-matrix analyses.

\item We observe a degeneracy between the chirp mass $M_{\rm ch}$ and the binary eccentricity $e_0$. This is manifested via an increasing bias in the chirp mass (lower-right panel of Fig.~\ref{fig:MCMCpdf-stat}) and via banana-shaped probability contours in the eccentricity-chirp mass plane (Fig.~\ref{fig:e0Mchellipse}).\footnote{This bias was also independently highlighted in the recent work of Ref.~\cite{OSheaKumar2021}, and is implicit in the results of \cite{lenon-nitz-brown2020MNRAS,wu-cao-zhu2020,romero-shaw-eeccGWTC2}.} We explain this degeneracy analytically via the introduction of an \emph{eccentric chirp mass} parameter $M_{\rm ch}^{\rm ecc}$, which is introduced and derived here. For low-eccentricity binaries, the eccentric chirp mass
\be
M_{\rm ch}^{\rm ecc} = \frac{M_{\rm ch}}{(1-\frac{157}{24} e_0^2)^{3/5}} 
\ee
approximately replaces the ``circular'' chirp mass $M_{\rm ch}$ as the primary ``effective parameter'' that governs the phase evolution of a compact object binary. Just as the precise measurement of $M_{\rm ch}$ for circular binaries leads to a degeneracy between the binary masses $m_1$ and $m_2$, the dominance of the eccentric chirp mass in low-eccentricity waveforms causes a degeneracy between $e_0$ and $M_{\rm ch}$. This is illustrated in Fig.~\ref{fig:m1m2chirp} and discussed in Sec.~\ref{sec:eccchirp}). We further explore the $e_0$-$M_{\rm ch}$ degeneracy via an examination of the error ellipses using both the MCMC and Fisher matrix approaches. In Sec.~\ref{sec:degen} we analytically illustrate how the $e_0$-$M_{\rm ch}$ correlation leads to a rotation of the corresponding error ellipse. 

\item For a range of compact binary systems, systematic errors from ignoring eccentricity become comparable to statistical errors for $e_0 \gtrsim 0.01 \mbox{--} 0.1$, with more massive systems at the higher end of this range (see Fig.~\ref{fig:SYSerrors}). When the detector low-frequency cutoff is at 35 Hz, neither GW150914 nor GW151226-like binaries experience any significant parameter bias for $e_0 <0.3$. Unless the eccentricity was greater than this value, circular templates were likely a very good approximation for those two LIGO detections. If the low-frequency limit is taken to 25 Hz, eccentricity-induced bias could play a role in GW151226-like systems if $e_0 \gtrsim 0.2$. At 10 Hz, parameter bias can be become an issue for both GW150914 or GW151226-like systems if $e_0 \gtrsim 0.07$. 

\item Systematic parameter bias is also investigated via a Bayesian MCMC approach for a GW151226-like system, showing a clear and significant bias for $e_0 \gtrsim 0.12$. The predicted bias agrees very well with calculations based on the Cutler-Vallisneri formalism~\cite{cutler-vallisneri-systematicerrors-PRD2007}.
\end{enumerate}

The remainder of this paper provides the details of the analysis and additional results. Section \ref{sec:waveform} discusses the details of our waveform model. It also provides explicit formulas for computing signal-to-noise ratios and frequency termination conditions. Section \ref{sec:eccchirp} introduces the notion of the eccentric chirp mass parameter and derives it from the GW phasing in the small-eccentricity limit.  Section \ref{sec:fisher} reviews the Fisher matrix parameter estimation formalism in detail, including the Cutler-Vallisneri \cite{cutler-vallisneri-systematicerrors-PRD2007} method for computing systematic errors. Approximate analytic scaling laws for both statistical and systematic errors are also provided. Section \ref{sec:statistical} presents our results for the statistical errors, focusing on the errors in the eccentricity parameter $e_0$. Errors and correlations in the other intrinsic system parameters are also discussed there. Section \ref{sec:systematic} discusses our Fisher-Cutler-Vallisneri (FCV) results for systematic parameter errors. Section \ref{sec:mcmc} reviews our Bayesian MCMC calculations, and discusses their application to computing statistical and systematic errors. We also provide a quantitative comparison between the Bayesian MCMC and FCV approaches.  Section \ref{sec:degen} explains the observed degeneracy between the chirp mass and eccentricity, via both analytical and numerical investigations using the Bayesian MCMC and Fisher matrix approaches. Section \ref{sec:concl} briefly discusses some implications and conclusions of our work. Four appendices discuss (i) the calculations shown in Table~\ref{tab:eccevol}, (ii) how cosmological redshift affects the waveform, (iii) formulas for the inspiral termination frequency, and (iv) some additional Fisher matrix results, including a discussion of parameter correlations. 

\section{Waveform model}\label{sec:waveform}
In this section we review the gravitational wave signal model that we apply in our parameter estimation calculations. Section \ref{subsec:waveformampphase} provides our waveform model in the stationary phase approximation, including the incorporation of eccentricity and spin effects. Appendix \ref{app:cosmo} reviews how cosmological effects enter the waveform. Section \ref{subsec:snr} provides some useful formulas for the signal-to-noise ratio (SNR), and Sec.~\ref{subsec:freq} discusses the termination condition for our waveforms at high frequencies. 

\subsection{\label{subsec:waveformampphase}Gravitational waveform model}
In terms of the GW polarizations $h_{+,\times}$ and the corresponding antenna pattern functions $F_{+,\times}$, the GW signal readout from the detector is
\bs
\be
\begin{split}
h(t) &= F_+ h_+(t) + F_{\times} h_{\times}(t) \\
&= A(t) \cos [2\phi(t) - 2\beta - 2\Phi_0],
\end{split}
\ee
where
\be
A(t) = -\frac{2 \eta M}{D} [v(t)]^2 \left[(1+C^2)^2 F_{+}^2 + 4 C^2 F_{\times}^2 \right]^{1/2} \, 
\ee
\be
\text{and} \;\;\;\;\; \Phi_0 = \frac{1}{2} \arctan\left[ \frac{2 F_{\times} C}{F_+ (1+C^2)} \right] \,.
\ee
\es
Here $C\equiv \cos \iota$, with $\iota$ the binary inclination angle (the angle of the Newtonian orbital angular momentum direction relative to the line from source to detector), $\beta$ specifies the azimuth angle of the Newtonian orbital angular momentum relative to a reference direction in the source frame, $D$ is the proper distance to the detector, $v(t)$ is the relative (orbit-averaged) speed of the binary, $\phi(t)$ is the orbital phase, $M=m_1+m_2$ is the sum of the component masses (we assume $m_1 \geq m_2$), and $\eta=m_1 m_2/M^2$ is the reduced mass ratio.  We will later make use of the chirp mass parameter, $M_{\rm ch}= \eta^{3/5} M$. In the above equations we are ignoring eccentricity corrections to the amplitude, which is effectively that of a circular binary and contains only one harmonic (at twice the orbital frequency). Eccentric corrections only enter the secular phase evolution $\varphi(t)$ via a low-eccentricity expansion accurate to $O(e_0^2)$ (described below). Except for the inclusion of spin effects, this waveform is equivalent to that developed in \cite{moore-etal-PRD2016}. 

To compute the Fourier transform (FT) of the signal,
\be
\tilde{h}(f) \equiv \int_{-\infty}^{\infty} h(t) e^{2\pi i f t} \, dt \,,
\label{eq:fourier2}
\ee
we use the stationary phase approximation (SPA). Following Sec.~VI E of \cite{moore-etal-PRD2016} the FT becomes
\bs
\label{eq:spawaveform}
\begin{align}
\tilde{h}(f) =& \;{\mathcal A} e^{i \Psi}\,, \;\;\;\; \text{where} \\
\label{eq:A_noz}
{\mathcal A} =& -M \sqrt{\frac{5\pi}{96}} \left(\frac{M}{D}\right) \sqrt{\eta} (\pi M f)^{-7/6} \nonumber \\
&\times \left[(1+C^2)^2 F_{+}^2 + 4 C^2 F_{\times}^2 \right]^{1/2}.
\end{align}
\es

The SPA phase can be written as a sum of several terms,
\begin{align}
\label{eq:Psiterms}
\Psi(f) = &\, \phi_c + 2\pi f t_c + \frac{3}{128 \eta v^5} \big(1 + \Delta \Psi_{\rm 3.5PN}^{\rm circ.}
+ \Delta \Psi_{\rm 4PN}^{\rm spin, \, circ.} \nonumber \\
& + \Delta \Psi_{\rm 3PN}^{\rm ecc.}  \big),
\end{align}
where $t_c$ and $\phi_c$ are the coalescence time and phase, and $v\equiv (\pi M f)^{1/3}$ is the PN orbital velocity parameter. Note that the angle $\beta$ is absorbed into a constant shift to $\phi_c$. 

The standard 3.5PN circular contribution is $\Delta \Psi_{\rm 3.5PN}^{\rm circ.} = \sum_{n=2}^{7} c_n(\eta) v^n$, where the $c_n(\eta)$ can be read off of Eq.~(3.18) of \cite{buonanno-iyer-oshsner-pan-sathya-templatecomparison-PRD2009}, and the 2.5PN and 3PN coefficients also depend on $\ln v$.

Spin effects to 4PN order are encapsulated in the term
\begin{align}
\Delta \Psi_{\rm 4PN}^{\rm spin, \, circ.} = &\; 4 \beta_{1.5} v^3 -10 \sigma v^4 \nonumber \\
&+ v^5 \ln v^3 \left[ \frac{40}{9} \beta_{2.5} - \beta_{1.5} \left( \frac{3715}{189} + \frac{220}{9}\eta \right) \right] \nonumber \\
&+ {\mathcal P}_6 v^6 + {\mathcal P}_7 v^7 +{\mathcal P}_8 v^8.
\end{align}
Here $\beta_{1.5}$ is the 1.5PN spin-orbit term \cite{kidderwillwiseman-spineffects,poisson-BHpertIV-slowrotPRD1993,kidder-spineffects}),
\be
\label{eq:beta15}
\beta_{1.5} = \sum_{i=1,2} \chi_i \kappa_i \left(\frac{113}{12} \frac{m_i^2}{M^2} + \frac{25}{4} \eta \right).
\ee 
The 2PN spin-spin term $\sigma = \sigma_{S_1 S_2} + \sigma_{\rm QM}+ \sigma_{\rm self\,spin}$  combines three effects [see Eq.~(9) of Ref.~\cite{gergely-selfspinPRD05}]. The first is the standard spin-spin interaction \cite{kidderwillwiseman-spineffects,kidder-spineffects},
\be
\sigma_{S_1 S_2} = \frac{1}{48} \eta \chi_1 \chi_2 (721 \kappa_1 \kappa_2 - 247 \gamma_{12})\;.
\ee
The second is the quadrupole-monopole term arising from corrections to the Newtonian potential caused by a spinning object's mass quadrupole moment \cite{poisson-quadrupolemonopoleterm-PRD1998},
\begin{align}
\sigma_{\rm QM} &= -\frac{5}{2} \sum_{i=1,2} p_i (3\kappa_i^2 -1) \nonumber \\
&= \frac{5}{2} \sum_{i=1,2} a_i \chi_i^2 \left( \frac{m_i}{M} \right)^2 (3\kappa_i^2 -1),
\end{align}
where $p_i=Q_i/(m_i M^2)$ for the quadrupole moment scalar $Q_i = -a_i \chi_i^2 m_i^3$, and where $a_i=1$ for BHs and $a_i \approx 4\mbox{--}8$ for neutron stars \cite{poisson-quadrupolemonopoleterm-PRD1998,laarakker-poisson-quadmomentApJ1999}.
The third term is the self-spin interaction arising from $(\text{current quadrupole})^2$ terms in the energy flux's multipole expansion \cite{gergely-selfspin-PRD2000,gergely-selfspinPRD05}:
\be
\sigma_{\rm SS-self} = \frac{1}{96} \sum_{i=1,2} \chi_i^2 \left(\frac{m_i}{M} \right)^2 (7-\kappa_i^2).
\ee
In these equations the dimensionless spin parameter $\chi_i$ is related to the individual compact object spin vectors via ${\bm S}_i = \chi_i m_i^2 \hat{{\bm s}}_i$, $\kappa_i$ is the cosine of the angle between the $i$th spin direction $\hat{{\bm s}}_i$ and the Newtonian orbital angular momentum ${\bm L}_N$ ($\kappa_i = \hat{{\bm s}}_i \cdot \hat{{\bm L}}_N$; hatted quantities denote unit vectors\footnote{Note that Ref.~\cite{gergely-selfspinPRD05} uses the notation $\kappa_i \rightarrow \cos\kappa_i$.}), and $\gamma_{12} = {\hat{\bm s}}_1 \cdot {\hat{\bm s}}_2$. All the spins used here refer to a spin-supplementary condition in which the magnitudes of the spin vectors are constant (see, e.g., \cite{faye-buonanno-luc-higherorderspinI,faye-buonanno-luc-higherorderspinII,*faye-buonanno-luc-higherorderspinIIerratum,*faye-buonanno-luc-higherorderspinIIerratum2}, or Ref.~\cite{favata-PNspinisco} where these spin vectors are denoted ${\bm S}_i^{\rm c}$).

The 2.5PN spin-orbit term $\beta_{2.5}$  is  \cite{faye-buonanno-luc-higherorderspinII,*faye-buonanno-luc-higherorderspinIIerratum,*faye-buonanno-luc-higherorderspinIIerratum2}:
\begin{align}
\label{eq:beta25SO}
\beta^{\rm SO}_{2.5} = & \sum_{i=1,2} \chi_i \kappa_i \Bigg[ \frac{m_i^2}{M^2} \left( -\frac{31319}{1008} + \frac{1159}{24}\eta \right) \nonumber \\ 
 &+ \eta \left( -\frac{809}{84} + \frac{281}{8}\eta  \right) \Bigg] \, ,
\end{align}
where we neglect BH absorption terms \cite{alvi-BHabsorptionPRD2001}. 
The 3PN, 3.5PN, and 4PN terms ${\mathcal P}_6$, ${\mathcal P}_7$, ${\mathcal P}_8$, are taken from Eqs.~(5) and (6) of Ref.~\cite{mishra-etal-spinsPRD2016}.
This analysis assumes nonprecessing (aligned) spins, so all the $\beta_{(\cdots)}$ and $\sigma_{(\cdots)}$ parameters are functions of $\chi_i$ and constant in time. Similarly, at 3PN order the spin terms contain spin-orbit effects arising from the tail contribution~\cite{BBF2011} and the 1PN correction to the leading spin-spin term~\cite{BFMP2015}.  The 3.5PN term contains the spin-orbit and cubic spin interactions in the spin dynamics~\cite{BMB2013,MarsatCubic}; the  4PN spin terms contain only spin-orbit interactions~\cite{M3B} and no other contributions. (See similar computations within the effective field theory approach~\cite{PortoEFTrev,EFTSpinOrbit,EFTSpinSpin}.) 

Leading-order in eccentricity corrections to the SPA phase were derived to 3PN order in Eq.~(6.26) of \cite{moore-etal-PRD2016}, making use of the results of prior PN modeling of eccentric binaries \cite{petersmathews,junker-schafer,gopakumar-iyer-iyer-PRD1997,arun-eccentrictailsEflux,arun-eccentricEflux3PN,arun-etal-eccentric-orbitalelements-PRD2009}. (See \cite{khalil-etalEOBeccPRD2021} and the references therein for recent work in the development of eccentric waveforms.) To display the structure we show here only the 1PN-order corrections (but we use the full 3PN expression in all our calculations):
\begin{align}
    \label{eq:Psiecc}
\Delta \Psi_{\rm 3PN}^{\rm ecc.} = & -\frac{2355}{1462} e_0^2 \left( \frac{v_0}{v} \right)^{19/3} \nonumber \\
& \times \Bigg[ 1 + v^2 \left( \frac{299\,076\,223}{81\,976\,608}  + \frac{18\,766\,963}{2\,927\,736}\eta \right) \nonumber \\
& + v_0^2 \left( \frac{2833}{1008} - \frac{197}{36} \eta \right) + \cdots + O(v^6) \Bigg].
\end{align}
Here, $e_0$ is the eccentricity at a reference frequency $f_0$, and $v_0 \equiv (\pi M f_0)^{1/3}$. The choice of $f_0$ is arbitrary; through most of this paper we set $f_0= 10$ Hz except where otherwise noted.

The $\Delta \Psi_{\rm 3PN}^{\rm ecc.}$ phase correction above ignores periodic oscillations in the phase that occur on the orbital timescale. We are also ignoring eccentricity-induced harmonics of the GW signal at frequencies other than twice the orbital frequency (these are small for low eccentricity). At large frequencies, this waveform correction (like all PN waveforms) will begin to violate the assumptions inherent in the PN expansion. Similarly, as the eccentricity becomes large, this correction will become inaccurate. These issues are discussed quantitatively in \cite{moore-etal-PRD2016} and addressed again in Sec.~\ref{sec:statistical} below. 

It is helpful to note that while the evolution of the eccentricity variable $e_t$ with time or frequency does not have a general analytic solution, an analytic solution can be found in the small-eccentricity limit:
\be
\label{eq:et-f}
e_t = e_0 \left(\frac{f_0}{f} \right)^{19/18} \left[1 + O(\pi M f)^{2/3} + O(\pi M f_0)^{2/3} \right] \;,
\ee
where the full expression with corrections to 3PN order is found in Eq.~(4.17) of \cite{moore-etal-PRD2016}. A similar expression giving the time-dependence $e_t(t)$ is
\be
\label{eq:et-t}
e_t = e_0 \left(\frac{t_c-t}{t_c-t_0} \right)^{19/48} \;,
\ee
where $t_c$ is the coalescence time, $t_0$ is the time when $e_t=e_0$, and the 3PN corrections to this formula are found in Eq.~(4.23) of \cite{moore-etal-PRD2016}. The above are helpful for understanding the decay of binary eccentricity in the small $e_0$ limit. Note that Eq.~\eqref{eq:Psiecc} already accounts for the variation of the eccentricity with time (i.e., $e_0$ is a constant parameter specifying the value of $e_t$ at the reference frequency $f_0$). 

We refer to circular waveforms in the SPA approximation as discussed above as {\tt TaylorF2} waveforms; when the eccentric correction in Eq.~\eqref{eq:Psiecc} is included, we refer to this as the {\tt TaylorF2Ecc} waveform. The {\tt TaylorF2Ecc} waveform has been coded into {\tt LALSuite} \cite{LALSuite} by the present authors \cite{TaylorF2EccURL} and reduces to the {\tt TaylorF2} waveform in the $e_0 \rightarrow 0$ limit. We note that {\tt TaylorF2Ecc} has been used for parameter estimation on LIGO data to look for evidence of eccentricity \cite{nitz-lenon-brown2020ApJ,lenon-nitz-brown2020MNRAS}.

Cosmological corrections to the waveform become important at large distances, and the incorporation of these effects is discussed in Appendix \ref{app:cosmo}. The appropriate correction to Eq.~\eqref{eq:spawaveform} is obtained by replacing $D$ with the redshift-dependent luminosity distance $d_L(z)$ [Eq.~\eqref{eq:dLz}], and also by replacing the total mass with the detector-frame mass $(1+z) M$. To emphasize these redshift-dependent corrections we redefine the FT of the GW signal in the observer's (i.e., detector's) frame as
\be
\label{eq:waveform-z-pre}
\tilde{h}(f) = {\mathcal A}_z e^{i \Psi(f)} = \hat{{\mathcal A}}_z f^{-7/6} e^{i \Psi(f)}\,, \;\;\;\; \text{where}
\ee
$\hat{\mathcal A}_z \equiv {\mathcal A}_z f^{7/6}$, with $\Psi(f)$ and ${\mathcal A}_z$ defined via making the above replacements in Eq.~\eqref{eq:spawaveform} [e.g., Eq.~\eqref{eq:Az}]. See Appendix \ref{app:cosmo} for details. Note that throughout this paper, the symbol $M$ refers to the total mass in the source frame (the frame of the binary) and $f$ refers to the frequency of the GWs measured at the detector.  

\subsection{\label{subsec:snr}Signal-to-noise ratio}
The signal-to-noise ratio (SNR) $\rho$ is defined via
\be
\rho^2 = 4 \int_{0}^{\infty} \frac{|\tilde{h}(f)|^2}{S_n(f)} \, df \,,
\ee
where $S_n(f)$ is the one-sided noise power spectral density, and all quantities are defined in the observer frame. Using Eq.~\eqref{eq:waveform-z-pre} allows us to rewrite $\rho$ as
\begin{align}
\label{eq:snrsq}
\rho^2 = &\; 4 \hat{{\mathcal A}}_z^2 \int_{f_{\rm low}}^{f_{\rm high}} \frac{f^{-7/3}}{S_n(f)} df\,, \;\;\;\; \text{where} \\
|\hat{{\mathcal A}}_z|^2 = &\; \frac{5}{96 \pi^{4/3}}  (1+z)^{5/3} \eta \frac{M^{5/3}}{d_L^2} \nonumber \\
& \times \left[ (1+C^2)^2 F_+^2 + 4 C^2 F_{\times}^2 \right] \,,
\end{align}
and we have replaced the integration limits with the low-frequency limit of  
the detector and a high-frequency cutoff determined by a waveform termination condition (see Sec.~\ref{subsec:freq} below). 
For antenna patterns $F_{+,\times}(\theta,\varphi,\psi)$ depending on sky-position angles $(\theta, \varphi)$ and a polarization angle $\psi$, and for GW polarizations $h_{+,\times}(\iota, \beta)$ depending on the direction of the Newtonian orbital angular momentum to the observer $(\iota, \beta)$ in the source frame, we can compute the angle-averaged SNR via
\be
\langle \rho^2 \rangle =  \int_0^{2 \pi} \int_0^{\pi} \frac{\sin\theta \,d\theta \, d\varphi}{4\pi}  \int_0^{\pi} \frac{d\psi}{\pi}  \int_0^{2 \pi} \int_0^{\pi} \frac{\sin\iota \,d\iota \,d\beta}{4\pi}  \rho^2.
\ee
Using $\langle F_{+,\times}^2 \rangle = 1/5$ (for interferometers with $90^{\circ}$ arms),  $\langle C^2 \rangle=1/3$, and  $\langle (1+C^2)^2 \rangle=28/15$, we find 
\begin{align}
\label{eq:snr}
\langle \rho^2 \rangle = &\; \frac{2}{15 \pi^{4/3}} (1+z)^{5/6} \frac{\eta M^{5/3}}{d_L^2} {\mathcal F}_{7/3} \,, \;\;\;\; \text{where} \\
\label{eq:F73}
{\mathcal F}_{7/3} \equiv & \;  \int_{f_{\rm low}}^{f_{\rm high}} \frac{f^{-7/3}}{S_n(f)} df \,.
\end{align}
(Since dimensionally $[S_n]= 1/{\rm Hz}$, $[{\mathcal F}_{7/3}] = 1/{\rm Hz}^{1/3}$.) 
For an ``optimally oriented'' and ``optimally located'' binary  (e.g., $\iota=\theta=\varphi=\psi=0$, $C=F_{+}=1, F_{\times}=0$), $(1+C^2)^2 F_+^2 + 4 C^2 F_{\times}^2=4$, and $\rho^2$ becomes 
\be
\rho_{\rm opt}^2 = \frac{5}{6 \pi^{4/3}} (1+z)^{5/6} \frac{\eta M^{5/3}}{d_L^2} {\mathcal F}_{7/3} \,.
\ee
Note that $\rho_{\rm opt} = \frac{5}{2} \sqrt{\langle \rho^2 \rangle}$.

\subsection{\label{subsec:freq}Frequency range}
The limits of integration in ${\mathcal F}_{7/3}$ [Eq.~\eqref{eq:F73}] and in our Fisher matrix calculation below are computed as follows: $f_{\rm low}$ is generally taken to be the fiducial seismic cutoff for the Advanced LIGO design, $f_{\rm low}=10 {\rm Hz}$, although we also consider other values as discussed below. For the high-frequency cutoff we investigated two choices: (i) a conservative choice given by the validity limit of the quasi-Keplerian approximation estimated in Eq.~(3.22) of \cite{moore-etal-PRD2016}: 
\be
\label{eq:fqK}
f_{\rm qK} = \frac{2585}{1+z} \left(\frac{1 M_{\odot}}{M} \right) \; {\rm Hz} \,,
\ee
where the factor $(1+z)$ is needed to provide the cutoff frequency at the detector in terms of the source-frame mass. (ii) As a less-conservative alternative---and in the spirit of pushing PN expansions to their limit (with the understanding that actual LIGO measurements will make use of waveforms calibrated to numerical relativity)---we choose either (a) $f_{\rm high}=1000\, {\rm Hz}$ for BNS\footnote{This value is motivated by Figs.~5 and 7 of Ref.~\cite{bernuzzi-nagar-brugmann-PRD2012-tidalNS}, which indicate that PN waveforms with tidal corrections and EOB point-particle waveforms start to develop large phase errors past frequencies $M\omega \approx 0.05$. In any case, because the LIGO sensitivity is poor close to $1000 \, {\rm Hz}$, moderate perturbations about this upper frequency cutoff do not affect our results substantially. For this reason we ignore the redshift correction in computing $f_{\rm high}$ for the BNS case.}, or (b) the ISCO frequency corresponding to the final BH formed following the merger for BBH or NS/BH systems. In case (b) we do not distinguish between BBH and NS/BH systems, treating both objects as point masses. The frequency we use is given by
\be
\label{eq:fisco}
f_{{\rm isco}, z} = \frac{1}{1+z} \frac{\hat{\Omega}_{\rm isco}(\chi_f)}{\pi M_f}\,,
\ee
where $\hat{\Omega}_{\rm isco}(\chi) \equiv M_{\rm kerr} \Omega_{\rm isco} $ is the dimensionless angular frequency for a circular-equatorial orbit around a Kerr BH with mass $M_{\rm kerr}$ and spin parameter $\chi$ \cite{bptkerr}. The dimensionless frequency $\hat{\Omega}_{\rm isco}$ depends only on a spin parameter $\chi$, which is taken to be the final spin parameter $\chi_f$ of the BH merger remnant. Similarly, the mass parameter $M_f$ entering Eq.~\eqref{eq:fisco} is the final mass of the BH merger remnant. Explicit formulas for $\hat{\Omega}_{\rm isco}$, $M_f$, and $\chi_f$ are given in Appendix \ref{app:iscoeqs}, with the latter two quantities determined by fits to numerical relativity simulations.

\section{\label{sec:eccchirp}The eccentric chirp mass}
Before introducing our parameter estimation formalism, we first discuss an important feature of eccentric waveforms which has, to our knowledge, not been previously highlighted. This concerns the identification of the appropriate ``effective mass'' parameter that enters the waveform for a coalescing compact binary. 

It is well known that for circular binaries the chirp mass $M_{\rm ch} \equiv \eta^{3/5} M = (m_1 m_2)^{3/5} (m_1+m_2)^{-1/5}$ is the combination of intrinsic system parameters that most directly governs the strength and evolution of the GW signal.  This is easily seen by examining the functional form of key quantities such as the GW luminosity (energy flux), frequency evolution, and the waveform SPA amplitude and phase at leading (0PN) order:
\bs
\label{eq:Mchdepend}
\begin{align}
&{\mathcal L}_{\rm gw}^{\rm circ, 0PN}  = \frac{32}{5} M_{\rm ch}^{10/3} (\pi f)^{10/3} \;,\\
&\frac{df}{dt} \bigg|_{\rm circ, 0PN}   = \frac{96}{5\pi} M_{\rm ch}^{5/3} (\pi f)^{11/3} \;,\\
&{\mathcal A}_{\rm circ, 0PN} \propto -\sqrt{\frac{5\pi}{96}} \frac{M_{\rm ch}^{5/6}}{D} (\pi f)^{-7/6} \;, \\
&\Psi_{\rm circ, 0PN} = \phi_c + 2\pi f t_c + \frac{3}{128 M_{\rm ch}^{5/3} (\pi f)^{5/3}}  \;. 
\end{align}
\es
For this reason the chirp mass is the parameter that is measured with the highest accuracy in systems that are inspiral dominated. For example, the detector frame chirp mass for GW170817 was measured with a fractional precision $\sim 0.05\%$ \cite{GW170817}. Because the chirp mass---not the individual system masses ($m_1, m_2)$---is the dominant parameter, this leads to a degeneracy in the $m_1$-$m_2$ plane: the chirp mass is measured precisely, but the individual system masses are not. This results in the famous chirp mass ``banana'' (see, e.g., the left panel of Figure \ref{fig:m1m2chirp}).

We extend this line of reasoning to the case of eccentric binaries. It is important to note that eccentricity is not a high-order PN effect: the effects of eccentricity modify the quantities in Eq.~\eqref{eq:Mchdepend} at leading (0PN/Newtonian) order:
\bs
\label{eq:Mchetdepend}
\begin{align}
\label{eq:dLdtecc}
&{\mathcal L}_{\rm gw}^{\rm ecc, 0PN} = {\mathcal L}_{\rm gw}^{\rm circ, 0PN}  \times \left[ \frac{1+\frac{73}{24}e_t^2 + \frac{37}{96} e_t^4}{(1-e_t^2)^{7/2}} \right]  \;,\\
\label{eq:dfdtecc}
&\frac{df}{dt}\bigg|_{\rm ecc, 0PN}  = \frac{df}{dt}\bigg|_{\rm circ, 0PN} \times \left[ \frac{1+\frac{73}{24}e_t^2 + \frac{37}{96} e_t^4}{(1-e_t^2)^{7/2}} \right]  \;,\\
\label{eq:dAdtecc}
&{\mathcal A}_{\rm ecc, 0PN} \propto {\mathcal A}_{\rm circ, 0PN} \times \left[ \frac{(1-e_t^2)^{7/4}}{\left(1+\frac{73}{24}e_t^2 + \frac{37}{96} e_t^4 \right)^{1/2}} \right]  \;, \\
\label{eq:Psiecc0PN}
&\Psi_{\rm ecc, 0PN} \propto  \frac{3}{128 M_{\rm ch}^{5/3} (\pi f)^{5/3}} \left[ 1 -\frac{2355}{1462} e_0^2 \left(\frac{f_0}{f}\right)^{19/9} \right] \;, 
\end{align}
\es
where Eqs.~\eqref{eq:dLdtecc} and Eqs.~\eqref{eq:dfdtecc} come from Eqs.~(6.7) and (6.10) of \cite{arun-etal-eccentric-orbitalelements-PRD2009} and are implicit in the work of \cite{petersmathews,peters}. The correction in Eq.~\eqref{eq:dAdtecc} comes from Eq.~(4.23) of \cite{yunes-arun-berti-will-eccentric-PRD2009} and includes additional eccentric corrections not displayed here. Note that the $e_t$ appearing in the above equations is itself a function of the frequency $e_t = e_0 F(f/f_0)$, where the behavior of the function $F(f/f_0)$ in the low-eccentricity limit is shown in Eq.~\eqref{eq:et-f}. Here $f$ represents the GW frequency of the harmonic at twice the orbital frequency. The last equation shows the 0PN SPA phase $\Psi$ from Eqs.~\eqref{eq:Psiterms} and \eqref{eq:Psiecc}.\footnote{Unlike the expressions in Eqs.~\eqref{eq:dLdtecc}\mbox{--}\eqref{eq:dAdtecc}, the 0PN SPA phase cannot be written as a \emph{relatively simple} closed-form analytic function valid for arbitrary $e_t$. An exact expression for the SPA phasing is possible at 0PN order while also including the 1PN periastron precession \cite{mikoczi-etal2012PhRvD}, but it requires the evaluation of hypergeometric functions to obtain the phasing as a function of $e_0$ and $f$. See also Eq.~(4.28) of \cite{yunes-arun-berti-will-eccentric-PRD2009} for a higher-order expansion in the small $e_0$ limit.} Clearly, these expressions show that the GW signal now depends (at leading order) on a combination of the chirp mass \emph{and} the eccentricity. 

Since most of the information in a coalescing binary's GW signal comes from the variation in the signal's phase, we focus on the parameter dependence in $\Psi(f)$ above. There we see that the eccentric correction is most important near $f\approx f_0=10 {\rm Hz}$; it becomes even larger at lower frequencies, but these are typically below the LIGO band. The size of the eccentric correction then rapidly decays for $f > f_0$. If we series expand $\Psi(f)$ near $f=f_0$, the SPA phase can be written as
\be
\begin{split}
    \label{eq:Psiexpand}
\Psi(f) \approx &\: \Psi(f_0) + (2\pi t_c - A) (f-f_0) + B (f-f_0)^2 \\ 
&+ O[(f-f_0)^3]\,,
\end{split}
\ee
where 
\begin{align}
A &= \frac{5}{128 M_{\rm ch}^{5/3} (\pi f_0)^{5/3} f_0} \left(1-\frac{157}{43}e_0^2 \right)\,, \;\;\; \text{and}\\
B &= \frac{5}{96 M_{\rm ch}^{5/3} (\pi f_0)^{5/3} f_0^2} \left(1-\frac{157}{24}e_0^2 \right)\,.
\end{align}
The first two terms in Eq.~\eqref{eq:Psiexpand} [$\propto (f-f_0)$] shift the values of $t_c$ and the phase constant $\phi_c$. An examination of the $B$-term above suggests that the ``effective mass'' parameter that dominates the phase evolution near $f_0$ is an ``eccentric chirp mass'' parameter 
\bs
\begin{align}
\label{eq:eccchirp}
M_{\rm ch}^{\rm ecc} &\equiv \frac{M_{\rm ch}}{(1-\frac{157}{24} e_0^2)^{3/5}} \;, \\
\label{eq:eccchirpb}
&\approx M_{\rm ch} \left( 1+\frac{157}{40} e_0^2 \right) \;.
\end{align}
\es
This effective parameter is approximate as it relies on the small $e_0$ limit and an expansion of the 0PN phasing near $f=f_0$.\footnote{Unlike the circular case, this approximate combination does not appear in the amplitude, where a similar series expansion near ${f \approx f_0}$ of Eq.~\eqref{eq:dAdtecc} yields ${{\mathcal A}_{\rm ecc, 0PN} \propto M_{\rm ch}^{5/6} f_0^{-7/6} \left( 1- \frac{157}{48} e_0^2  \right) + O(f-f_0)}$, suggesting ${M_{\rm ch}^{\rm ecc} = M_{\rm ch}  \left( 1- \frac{157}{48} e_0^2  \right)^{6/5} \approx M_{\rm ch}  \left( 1- \frac{157}{40} e_0^2  \right)}$. Note the sign difference relative to Eq.~\eqref{eq:eccchirpb}.} Still, it is a helpful conceptual tool for understanding the parameter dependence of eccentric waveforms. 

Just as an accurate measurement of $M_{\rm ch}$ in a circular binary yields a comparatively larger spread in the range of $m_1$ and $m_2$ (see the left panel of Fig.~\ref{fig:m1m2chirp}), $M_{\rm ch}^{\rm ecc}$ plays an analogous role for signals with small eccentricity: The parameter $M_{\rm ch}^{\rm ecc}$ effectively governs the phase evolution and is precisely measured, suggesting a degeneracy between the ``circular chirp mass'' $M_{\rm ch}$ and $e_0$. This is illustrated in the right panel of Fig.~\ref{fig:m1m2chirp}. There we see that contours of constant $M_{\rm ch}^{\rm ecc}$ form a parabolic shape, $M_{\rm ch} \approx M_{\rm ch}^{\rm ecc}  \left( 1- \frac{157}{40} e_0^2  \right)$, analogous to the circular chirp mass ``banana'' plot. In Sec.~\ref{sec:degen} below we will return to this as a means of explaining the increasing chirp mass bias seen in the lower-right panel of Fig.~\ref{fig:MCMCpdf-stat}.  

While this paper was being finalized, we learned of an independent work~\cite{EccMcBurst} that also discusses the notion of an effective chirp mass for eccentric binaries, but in the context of burst searches using time-frequency maps. Their effective parameter is a phenomenological polynomial function to $O(e^6)$, based on fitting the leading order (0PN) frequency evolution. Our eccentric chirp mass parameter is analytically derived directly from the $O(e^2)$ corrections to the SPA phasing, which is the function more directly relevant for parameter estimation.

\section{\label{sec:fisher}Fisher Matrix Formalism for Statistical and Systematic Parameter Estimation}
The Fisher information matrix is widely discussed in the GW literature (for a small selection of examples, see e.g., Refs.~\cite{finn-PRD1992,finn-chernoff-PRD1993,flanagancutler,poisson-will-2PNparameterestimate,arun-etal-PRD2005-35PNparameterestimation,*arun-etal-PRD2005-35PNparameterestimation-errata,berti-buonanno-will-PRD2005,vandenbroeck-sengupta-BBHspectro-CQG2007,ajith-bose-PRD2009}). It yields accurate results only in the high SNR limit. (For a discussion of its limitations see Refs.~\cite{vallisneri-fisherabuse-PRD2008,rodriguez-farr-farr-mandel-fisherinadequacies}.) Here we apply the Fisher matrix as a way to make fast but somewhat crude assessments of the parameter estimation capabilities of ground-based detectors. A comparison between the Fisher matrix approach and parameter estimation via {\tt LALInferenceMCMC} is performed in Sec.~\ref{sec:mcmc}. Our goal in this section is to provide a brief but clear and explicit review of the Fisher matrix approach, including a discussion of how systematic parameter biases can be computed. We also derive some simple analytic scaling estimates for statistical and systematic parameter errors.  

\subsection{\label{subsec:Fisherstat}Computing statistical errors with the Fisher matrix}
Here we largely follow the presentation in Ref.~\cite{poisson-will-2PNparameterestimate}. For stationary, Gaussian noise and in the limit of large SNR, the probability of detecting the parameter set ${\bm \theta} =[\theta^a]$ given some detector data $d(t)$ is
\be
p({\bm \theta}|d) \propto p_0({\bm \theta}) \exp\left[ -\frac{1}{2} \Gamma_{ab} (\theta^a -\hat{\theta}^a) (\theta^b -\hat{\theta}^b) \right],
\ee
where $\Delta \theta^a \equiv \theta^a -\hat{\theta}^a$, $\hat{\theta}^a$ are the parameter values that maximize the probability distribution function (PDF), and $p_0({\bm \theta})$ is the prior probability that the signal is characterized by the values ${\bm \theta}$. The Fisher matrix is given by
\be
\Gamma_{ab} = \left(\frac{\partial h}{\partial \theta^a} \bigg| \frac{\partial h}{\partial \theta^b} \right)
\ee
and is evaluated at the maxima $\hat{\theta}^a$. The $(\cdots|\cdots)$ refers to the standard waveform inner product weighted by the detector noise $S_n$,
\be
\label{eq:innerprod}
\left(a | b \right) = 2 \int_{f_{\rm low}}^{f_{\rm high}} \frac{df}{S_n(f)} \left[ \tilde{a}(f) \tilde{b}^{\ast}(f) +  \tilde{a}^{\ast}(f) \tilde{b}(f)  \right] \,,
\ee
where $\ast$ denotes complex conjugation.

We assume that our prior knowledge of the model parameters corresponds to a Gaussian distribution about values $\bar{\theta}^a$,
\be
 p_0({\bm \theta}) \propto \exp\left[ -\frac{1}{2} \Gamma^0_{ab} (\theta^a-\bar{\theta}^a) (\theta^b - \bar{\theta}^b) \right],
\ee
where the $\bar{\theta}^a$ need not be the same as $\hat{\theta}^a$. If the difference is negligible ($\bar{\theta}^a \approx \hat{\theta}^a$, which we assume), then the posterior distribution $p({\bm \theta}|d)$ is peaked at $\hat{\theta}^a$ and the covariance matrix is given by the matrix inverse of the sum of the matrices: 
\be
\Sigma_{ab} \equiv {\rm E}[ (\theta_a -\hat{\theta}_a) (\theta_b -\hat{\theta}_b) ] = (\Gamma_{ab}+\Gamma_{ab}^0)^{-1} \,,
\ee
where ${\rm E}[\;]$ denotes the expectation value.
The $1$-sigma statistical measurement error in the parameter $\theta^a$ is then
\be
\sigma_a = \sqrt{\Sigma_{aa}}
\ee
(with no summation over repeated indices). (More accurately, $\sqrt{\Sigma_{aa}}$ represents the Cram\'{e}r-Rao lower-bound on the $1$-sigma error.) 
The correlation between the parameters $\theta^a$ and $\theta^b$ is given by the correlation matrix
\be
\label{eq:cAB}
c_{ab} = \frac{\Sigma_{ab}}{\sigma_a \sigma_b} \,.
\ee
The $n$-sigma error ellipsoid is given by the equation
\be
\label{eq:errorellipsoid}
(\Gamma_{ab} + \Gamma^0_{ab}) (\theta^a -\hat{\theta}^a) (\theta^b -\hat{\theta}^b) = n^2 \,.
\ee

For waveforms in the restricted SPA form [i.e., Eq.~\eqref{eq:waveform-z-pre}, but dropping the $z$ label below for simplicity] and for which ${\mathcal A}\equiv \hat{\mathcal A} f^{-7/6}$ and $\Psi$ are real, the Fisher matrix simplifies to:
\begin{align}
\label{eq:fisher2}
&\Gamma_{ab} = 4 \int_{f_{\rm low}}^{f_{\rm high}} \frac{  f^{-7/3} df}{S_n(f)} \left( \partial_a \hat{{\mathcal A}} \partial_b \hat{{\mathcal A}} + \hat{{\mathcal A}}^2 \partial_a \Psi \partial_b \Psi \right), \\
&= \frac{\rho^2}{{\mathcal F}_{7/3}} 
 \int_{f_{\rm low}}^{f_{\rm high}} \frac{ f^{-7/3} df}{S_n(f)} \left( \delta_{a, \ln \hat{\mathcal A}} \delta_{b, \ln \hat{\mathcal A}}  + \partial_a\Psi \partial_b \Psi \right),
\end{align}
where $\partial_a \equiv \partial/\partial \theta^a$, and in the second line we used Eq.~\eqref{eq:snrsq} (taking $\ln \hat{\mathcal A}$ to be one of our parameters). For our purposes here (where we are not interested in constraining the distance to the source or the sky position), $\hat{\mathcal A}$ is not a parameter of interest. Since it decouples completely from the rest of the Fisher matrix we need only consider the inner block of the matrix that depends on the derivative of $\Psi$,
\be
\label{eq:GammaAB}
\Gamma_{AB} = \frac{\rho^2}{{\mathcal F}_{7/3}} 
 \int_{f_{\rm low}}^{f_{\rm high}} \frac{ f^{-7/3} df}{S_n(f)}  \partial_A \Psi \partial_B \Psi \,,
\ee 
where the capital indices span the parameters $\theta_a$ but excluding $\ln \hat{\mathcal A}$. (Note that we raise and lower indices via the Kronecker delta, so $\theta_a = \theta^a$ and repeated indices denote summation except where stated otherwise.)

We incorporate Gaussian priors by adding to the diagonal elements of our Fisher matrix terms of the form $\Gamma^0_{AA} = 1/(\delta \theta_A)^2$. The $\delta \theta_A$ are assigned a size corresponding to the parameter's maximum variation from zero. Our parameter set consists of
\be
\label{eq:theta}
\theta_A = (t_c, \phi_c, \ln M, \ln \eta, \chi_1, \chi_2, \ln e_0)\,.
\ee
In our analysis below we use priors on the parameters\footnote{To apply a sensible prior to $e_0$, we actually use $e_0$ (rather than $\ln e_0$) as the parameter in our Fisher matrix, and later compute the fractional error as $\sigma_{\ln e_0} = \sigma_{e_0}/e_0$. Removing the prior on $e_0$ has very little effect on our results. A similar approach can be used to apply a prior on $\eta \in [0,0.25]$: $\delta \eta = 0.25$. Doing so results in a very small improvement to our reported errors (in the third or fourth digit), so we leave out this prior for simplicity.}  $\phi_c \in [-\pi, \pi]$, $\chi_{1,2} \in [-1,1]$, and $e_0 \in [0,1]$, corresponding to
\be
\label{eq:priors}
\delta \phi_c = \pi\,, \;\;\; \delta \chi_{1,2} = 1\,, \;\;\; \delta e_0 =1\,.
\ee
The prior on $\phi_c$ has the most significance on reducing our errors, with the $\chi_{1,2}$ priors having a smaller effect (and little effect on $\sigma_{e_0}$). We note that the statistical errors on the remaining parameters do not change if we replace the $\ln M$ parameter with $\ln M_{\rm ch}$.

In all our calculations we use a fit to the zero-detuned high-power LIGO sensitivity found in Eq.~(4.7) of Ref.~\cite{ajith-spin-PRD2011}. Cosmological effects are incorporated by using expressions for $\rho$, $\Psi$, and $f_{\rm high}$ corrected by the appropriate factors of $(1+z)$ as discussed in Appendix \ref{app:cosmo}. Mass parameters listed in this paper refer to the source-frame masses unless otherwise noted.

\subsection{\label{subsec:sys}Systematic error formalism}
The Fisher matrix allows us to determine the parameter errors due to the random (statistical) error associated with the detector noise. In addition to this source of error, there is also a systematic error due to possible inaccuracies in our waveform model. Cutler and Vallisneri~\cite{cutler-vallisneri-systematicerrors-PRD2007} developed a formalism to compute these systematic errors, which we apply to our Fisher matrix analysis here (and denote as the FCV formalism).  

Working in the restricted SPA, consider an \emph{approximate} waveform 
\be
\tilde{h}_{\rm AP} = {\mathcal A}_{\rm AP} e^{i\Psi_{\rm AP}} = \hat{\mathcal A}_{\rm AP} f^{-7/6} e^{i\Psi_{\rm AP}}\,
\ee
and a \emph{true} waveform 
\be
\tilde{h}_{\rm T} = ({\mathcal A}_{\rm AP} + \Delta {\mathcal A}) e^{i(\Psi_{\rm AP} + \Delta \Psi)}
\ee
that differs from the approximate model by small corrections to the amplitude $\Delta {\mathcal A}$ and to the phase $\Delta \Psi$. The systematic error $\Delta \theta_a \equiv \theta^{\rm T}_a - \hat{\theta}_a$ due to waveform modeling uncertainty in the parameter $\theta_a$ is the difference between the ``true'' value of the parameter ($\theta^{\rm T}_a$, i.e., the value given a waveform without modeling errors) and the recovered parameter $\hat{\theta}_a$ (i.e., the peak of the recovered Gaussian PDF).
From Eq.~(29) of \cite{cutler-vallisneri-systematicerrors-PRD2007}, the systematic error $\Delta \theta_a$ is given by
\be
\label{eq:Deltatheta}
\Delta \theta_a \approx \Sigma_{ab} \left([\Delta {\mathcal A} + i {\mathcal A}_{\rm AP} \Delta \Psi] e^{i\Psi_{\rm AP}} | \partial^b h_{\rm AP} \right) \,,
\ee
where $\Sigma_{ab}$ is computed from $h_{\rm AP}$ and all terms on the right-hand side are evaluated at $\hat{\theta}_a$ (the best fit parameter values determined using the approximate waveform $h_{\rm AP}$). Ignoring any systematic errors that enter the amplitude ($\Delta {\mathcal A}=0$) and using Eq.~\eqref{eq:snrsq}, Eq.~\eqref{eq:Deltatheta} simplifies to
\be
\Delta \theta_A = \frac{\rho^2}{{\mathcal F}_{7/3}} \Sigma_{AB}  \int_{f_{\rm low}}^{f_{\rm high}} \frac{f^{-7/3} df}{S_n(f)}  \Delta \Psi \partial^B \Psi_{\rm AP},
\ee
where we have replaced lower-case Latin indices with capital letters to emphasize (as before) that we do not include the amplitude as a parameter. Notice that since $\Sigma_{AB} \propto \rho^{-2}$, $\Delta \theta_A$ is independent of the SNR $\rho$.

\subsection{\label{subsec:scaling}Scaling estimates}
Ignoring parameter correlations, we can make simple scaling estimates of the statistical errors using the crude approximation $\sigma_a \approx 1/\sqrt{\Gamma_{aa}} \sim 1/(\rho \partial_a \Psi)$. Dropping the numerical coefficients arising from the integration, we find the scalings:
\bs
\begin{align}
\frac{\sigma_M}{M} &\sim \frac{\sigma_{\eta}}{\eta} \propto \frac{\eta}{\rho} (M f_c)^{5/3} \,, \\
\label{eq:sigmae0}
\frac{\sigma_{e_0}}{e_0} &\propto \frac{\eta}{\rho}(M f_c)^{5/3} \frac{1}{e_0^2} \left(\frac{f_c}{f_0}\right)^{19/9}  \,,
\end{align}
\es
where $f_c$ is a characteristic frequency scale that enters on dimensional grounds (but does not appear explicitly when the full numerical integration is performed). We note that the number of wave cycles in the detector band scales as 
\be
\Delta N_{\rm cyc} \equiv \frac{1}{\pi} [ \phi(f_{\rm high}) - \phi(f_{\rm low}) ] \sim \frac{1}{\eta (M f_{\rm low})^{5/3}}
\ee
at leading order and 
\be
\Delta N_{\rm cyc, ecc} \sim  -\frac{e_0^2 (f_0/f_{\rm low})^{19/9}}{\eta (M f_{\rm low})^{5/3}}
\ee
for eccentric terms. If $f_c \approx f_{\rm low}$, this implies that statistical errors scale like
\bs
\label{eq:sigmae0-scaling}
\begin{align}
\frac{\sigma_{a}}{\theta_{a}} &\sim \frac{1}{\rho \Delta N_{\rm cyc}} \,, \;\; (a = M, \eta) \\
\frac{\sigma_{e_0}}{e_0} &\sim \frac{1}{\rho \Delta N_{\rm cyc, ecc}} \sim \frac{\sigma_a/\theta_a}{e_t(f_c)^2} \,.
\end{align}
\es 
[In the last term we made use of Eq.~\eqref{eq:et-f}.] Hence, larger SNR and more cycles in band suggests smaller statistical errors (as is widely understood).

Analogous estimates of the systematic error on a parameter $\theta_a$ can be crudely approximated via
\begin{align}
\Delta \theta_a &\sim \frac{\rho^2}{{\mathcal F}_{7/3}} \Sigma_{aa}  \int_{f_{\rm low}}^{f_{\rm high}} \frac{f^{-7/3} df}{S_n(f)}  \Delta \Psi \partial^a \Psi_{\rm AP}\,, \nonumber \\
&\sim \rho^2 \sigma_a^2 \Delta \Psi {\partial_a \Psi_{\rm AP}}  \sim \frac{\Delta \Psi}{\partial_a \Psi_{\rm AP}} \sim \rho \sigma_a \Delta \Psi\,,
\end{align}
where there is no sum on the index $a$ above.
If systematic errors result from ignoring eccentricity [$\Delta \Psi \propto \Delta \Psi^{\rm ecc}_{\rm 3PN}/(\eta v^5)$], then we can crudely approximate the fractional bias in the mass parameters via
\begin{align}
\label{eq:sys-scaling}
\frac{\Delta M}{M} &\sim \frac{\Delta \eta}{\eta} \sim  e_0^2 \left( \frac{f_0}{f_c} \right)^{19/9} \sim e_t(f_c)^2 \\
&\sim \frac{\sigma_a/\theta_a}{\sigma_{e_0}/e_0} \sim \frac{\Delta N_{\rm cyc, ecc}}{\Delta N_{\rm cyc}} \,.
\end{align}
This is clearly SNR independent and scales like the fractional number of cycles due to the eccentric corrections. In the next two sections we explicitly evaluate and compare these statistical and systematic errors.

\section{\label{sec:statistical} Results: Statistical errors on eccentricity via the Fisher matrix }
Using the Fisher matrix formalism discussed above, we compute the $1$-sigma parameter estimation errors for the parameter set given in Eq.~\eqref{eq:theta}. We do this for a variety of ``fiducial'' GW sources and show the results for the fractional error in $e_0$ in the left panel of Figure \ref{fig:PEe0results}. For the systems with neutron stars, we take the NS dimensionless spin to be $\chi_i = 0.01$ and its quadrupole-monopole parameter to be $a_i=5$. For all BHs we take the dimensionless BH spins to be $\chi_i = 0.5$ and $a_i=1$. Throughout this paper we assume that all spins are aligned with each other and with the angular momentum axis ($\kappa_i=\gamma_{12}=1$). We consider one BNS system, one NS/BH system, and four binary BH (BBH) systems, and we summarize their parameters as follows:
\begin{enumerate}[(i)]
    \item NS/NS (BNS): $m_1=1.4 M_{\odot}$, $m_2=1.25 M_{\odot}$ at $100$ Mpc ($z=0.02227$). We integrate to $1000$ Hz and obtain an angle-averaged SNR [via Eq.~\eqref{eq:snr}] of $13.85$. 
    \item NS/BH:  $m_1=15 M_{\odot}$, $m_2=1.4 M_{\odot}$ at $200$ Mpc ($z=0.04384$). We integrate to $493.5$ Hz [following from Eq.~\eqref{eq:fisco}] and obtain an SNR  of $17.87$. 
    \item BBH1:  $m_1=10 M_{\odot}$, $m_2=5 M_{\odot}$ at $500$ Mpc ($z=0.1051$). We integrate to $723.3$ Hz [again following from Eq.~\eqref{eq:fisco}] and obtain an SNR  of $11.81$.
    \item BBH2:  $m_1=15 M_{\odot}$, $m_2=10 M_{\odot}$ at $500$ Mpc. We integrate to $455.6$ Hz and obtain an SNR of $18.64$.
    \item BBH3:  $m_1=25 M_{\odot}$, $m_2=15 M_{\odot}$ at $500$ Mpc. We integrate to $280.3$ Hz and obtain an SNR of $26.76$.
    \item BBH4:  $m_1=35 M_{\odot}$, $m_2=25 M_{\odot}$ at $500$ Mpc. We integrate to $191.5$ Hz and obtain an SNR of $37.18$.    
\end{enumerate}
In all cases above we assume a single detector, and we start our integration at $10$ Hz. The masses listed above are source-frame masses. 

From the left-panel of Figure \ref{fig:PEe0results} we make the following observations:
\begin{enumerate}[(a)]
    \item First, the fractional error in $e_0$ improves (i.e., decreases) for systems with more cycles in band. E.g., we see that BNS and NS/BH binaries have the smallest $\sigma_{e_0}/e_0$, while the more massive BBHs have the largest. This agrees qualitatively with the $\sigma_{e_0}/e_0 \sim 1/\Delta N_{\rm cyc}$ scaling predicted in Eq.~\eqref{eq:sigmae0-scaling}. As predicted by Eq.~\eqref{eq:sigmae0}, the slope of the lines in that figure are very close to $-2$.
    \item Second, to obtain meaningful constraints on eccentricity with LIGO (e.g., $\sigma_{e_0}/e_0 \lesssim 0.3$), the eccentricity at 10 Hz must be in the range $e_0 \gtrsim 0.01$ to $0.1$, depending on the system masses. In particular, we find that the fractional error in eccentricity decreases below $20\%$ when $e_0$ exceeds $0.013$ for BNS, $0.028$ for NS/BH binaries, $0.05$ for BBH1, $0.065$ for BBH2, $0.08$ for BBH3, and $0.1$ for BBH4.  
\end{enumerate}

Note that the statistical parameter errors discussed here (and throughout this paper) can be scaled to other SNR values using the fact that $\sigma_{\theta_i} \propto 1/\rho$. Those errors can be \emph{approximately} scaled to other distances using $\sigma_{\theta_i} \propto d_L$, however this scaling is not exact as it ignores the redshift dependence of the masses and the upper-frequency cutoff.

Appendix \ref{app:otherstaterrors} shows additional parameter estimation results for the systems shown in the left panel of Fig.~\ref{fig:PEe0results}. Figure \ref{fig:PEotherresults} shows that the statistical errors on $M$, $\eta$, and $\chi_{1,2}$ are largely independent of $e_0$, except for $\chi_2$ (the spin parameter of the secondary) for systems with unequal mass ratios. This is expected as the circular piece of the waveform phasing largely determines the mass parameter errors in the low-eccentricity limit, and our phasing does not contain ``mixed'' terms of the form $\sim O(e_0^n \chi_{1,2}^m)$. Figure \ref{fig:corr} shows the correlation coefficient $c_{AB}$ [Eq.~\eqref{eq:cAB}] of $e_0$ with the other intrinsic system parameters. Those correlations are weak, with $|c_{e_0 \theta_A}|\lesssim 0.5$ in most cases.

In the right panel of Fig.~\ref{fig:PEe0results} we show the fractional error in $e_0$ for two systems with parameters similar to the first two LIGO detections, GW150914 \cite{detectionPRL2016} and GW151226 \cite{gw151226-PRL2016}. These systems are chosen as their masses are representative examples of the high mass and low mass LIGO-Virgo BBHs. (Systems with masses higher than GW150914 will have very few GW cycles in the inspiral phase and are less amenable to an analysis of binary eccentricity.)  For these two systems we use the following parameters consistent with the observations as reported in Table IV of \cite{O1BBH}: 
\begin{enumerate}[(i)]
    \item GW150914: $m_1^{\rm det} = 39.4 M_{\odot}$, $m_2^{\rm det} = 31.7 M_{\odot}$, $\chi_1 = 0.32$, $\chi_2 = 0.48$, $\rho = 23.6$. We integrate to $f_{\rm isco} = 166.2$ Hz.
    \item GW151226: $m_1^{\rm det} = 15.6 M_{\odot}$, $m_2^{\rm det} = 8.2 M_{\odot}$, $\chi_1 = 0.49$, $\chi_2 = 0.52$, $\rho = 13.0$. We integrate to $f_{\rm isco} = 506.7$ Hz.
\end{enumerate}    
In both cases we use the LIGO noise curve as before (not the actual detector noise at the time of detection), and we vary the low-frequency cutoff as indicated in the figure. The masses given above are detector-frame masses. Rather than specify a distance, we assume SNRs (listed above) corresponding to the actual two-detector network SNR of the actual detections (see, e.g., Table I of \cite{O1BBH}).\footnote{For a network with $N$ detectors with identical orientations and sensitivities, the Fisher matrix parameter errors for the network is computed by evaluating the Fisher matrix for a single detector using the network SNR, where the network SNR is the quadrature sum of the individual SNRs, $\rho^2_{\rm network} = \sum_i^{N} \rho_i^2$.} The same waveform model is used as in the left panel of Fig.~\ref{fig:PEe0results}, again truncated at the ISCO corresponding to the final mass and spin as determined via the formulas in Sec.~\ref{subsec:freq} and Appendix \ref{app:iscoeqs} (setting $z=0$ when we make use of the detector-frame masses). 

We note that the low-frequency sensitivity at the time of the actual detections was close to $\approx 35$ Hz. As we are using the observed SNRs (not the larger SNRs that would be seen by an identical system at the same distance observed by LIGO at its final design sensitivity), the curves labeled $35$ Hz are approximate constraints on the measurement precision of $e_0$ that could be set at the time of detection assuming that the source had the value of $e_0$ given on the x-axis of Fig.~\ref{fig:PEe0results}. The curves labeled $25$ Hz and $10$ Hz show the same calculation using a lower value for the low-frequency limit of the integration in Eq.~\eqref{eq:GammaAB}. We clearly see that decreasing the low-frequency limit significantly increases the measurement precision of $e_0$. This naturally follows from the fact that $e_0$ is defined to be the value at $10$ Hz in all cases (regardless of the choice of $f_{\rm low}$). Since the instantaneous eccentricity varies as in Eq.~\eqref{eq:et-f}, a detector with a lower frequency limit is probing the waveform when the eccentricity (for a given value of $e_0$ fixed at 10 Hz) is higher. Since the measurement precision improves with larger instantaneous eccentricity [Eq.~\eqref{eq:sigmae0-scaling}], we expect that lowering the detector noise floor from $35$ Hz to $10$ Hz will increase the eccentricity measurement precision by a factor $\sim (35\, {\rm Hz}/10\, {\rm Hz})^{19/9} \approx 14$ (although the precise number will also depend on how the detector sensitivity varies at lower frequencies).

From the right panel of Fig.~\ref{fig:PEe0results} we see that both GW150914 and GW151226 have very poor measurement precision for $e_0$ in the 35 Hz case (or even the 25 Hz case, unless $e_0 \gtrsim 0.25$ for GW151226). However, if the detector sensitivity goes down to 10 Hz, then measurements with modest precision are possible if $e_0 \gtrsim 0.1$ for GW150914-like systems or $e_0 \gtrsim 0.05$ for GW151226-like systems. 
\begin{figure}[t]
\includegraphics[angle=0, width=0.48\textwidth]{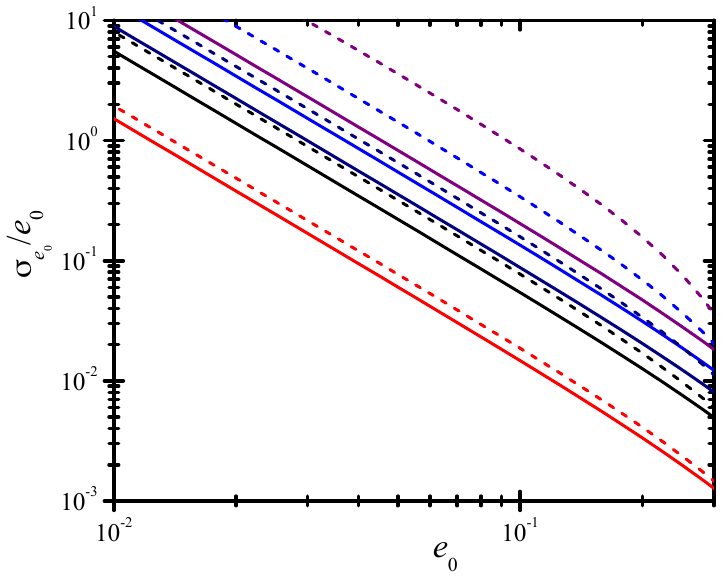}
\caption{\label{fig:PEfreqcutoff}Effect of changing the high-frequency cutoff of the integration entering the Fisher matrix. The color scheme is the same as the left panel of Fig.~\ref{fig:PEe0results}. The same systems are shown here (using the same color scheme), except for the NS/NS case. The solid curves are the same curves as in Figure \ref{fig:PEe0results}, which use $f_{\rm high}=f_{{\rm isco},z}$. The dashed curves use $f_{\rm high} = f_{\rm qK}$. From left to right the solid (or dashed) curves correspond to the systems NS/BH, BBH1, BBH2, BBH3, BBH4.}
\end{figure}
\begin{figure*}[th]
$
\begin{array}{cc}
\includegraphics[angle=0, width=0.48\textwidth]{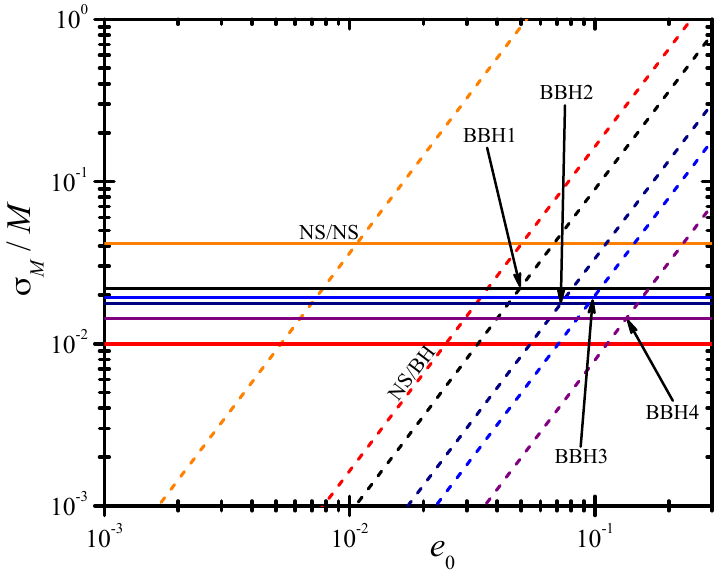} &
\includegraphics[angle=0, width=0.48\textwidth]{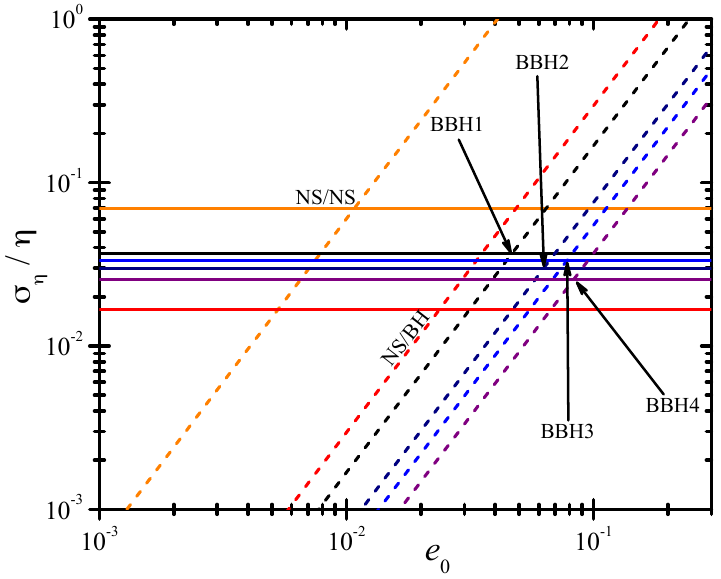} \\
\includegraphics[angle=0, width=0.48\textwidth]{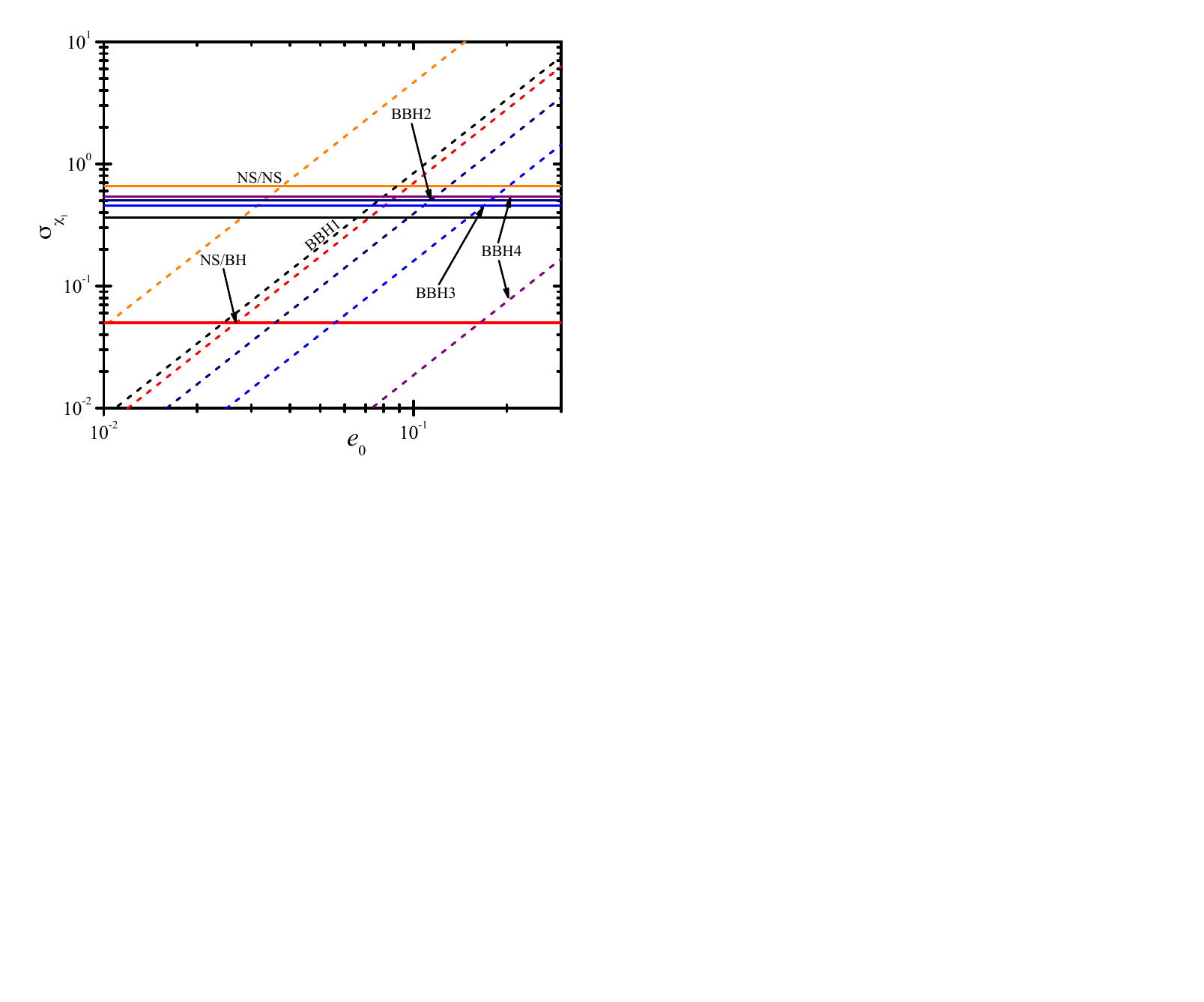} &
\includegraphics[angle=0, width=0.48\textwidth]{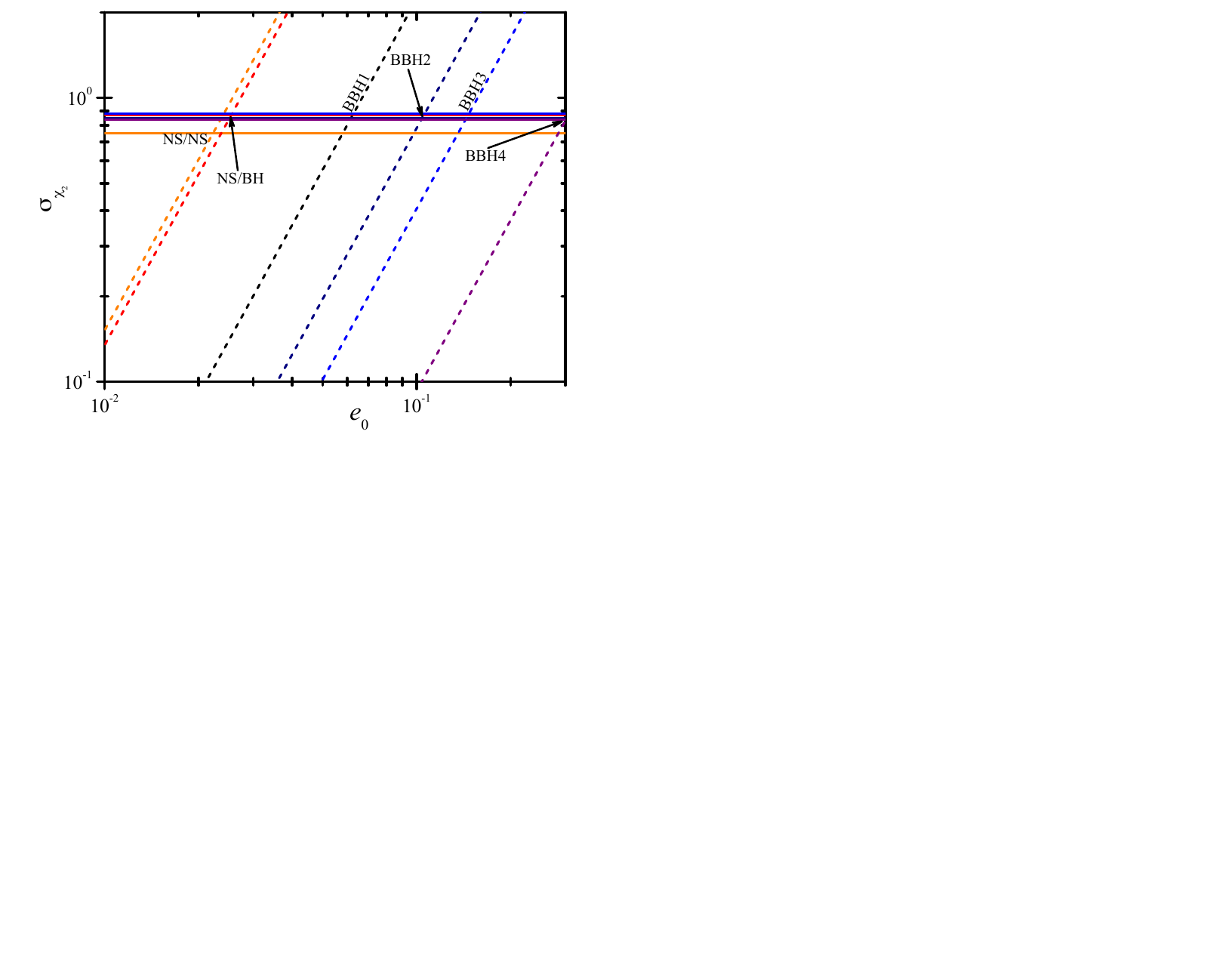}
\end{array}
$
\caption{\label{fig:SYSerrors}Systematic and statistical errors for fiducial binary systems considered in the left panel of Fig.~\ref{fig:PEe0results}. We show $1$-sigma fractional errors for $M$ and $\eta$, and $1$-sigma errors for $\chi_{1,2}$, both as a function of $e_0$. Statistical errors are shown as solid horizontal lines. Systematic errors are shown as upward-sloping dashed lines, with slopes $\approx 2$ consistent with the scaling in Eq.~\eqref{eq:sys-scaling}. The color scheme is the same as the left panel of Fig.~\ref{fig:PEe0results}. Labels for the various systems are placed such that they point to or are near the intersection of the statistical and systematic error curves for a given system.}
\end{figure*}
\begin{figure*}[t]
$
\begin{array}{cc}
\includegraphics[angle=0, width=0.48\textwidth]{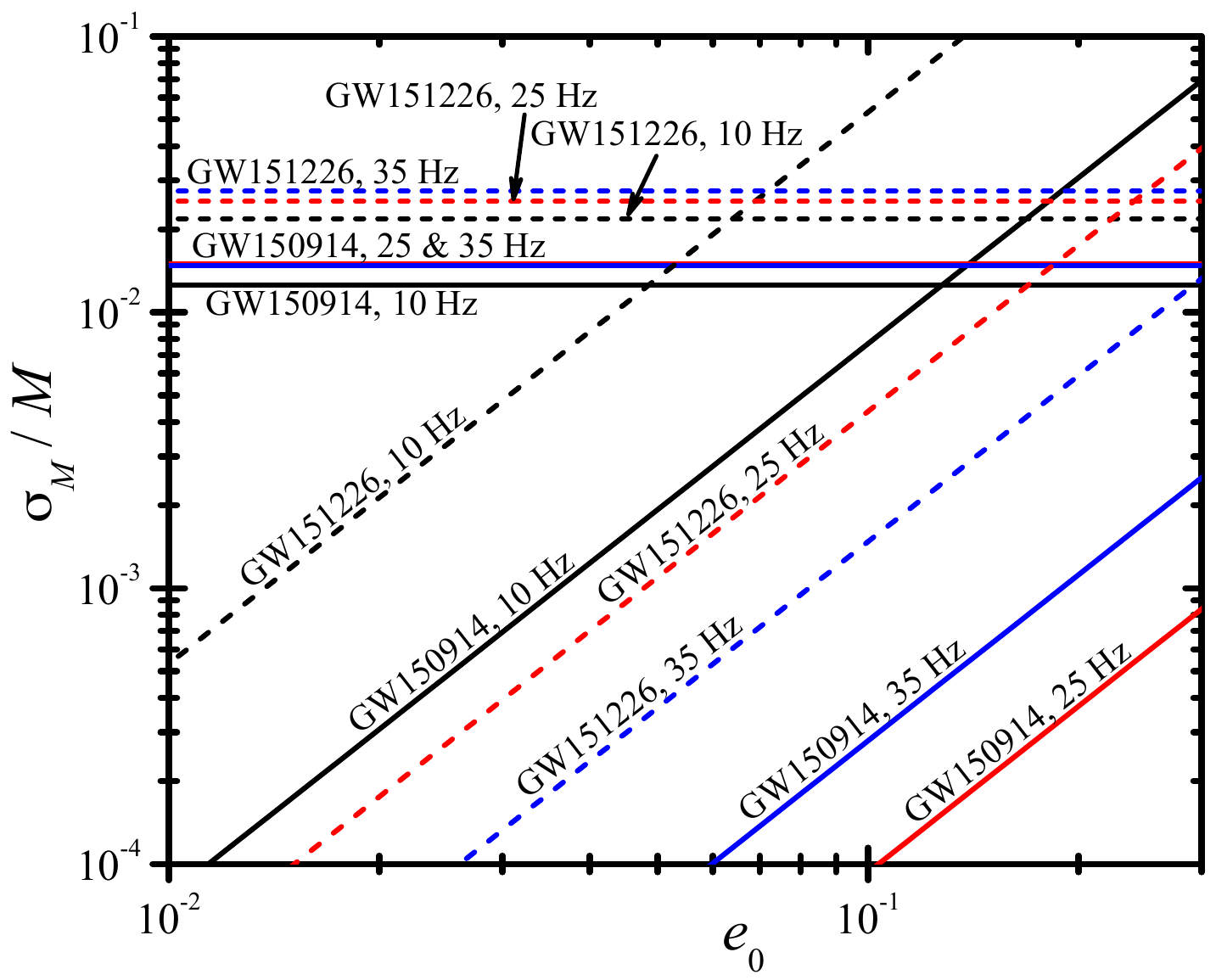} &
\includegraphics[angle=0, width=0.48\textwidth]{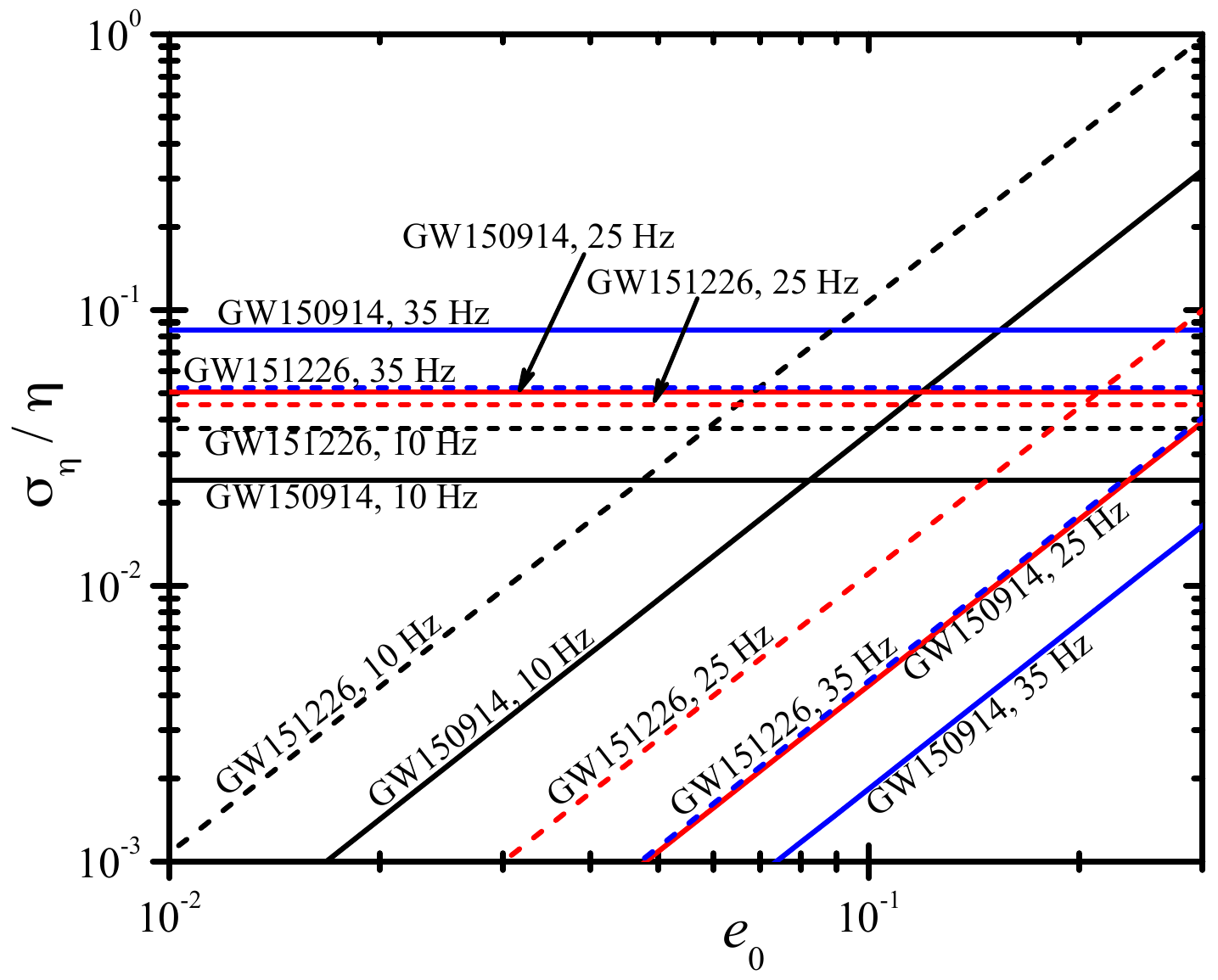} \\
\includegraphics[angle=0, width=0.48\textwidth]{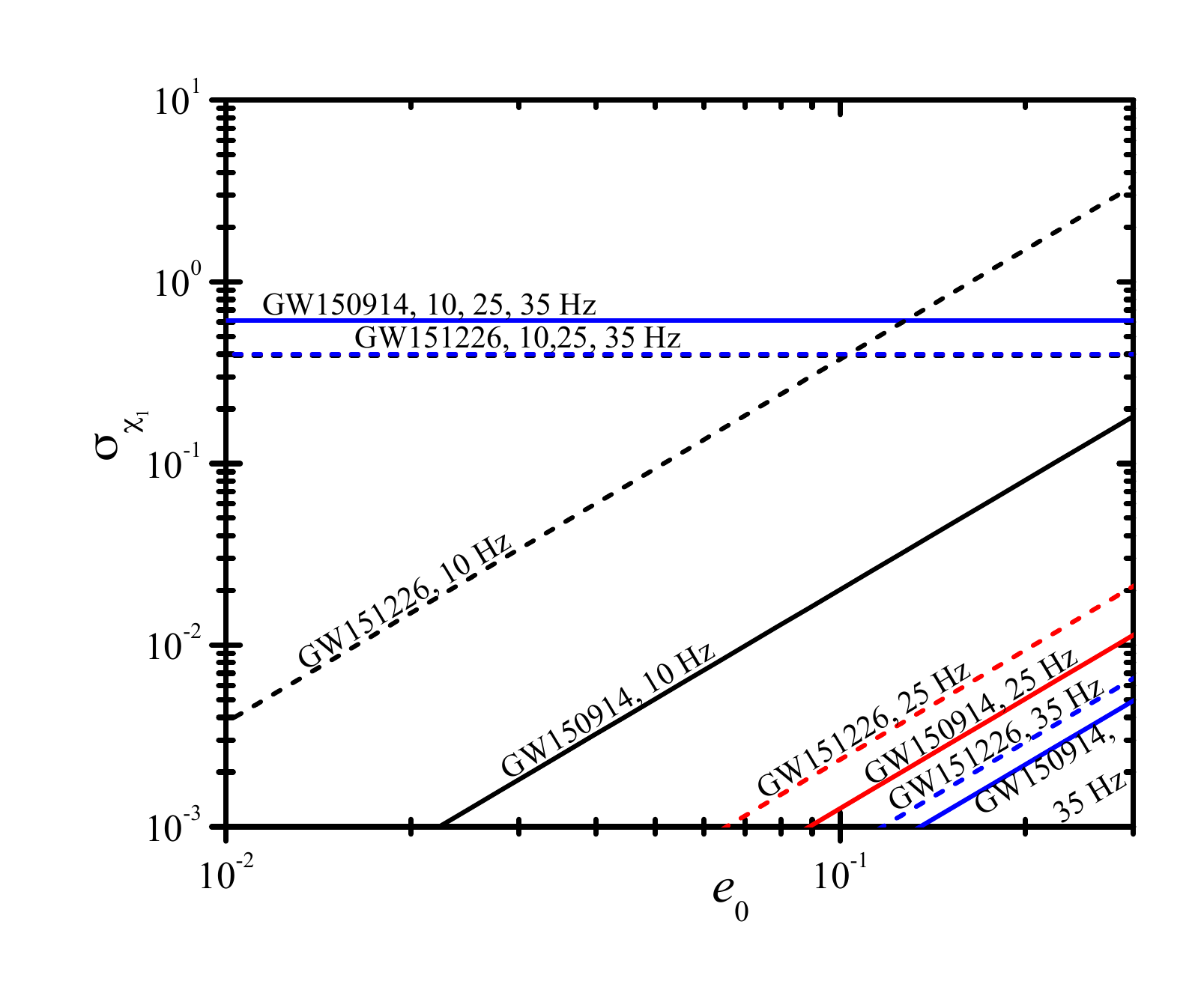} &
\includegraphics[angle=0, width=0.48\textwidth]{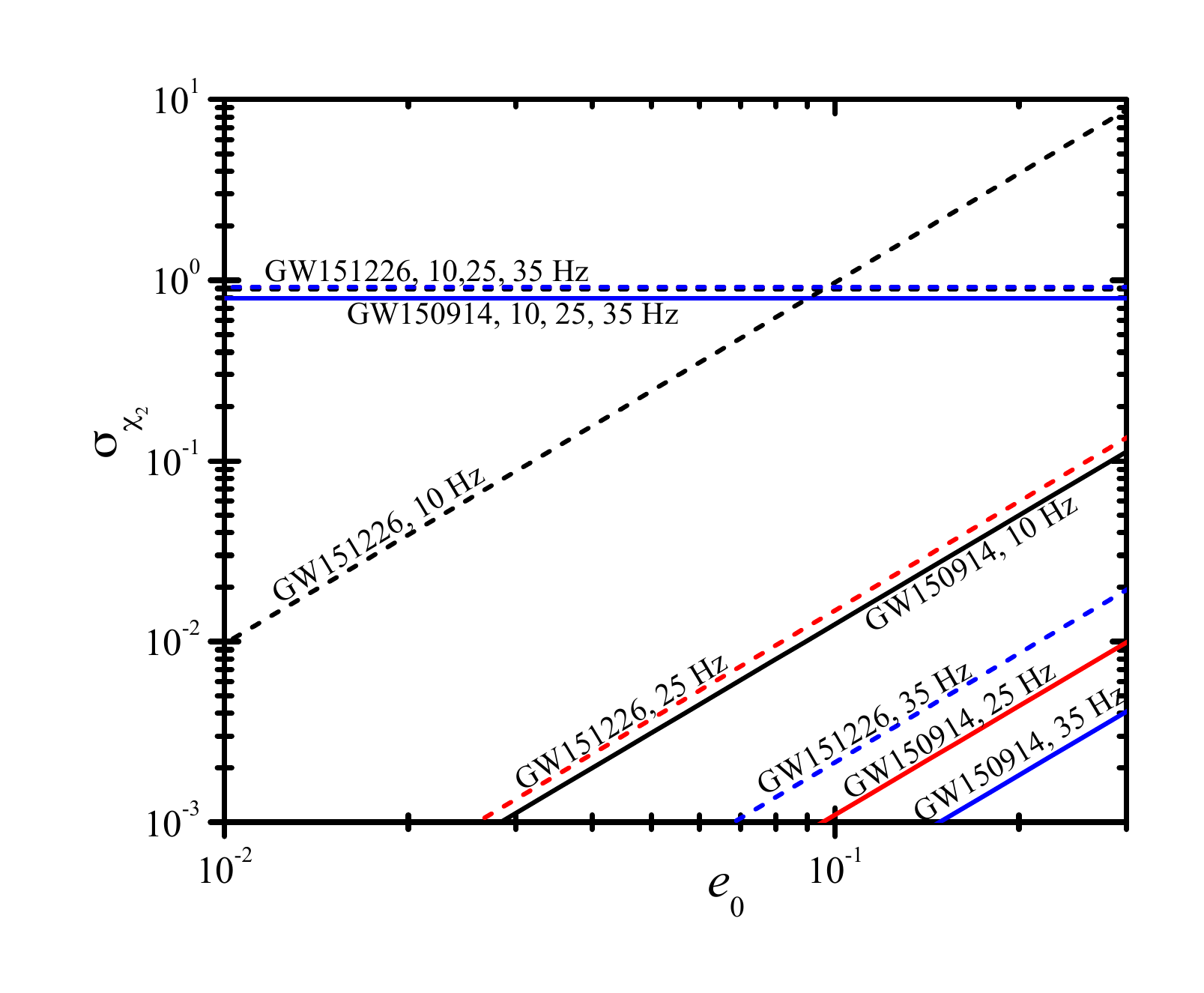}
\end{array}
$
\caption{\label{fig:SYSerrorsevents}Systematic and statistical errors for systems similar to GW150914 and GW151226 as considered in the right panel of Fig.~\ref{fig:PEe0results}. As in that figure, errors for GW150914 are shown with solid lines and errors for GW151226 are shown with dashed lines. Different choices for $f_{\rm low}$ are likewise labeled as in Fig.~\ref{fig:PEe0results}: black (10 Hz), red (25 Hz), and blue (35 Hz). As in Fig.~\ref{fig:SYSerrors}, statistical errors are horizontal lines and systematic errors are upward-sloping lines.}
\end{figure*}

In computing our results we have pushed our 3PN eccentric waveform [Eq.~\eqref{eq:Psiecc}] to values of $e_0$ as high as $0.3$. This is a bit past the range where the waveform remains accurate. In Sec.~VIII of \cite{moore-etal-PRD2016} it was estimated that phase accuracy ($\delta N_{\rm cyc} \lesssim 1$) is maintained for $e_0 \lesssim 0.06 \mbox{--} 0.15$. We estimate that higher-order eccentricity effects will produce corrections scaling like $\sim O(e_0)$ or higher; this corresponds to corrections  of order $\lesssim 30\%$ to our results (and typically much less for smaller values of $e_0$). Further, we have pushed our 3PN waveform to high frequencies, terminating them at the ISCO corresponding to the final mass and spin. However, in Ref.~\cite{moore-etal-PRD2016} we estimated an upper frequency $f_{\rm qK}$ [Eq.~\eqref{eq:fqK}] where the quasi-Keplerian approximation breaks down. In many cases this is significantly below the ISCO frequency.

To assess the impact of a more conservative high-frequency termination for our waveforms, Figure \ref{fig:PEfreqcutoff} examines the difference in the fractional error $\sigma_{e_0}/e_0$ between using $f_{\rm qK}$ and $f_{{\rm isco},z}$. For the case when $e_0=0.1$, the fractional error $\sigma_{e_0}/e_0$ is larger for the (lower-frequency) $f_{\rm qK}$ cutoff by a factor of $1.27$ (NS/BH), $1.42$ (BBH1), $1.80$ (BBH1), $2.53$ (BBH1), or $4.18$ (BBH1). While this is clearly significant for higher-mass BBHs, we note that our intention in this work is to provide representative estimates for eccentricity constraints that could be achieved with LIGO. By pushing our waveforms to high frequencies, we anticipate the development of improved waveform families that can accurately describe the regime close to merger. Considering the approximations involved in the Fisher formalism itself, these errors are in keeping with our desire to make crude estimates of the constraints achievable with LIGO.

\section{\label{sec:systematic}Results: Systematic Errors via the FCV formalism}
Having examined the precision with which eccentricity could be measured, we turn now to the question of the bias induced in the other parameters if eccentricity is neglected. Here we use the formalism summarized in Sec.~\ref{subsec:sys}.  We take as our parameter set  $\theta_A = (t_c, \phi_c, \ln M, \ln \eta, \chi_1, \chi_2)$ and use the priors as in Eq.~\eqref{eq:priors} for $\phi_c$ and $\chi_{1,2}$. Aside from the elimination of $e_0$ as a parameter, the calculation of the Fisher matrix and $\Sigma_{AB}$ is the same as described above. The systematic parameter error is computed via Eq.~\eqref{eq:Deltatheta}, taking $\Delta \Psi$ to be $\frac{3}{128\eta v^5} \Delta \Psi^{\rm ecc.}_{\rm 3PN}$ and $\Psi_{\rm AP}$ to be all terms in Eq.~\eqref{eq:Psiterms} except for $\Delta \Psi^{\rm ecc.}_{\rm 3PN}$. We compute the resulting statistical and systematic parameter errors for the same systems shown in Fig.~\ref{fig:PEe0results}. Figure \ref{fig:SYSerrors} shows those errors for the parameters $(M, \eta, \chi_{1,2})$ as a function of $e_0$. Figure \ref{fig:SYSerrorsevents} shows the corresponding errors for systems similar to GW150914 and GW151226. We make the same choices for the sensitivity curve, frequency range, and system parameters as in the previous section. 

Figure \ref{fig:SYSerrors} suggests that systematic errors begin to exceed statistical ones when $e_0 \gtrsim 0.01 \mbox{--} 0.1$, with the intersection point varying by system. Further, the intersection point (where statistical and systematic errors are equal) roughly increases as the system mass or number of cycles decreases. For our fiducial NS/NS binary, this intersection point occurs at $e_0 \approx 0.01$ for the mass parameters and $e_0 \gtrsim 0.022$ for the spin parameters. For our NS/BH binary the intersection point occurs near $e_0 \gtrsim 0.025$ for the mass and spin parameters. For BBH1, BBH2, BBH3, and BBH4, the approximate intersection points occur (respectively) at $e_0 \gtrsim 0.05$, $0.07$, $0.08$, and $0.09$ for the mass parameters, and $0.06$, $0.11$, $0.15$, and $0.3$ for the spin parameters. 

The systematic errors in Fig.~\ref{fig:SYSerrors} show a clear decreasing trend as the binary total mass increases. More massive and comparable-mass binaries are stronger emitters of GWs and will shed away eccentricity (circularize) more rapidly. Hence, only if they have a higher initial eccentricity will their parameters be biased by an amount comparable to that seen in lower mass systems (BNS or NS/BH).  

From Figure \ref{fig:SYSerrorsevents} we see that any eccentricity-induced systematic bias in the parameters is completely negligible for GW150914 and GW151226 when the low-frequency limit is taken to be 35 Hz (consistent with the actual observations). The use of circular templates for the analyses performed in \cite{detectionPRL2016,gw151226-PRL2016} is thus likely to be quite sufficient. However, as we go to smaller values of $f_{\rm low}$, eccentricity-induced bias can become more important. At 25 Hz, systematic errors can exceed statistical errors in the mass parameters if $e_0 \gtrsim 0.2$ for GW151226-like systems. Going to 10 Hz, systematic biases in the mass parameters become important for both GW151226 and GW150914-like systems for $e_0 \gtrsim 0.07$. Biases in the spin parameters become important only for GW151226-like systems for $e_0 \gtrsim 0.1$. 

\section{\label{sec:mcmc}Parameter estimation using Bayesian MCMC Inference}
\subsection{Overview of MCMC calculations}
In addition to our Fisher matrix study, we performed a limited investigation using \texttt{LALInferenceMCMC}, a parameter estimation pipeline included in LALSuite \cite{LALSuite} (and described in detail in \cite{veitch-etalPEpaper-PRD2015}; see also \cite{LIGO-PEpaper2013,detection-PEpaper2016}).  Unlike the Fisher matrix approach, the Bayesian inference approach does not assume that the SNR is large. It also allows for multi-modal and non-Gaussian posterior probability distributions, as well as more complex prior probability distributions. On the other hand, the resulting code is much more complex, and the significant increase in computational cost leads to a longer timescale for producing results (days to weeks vs.~seconds for the Fisher approach). For this reason we performed a limited investigation on a single binary system using the \texttt{LALInferenceMCMC} code.\footnote{Our version of the code was modified slightly to allow for the $e_0$ and $f_0$ parameters appearing in the {\tt TaylorF2Ecc} waveform, but is otherwise identical to that used by the LVC around August 2016 \cite{LALSuite}. The software features we use for MCMC parameter estimation here (e.g., likelihood computation, marginalization of a PDF) are essentially the same as in \cite{detection-PEpaper2016}.}

We focused on a single binary with parameters similar to GW151226, one of the lightest known BBHs from the first two LIGO/Virgo observing runs (O1 and O2). To reduce computation time, we imposed additional restrictions on our GW151226-like binary compared to the Fisher-based analysis considered in the previous sections. Specifically, we use the observed \emph{detector-frame} masses of $m_1^{\rm det}=15.6 M_{\odot}$ and $m_2^{\rm det}=8.2 M_{\odot}$ of GW151226, but here we ignore BH spins. This allows us to sample the (smaller) parameter space in less time. We also assume that the binary is located at a distance of $500$ Mpc, and is observed by a single LIGO detector (Hanford). The single-detector SNR is 20.12. In contrast to the case in Sec.~\ref{sec:statistical}, to speed up computation time we choose our low-frequency limit to be $25$ Hz. The upper-frequency limit is set to $184.75$ Hz  (corresponding to twice the Schwarzschild ISCO orbital frequency for GW151226). Note that we still define our eccentricity parameter $e_0$ at the frequency $f_0=10$ Hz. 

To directly compare with our {\tt LALInferenceMCMC} results, we reran our Fisher matrix code using parameters consistent with the MCMC calculation as described above. (Note that since we are now ignoring spins, our Fisher code is using only five parameters in this case, $[t_c,\phi_c,M_{\rm ch},\ln \eta, e_0]$, while the MCMC code is searching over a ten-parameter space.\footnote{The additional five parameters are $\theta_{JN}$ (inclination angle defined as the angle between the total angular momentum vector and the  direction to the detector), $d_L$ (luminosity distance), $\psi$ (polarization angle), and $(\alpha, \delta)$ (right ascension and declination of the source).})

Priors in our Fisher code are treated the same as previously discussed. In the MCMC code we assume that eccentricity is distributed uniformly between $0$ and $1$. (While the {\tt TaylorF2Ecc} waveform is not valid for large $e_0$ \cite{moore-etal-PRD2016}, we note that our posterior distributions show little support above $e_0=0.3$, due to the low $e_0$ values of our injected signals. Hence, the breakdown of {\tt TaylorF2Ecc} at higher eccentricities is unlikely to significantly alter our conclusions.) Priors on other parameters in the MCMC code are treated as in \cite{detection-PEpaper2016}.  Specifically, we use uniform priors on $t_c$ (with width 4 seconds) and uniform priors on $\phi_c$ over $[0,2\pi]$. We also use uniform priors in $m_{1,2} \in [1.0,100.0] M_{\odot}$ with $m_2 \leq m_1$. We assume that the sources are uniformly distributed on the sky with orientations distributed uniformly in cos(inclination angle). In the MCMC code, we marginalize over the coalescence time when computing the posterior probability distribution.
Both Fisher and MCMC codes were run using the LIGO design sensitivity.\footnote{The one-sided spectral density used in the MCMC code differed slightly near $25$ Hz from the analytic fit used in our Fisher matrix calculations (see Sec.~\ref{subsec:sys}). However, we checked that this had a negligible effect on our parameter estimates.} 

In order to perform MCMC parameter estimation for a {\it simulated} GW strain with eccentricity, we have used 16 seconds for the segment length, a low-frequency cut of $25$ Hz, a sampling rate of $2048$ Hz, and a reference frequency of $f_{\rm ref}=100$ Hz. All source angles are defined at this reference frequency. The eccentricity is defined at $f_0=10$ Hz as described earlier. 

An essential task of {\tt LALInference} and parameter estimation is to compute the log-likelihood, which depends on an inner product [defined in Eq.~\eqref{eq:innerprod}] between the detector data $d(t) = n(t) + h_{\rm true}(t; {\bm \theta}_{\rm true})$ and a template $h_T(t; {\bm \theta})$:
\be
\ln {\mathcal L} = - \frac{1}{2} \Big( d(t) - h_T(t; {\bm \theta}) \Big| d(t)-h_T(t; {\bm \theta}) \Big) \;.
\ee
Here $n(t)$ is the detector noise, $h_{\rm true}(t; {\bm \theta}_{\rm true})$ is the GW signal and depends on the ``true'' system parameters ${\bm \theta}_{\rm true}$, and the template $h_T(t; {\bm \theta})$ depends on parameters ${\bm \theta}$ (which can be thought of as independent variables that specify a particular template). Here, we assume ``zero noise'' (e.g., \cite{rodriguez-etal-aLIGO-PE-estimates}), which means the sample data generated within {\tt LALInferenceMCMC} uses the choice $d(t) = h_{\rm true}(t;{\bm \theta}_{\rm true})$ when computing the log-likelihood. [I.e., we assume the data contains only injected GW signals and $n(t)=0$.] However, a model for the detector noise is incorporated via the inner product [Eq.~\eqref{eq:innerprod}], which depends explicitly on a model for the detector's one-sided noise spectral density $S_n(f)$ \cite{finn-chernoff-PRD1993}.  In the limit of very large sampling and the \emph{zero noise} approximation, the resulting \emph{a priori} (i.e., before a prior probability is imposed) probability distribution should be peaked at ${\bm \theta}= {\bm \theta}_{\rm true}$, but with a spread (standard deviation) that is proportional to the detector's noise spectral density. 

We apply the ``zero noise'' approximation when performing Bayesian MCMC parameter estimation as the most optimistic realization to compare with Fisher matrix results. A more realistic parameter estimation study might include a particular noise realization $n(t)$. Aside from having the stationary (time-independent) and Gaussian spectral properties that are embedded in the noise spectral density $S_n(f)$, a chosen $n(t)$ might additionally contain time variations in the noise's spectral content or non-Gaussian features like ``glitches.'' These features (which are present in realistic detector data) may further alter the resulting \emph{a priori} distribution such that it will no-longer be maximized at ${\bm \theta}= {\bm \theta}_{\rm true}$. We expect realistic noise models to further limit (i.e., worsen) the measurement of binary eccentricity. 
\begin{figure*}[th]
$
\begin{array}{ccc}
\includegraphics[angle=0, width=0.3\textwidth]{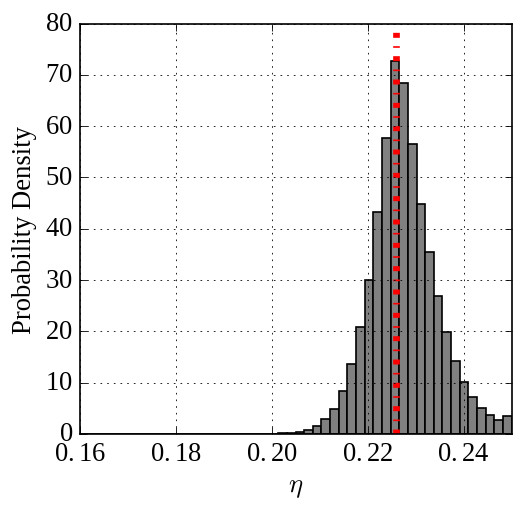} &
\includegraphics[angle=0, width=0.3\textwidth]{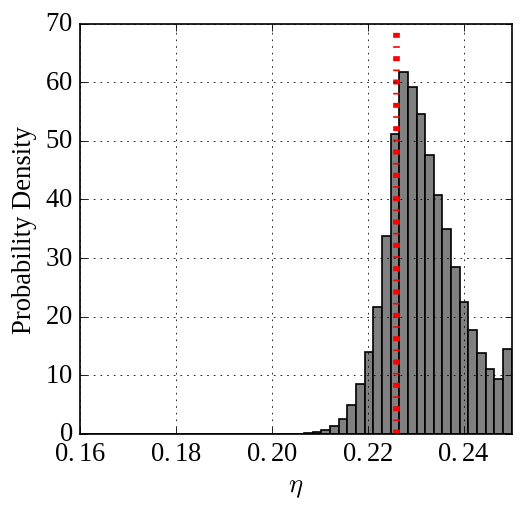} &
\includegraphics[angle=0, width=0.3\textwidth]{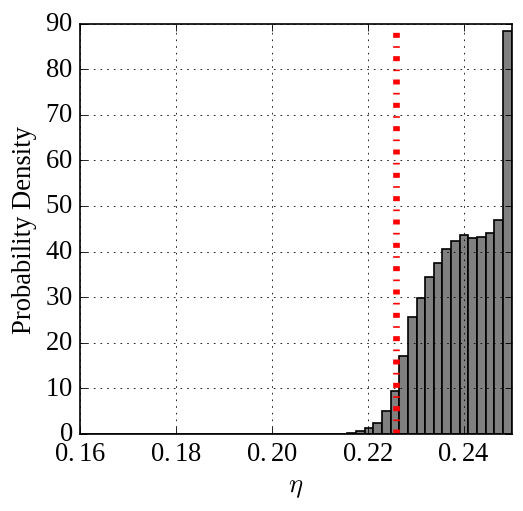} \\
\includegraphics[angle=0, width=0.3\textwidth]{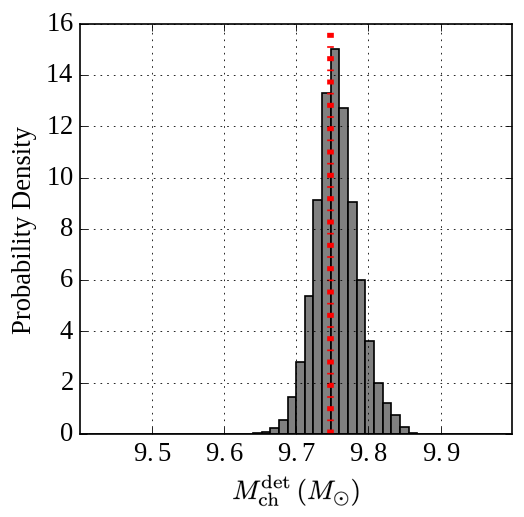} &
\includegraphics[angle=0, width=0.3\textwidth]{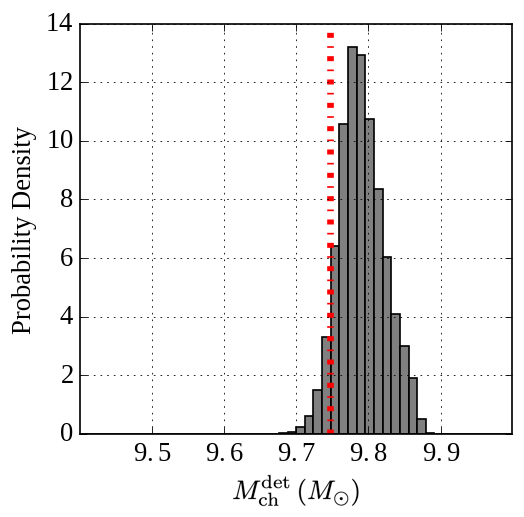} &
\includegraphics[angle=0, width=0.3\textwidth]{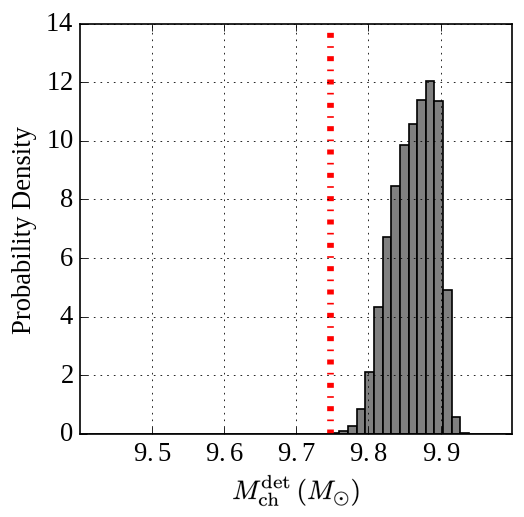}
\end{array}
$
\caption{\label{fig:ecccircMCMCpdf} Marginalized posterior probability distributions for the symmetric mass ratio $\eta$ (top row) and the detector-frame chirp mass $M_{\rm ch}^{\rm det}$ (bottom row), showing the systematic bias induced by the signal's unmodeled eccentricity. Eccentric signals are injected with the {\tt TaylorF2Ecc} waveform, but recovered using circular {\tt TaylorF2} templates. As in Fig.~\ref{fig:MCMCpdf-stat}, the vertical dotted lines indicate the injected values of $\eta$ and $M_{\rm ch}^{\rm det}$ (which are the same in each row). The injected eccentricity varies as $e_0^{\rm inj}= [0.04, 0.12, 0.2]$ from left to right.  A growing systematic bias is clearly seen as the eccentricity increases.} 
\end{figure*}

\subsection{\label{subsec:MCMCstat}MCMC results: Statistical errors}
In Fig.~\ref{fig:MCMCpdf-stat} we show the marginalized posterior probability distributions for $e_0$, $\eta$, and the detector-frame chirp mass $M_{\rm ch}^{\rm det}$ for three different values of the injected eccentricity. (The injected signal is generated with the {\tt TaylorF2Ecc} waveform with the specified value of $e_0$, and recovered with {\tt TaylorF2Ecc} waveform templates in which $e_0$ and the nine other parameters are allowed to vary.) Table \ref{tab:MCMCcompare_e0} quantifies the recovered parameter values for $M_{\rm ch}^{\rm det}$, $\eta$, and $e_0$ for a slightly larger selection of injected eccentricities, and also lists the statistical errors estimated from the Fisher-matrix calculation. 

Figure \ref{fig:MCMCpdf-stat} and Table \ref{tab:MCMCcompare_e0} indicate that $M_{\rm ch}^{\rm det}$ and $\eta$ are recovered with excellent accuracy (better than $1\%$ error in the maximum \emph{a posteriori} probability for $\eta$, with a fractional error $\sigma_{\eta}/\eta \sim 4\% \mbox{--} 5\%$; the equivalent numbers are smaller for $M_{\rm ch}^{\rm det}$). 
In the case of $e_0$ we see that the posterior distribution is much more broad, and is railing against $e_0=0$ when the eccentricity is small. This makes small eccentricity difficult to measure. However, as the injected $e_0$ is increased to $e_0 \ge 0.1$, the recovery accuracy increases significantly ($\sim$few to $10\%$), although the relative precision remains modest ($\sigma_{e_0}/e_0 \sim 38\% \mbox{--} 65\%$; consider the maP and $\sigma$ columns for the $e_0=0.1$ through $0.2$ cases in Table \ref{tab:MCMCcompare_e0}). 

While $e_0$ and $\eta$ are accurately recovered, we note an obvious bias in the recovery of the chirp mass $M_{\rm ch}^{\rm det}$ (Fig.~\ref{fig:MCMCpdf-stat}, bottom row) that grows with increasing $e_0$. We will return to this feature in Sec.~\ref{sec:degen} below.  While $\eta$ is recovered accurately even for $e_0=0.2$ (see Table \ref{tab:MCMCcompare_e0}), we do note that the posterior probability distribution has shifted slightly toward larger $\eta$ (middle-right panel of Fig.~\ref{fig:MCMCpdf-stat} and median values in Table~\ref{tab:MCMCcompare_e0}). 

We also note that the statistical error estimates ($\sigma$ values) between the MCMC and Fisher approaches agree to $\sim3\%$ to $13\%$ for $\eta$ and $\sim 2\%$ to $36\%$ for $M_{\rm ch}^{\rm det}$ (Table~\ref{tab:MCMCcompare_e0}). The statistical error estimates for $e_0$ show poor agreement for $e_0\leq 0.1$, but this improves to $\sim 20\%$ agreement for $e_0 \geq 0.15$. The top row of Fig.~\ref{fig:MCMCpdf-stat} suggests that this poor agreement with the Fisher calculation is due to the prior distribution for $e_0$ considered in each approach. The MCMC calculation assumes $e_0$ is uniform in the range $[0,1]$, while the Fisher matrix approach only allows for a Gaussian prior which has nonzero support for $e_0<0$. 


\setlength{\tabcolsep}{1pt}
\addtolength{\tabcolsep}{4pt} 
\begin{table*}[p]
\caption{\label{tab:MCMCcompare_e0} Comparison of statistical errors between MCMC and Fisher matrix calculations for a GW151226-like binary black hole system. We show results for the parameters $M_{\rm ch}^{\rm det}$, $\eta$ and $e_0$ for selected injected values of the eccentricity parameter ($e_0^{\rm inj}$). The injected values for the mass parameters are $M_{\rm ch, inj}^{\rm det}=9.746 M_{\odot}$ and $\eta_{\rm inj}=0.2258$. See also Fig.~\ref{fig:MCMCpdf-stat}. The different columns refer to the following quantities computed via the MCMC code: the maximum \emph{a posteriori} probability (maP, most-likely value accounting for the prior; this is generally different from the maximum value of the marginalized 1D PDFs shown in Fig.~\ref{fig:MCMCpdf-stat}), the median of the probability density (med.), and the standard deviation. The last column for each parameter shows the standard deviation computed using the Fisher matrix approach.}
\begin{tabular}{|l|llll|llll|llll|}
	\hline 
 & \multicolumn{4}{c|}{Parameter $M_{\rm ch}^{\rm det} \,(M_{\odot})$} & \multicolumn{4}{c|}{Parameter $\eta$} & \multicolumn{4}{c|}{Parameter $e_0$} \\
\hline
\multicolumn{1}{|c|}{$e_0^{\rm inj}$} & \multicolumn{1}{c}{maP} & \multicolumn{1}{c}{med.}  & \multicolumn{1}{c}{$\sigma_{\rm MCMC}$} & \multicolumn{1}{c|}{$\sigma_{\rm Fisher}$} & \multicolumn{1}{c}{maP} & \multicolumn{1}{c}{med.}  &  \multicolumn{1}{c}{$\sigma_{\rm MCMC}$} & \multicolumn{1}{c|}{$\sigma_{\rm Fisher}$}  & \multicolumn{1}{c}{maP} & \multicolumn{1}{c}{med.}  &  \multicolumn{1}{c}{$\sigma_{\rm MCMC}$} & \multicolumn{1}{c|}{$\sigma_{\rm Fisher}$}  \\
	\hline
	0.04 & 9.744 & 9.730 & 0.04969 & 0.06794 & 0.2263 & 0.2245 & 0.008043 & 0.008744 & 0.07384  & 0.08533  & 0.06126 & 0.3036\\ 
	0.08 & 9.760 & 9.740 & 0.05139 & 0.07000 & 0.2268 & 0.2254 & 0.008212 & 0.008906 & 0.005230 & 0.09030  & 0.06295 & 0.1570\\ 
	0.10 & 9.752 & 9.747 & 0.05333  & 0.07025 & 0.2273 & 0.2260 & 0.008402 & 0.008926 & 0.09802  & 0.09543  & 0.06473 & 0.1260 \\
	0.12 & 9.747 & 9.755 & 0.05541 & 0.07037 & 0.2261 & 0.2267 & 0.008651 & 0.008937 & 0.1131  & 0.1017  & 0.06641 & 0.1051 \\
	0.15 & 9.741 & 9.770 & 0.06029 & 0.07046 & 0.2257 & 0.2282 & 0.009206 & 0.008946 & 0.1651  & 0.1152  & 0.06989 & 0.08393 \\
	0.20 & 9.748 & 9.794 & 0.07204 & 0.07047 & 0.2253 & 0.2308 & 0.01027 & 0.008952 & 0.2020 & 0.1529  & 0.07631 & 0.06257 \\	
		\hline 
\end{tabular}
\end{table*}
\addtolength{\tabcolsep}{-4pt} 
\addtolength{\tabcolsep}{4pt} 
\begin{table*}[p]
\caption{\label{tab:MCMC-systematic_eta} Comparison of systematic bias in the reduced mass ratio $\eta$ between the MCMC and Fisher-Cutler-Vallisneri (FCV) methods. Parameters are as in Table \ref{tab:MCMCcompare_e0}, except here signals are injected with the {\tt TaylorF2Ecc} waveform and recovered with the (circular) {\tt TaylorF2} waveform.  Columns 2 and 3 show the difference between the estimators $x_{\eta}$ (maP, median) and the injected (true) value of $\eta$. Column 4 shows the parameter bias predicted by the FCV approach. Columns 5 and 6 show the relative error between the indicated MCMC estimators and the FCV approach. The last two columns show the standard deviations (statistical errors) computed via the two methods. The MCMC and FCV approaches generally show consistent agreement.} 
\begin{tabular}{|l|lll|ll|ll|}
	\hline 
 & \multicolumn{3}{c|}{$x_{\eta}- \eta_{\rm inj}$} & \multicolumn{2}{c|}{$|x_{\eta}-FCV_{\eta}|/\eta_{\rm inj}$} & \multicolumn{2}{c|}{$\sigma_{\eta}$} \\[1pt]
\hline
\multicolumn{1}{|c|}{$e_0^{\rm inj}$} & \multicolumn{1}{c}{maP} & \multicolumn{1}{c}{median}  & \multicolumn{1}{c|}{FCV} & \multicolumn{1}{c}{maP} & \multicolumn{1}{c|}{median} & \multicolumn{1}{c}{(MCMC)} & \multicolumn{1}{c|}{(Fisher)} \\
	\hline
	0.04 & 0.0001242 & 0.001297 & 0.0004082  & 0.001257 & 0.003937 & 0.006848 & 0.006164 \\ 
    0.08 & 0.0003878 & 0.002550 & 0.001633 & 0.005514 &	0.004060 & 0.007036 & 0.006164 \\ 
	0.10 & 0.001676 & 0.003613 & 0.002551 & 0.003875 & 0.004701 & 0.007185 & 0.006164 \\
	0.12 & 0.003078 & 0.004929 & 0.003674 & 0.002640 & 0.005558 & 0.007377 & 0.006164 \\
	0.15 & 0.001829 & 0.007708 & 0.005740 & 0.01732 & 0.008714 & 0.007629 & 0.006164 \\
	0.20 & 0.006579 & 0.01466 & 0.01021 & 0.01608 & 0.01972 & 0.007197 & 0.006164 \\	
		\hline 
\end{tabular}
\end{table*}
\addtolength{\tabcolsep}{-4pt}
\addtolength{\tabcolsep}{4pt} 
\begin{table*}[p]
\caption{\label{tab:MCMC-systematic_chirp} Same as Table \ref{tab:MCMC-systematic_eta} except here we show results for the systematic bias in the detector-frame chirp mass $M_{\rm ch}^{\rm det}$.}
\begin{tabular}{|l|lll|ll|ll|}
	\hline 
 & \multicolumn{3}{c|}{$x_{M_{\rm ch}^{\rm det}}- M_{\rm ch, inj}^{\rm det}$ ($M_{\odot}$)} & \multicolumn{2}{c|}{$\frac{|x_{M_{\rm ch}^{\rm det}}-FCV_{M_{\rm ch}^{\rm det}}|}{M_{\rm ch, inj}^{\rm det}}$} & \multicolumn{2}{c|}{$\sigma_{M_{\rm ch}^{\rm det}}$ ($M_{\odot}$)} \\[6pt]
\hline
\multicolumn{1}{|c|}{$e_0^{\rm inj}$} & \multicolumn{1}{c}{maP} & \multicolumn{1}{c}{median}  & \multicolumn{1}{c|}{FCV} & \multicolumn{1}{c}{maP} & \multicolumn{1}{c|}{median} & \multicolumn{1}{c}{(MCMC)} & \multicolumn{1}{c|}{(Fisher)} \\
	\hline
	0.04 & 0.0005586 &0.008325  & 0.004089 & 0.0003622 & 0.0004347 & 0.02986 & 0.02752 \\ 
	0.08 & 0.01331  & 0.02070  & 0.01636 & 0.0003127  &  0.0004454 & 0.03041 & 0.02752 \\ 
	0.10 & 0.02055  & 0.03059  & 0.02556 & 0.0005145  & 0.0005162 & 0.03084 & 0.02752 \\
	0.12 & 0.03358  & 0.04245  & 0.03680 & 0.0003304  &	0.0005792 & 0.03134 & 0.02752 \\
	0.15 & 0.03971  & 0.06563   & 0.05750 & 0.001825  &	0.0008340 & 0.03212 & 0.02752 \\
	0.20 & 0.08470  & 0.1193   & 0.1022  & 0.001796  &	0.001753 & 0.03004 & 0.02752 \\	
		\hline 
\end{tabular}
\end{table*}
\addtolength{\tabcolsep}{-4pt} 

\begin{figure*}[htp!]
$
\begin{array}{ccc}
\includegraphics[angle=0, width=0.32\textwidth]{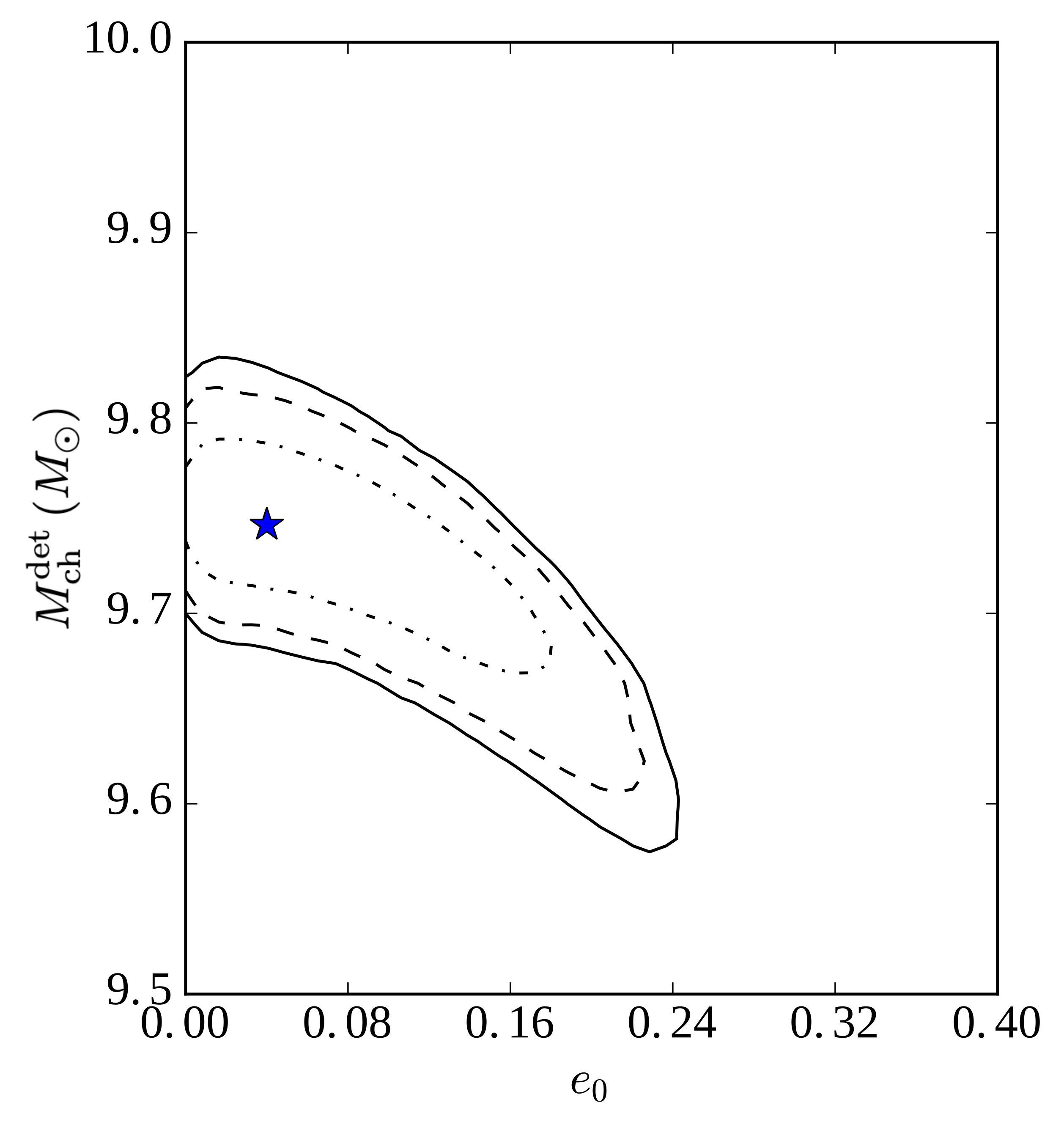} &
\includegraphics[angle=0, width=0.32\textwidth]{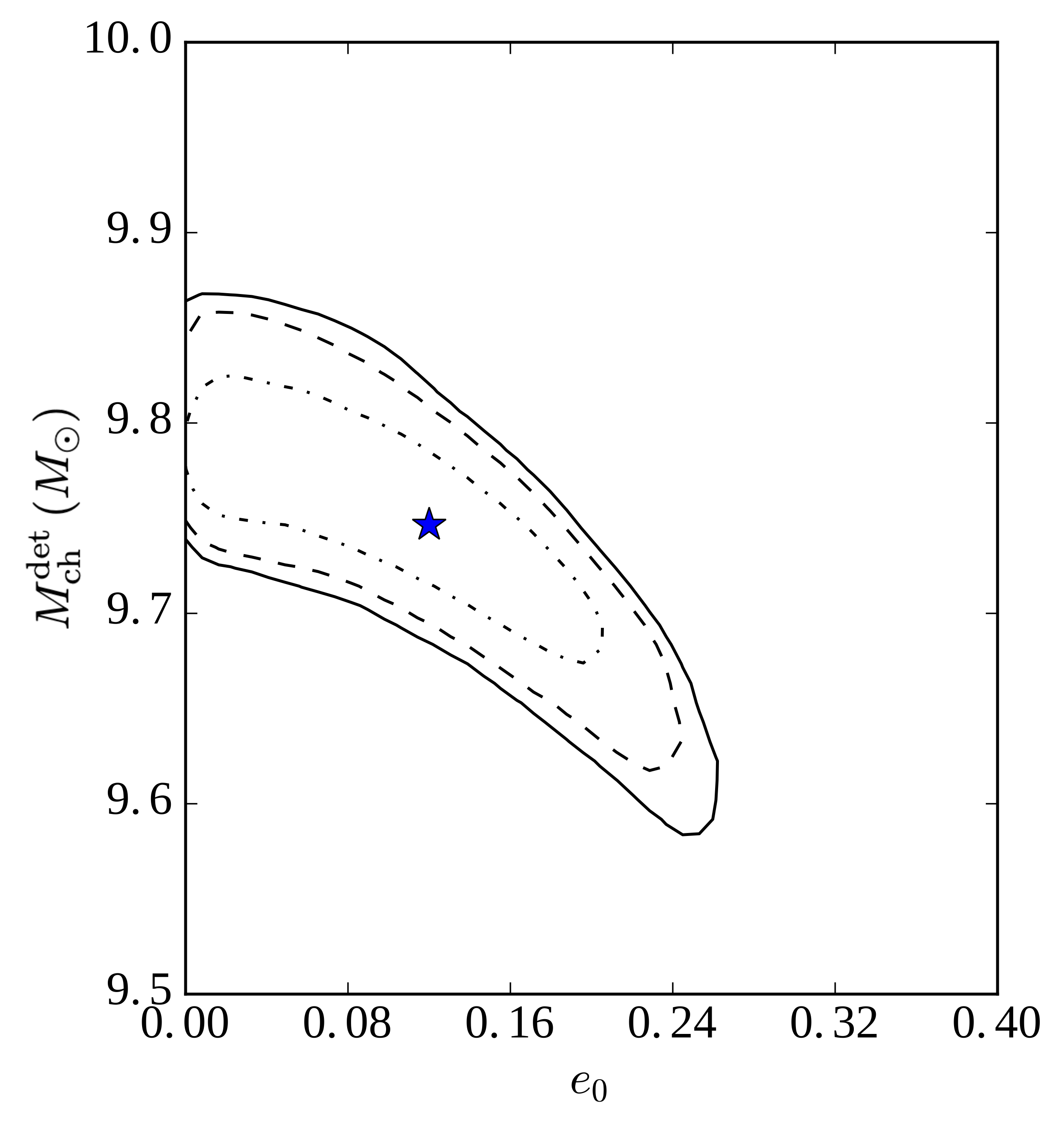} &
\includegraphics[angle=0, width=0.32\textwidth]{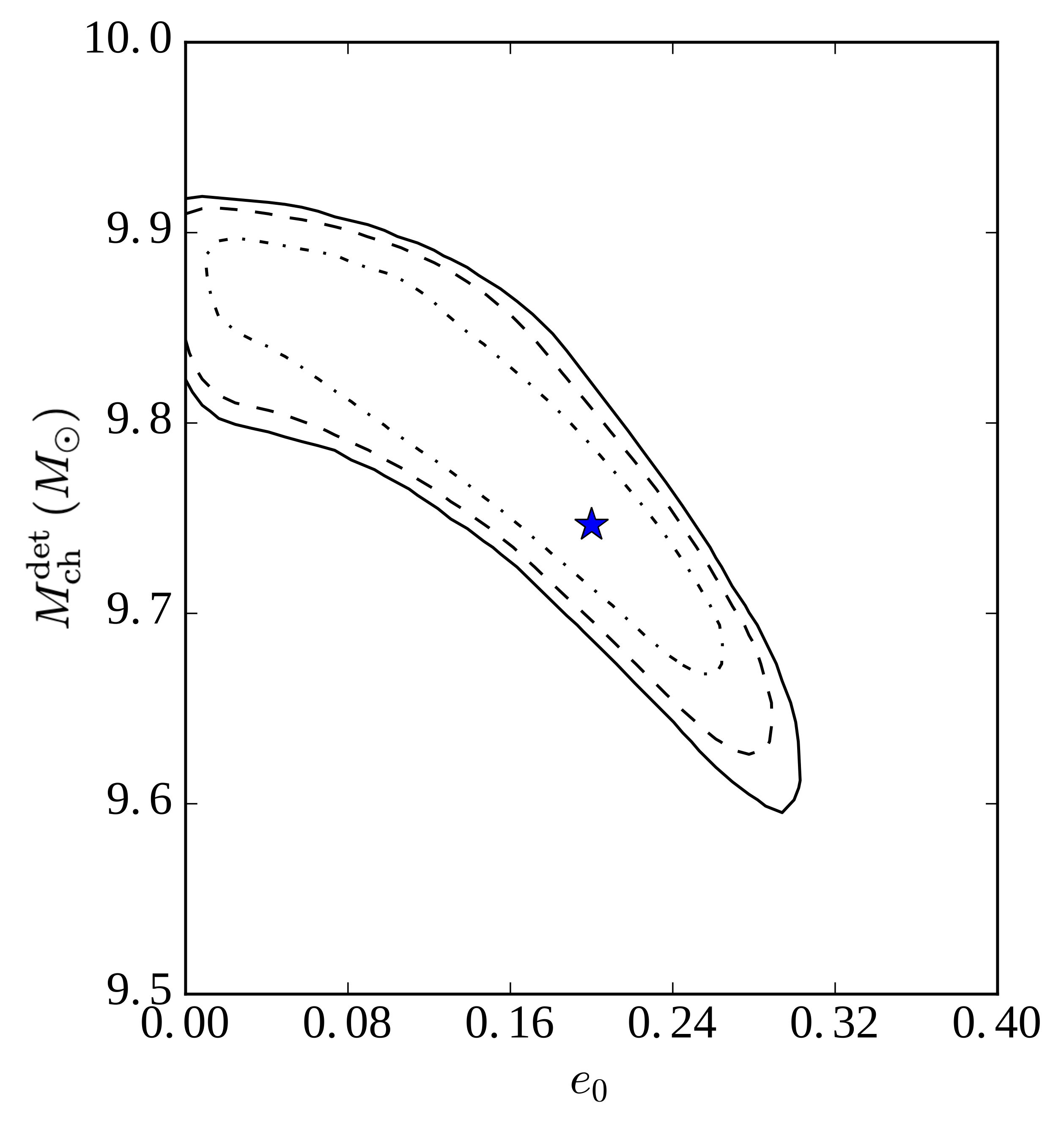}
\end{array}
$
\caption{\label{fig:banana} 
Posterior probability contours in the $e_0$-$M_{\rm ch}^{\rm det}$ plane computed via {\tt LALInferenceMCMC}. Each panel shows a different value of the injected initial eccentricity, $e_0^{\rm inj}=[0.04,0.12, 0.2]$ from left to right, for the same value of injected chirp mass $M_{\rm ch, inj}^{\rm det}=9.746 M_{\odot}$. The three contours represent $67\%$, $90\%$, and $95\%$ confidence intervals, and the star indicates the injected parameter values. As $e_0^{\rm inj}$ increases, we see that the error ellipse tightens across its narrow dimension near the injected value and rotates slightly. The ellipse angle is due to the correlation between the two parameters, and roughly follows the behavior shown in a Fisher matrix analysis (see discussion in Sec.~\ref{sec:degen} and Figs.~\ref{fig:m1m2chirp} and \ref{fig:e0Mchellipse}). The increasing correlation-induced rotation of the error ellipse seen here, combined with the prior on $e_0$, is ultimately responsible for the growing bias in the chirp mass seen in the bottom row of Fig.~\ref{fig:MCMCpdf-stat}. One clearly sees here that the projection of the probability density on the $M_{\rm ch}^{\rm det}$-axis skews the PDF of $M_{\rm ch}^{\rm det}$ to values $>M_{\rm ch, inj}^{\rm det}$, illustrating the behavior seen in the bottom-right panel of Fig.~\ref{fig:MCMCpdf-stat}.} 
\end{figure*}
\begin{figure*}[htp!]
$
\begin{array}{ccc}
\includegraphics[angle=0, width=0.32\textwidth]{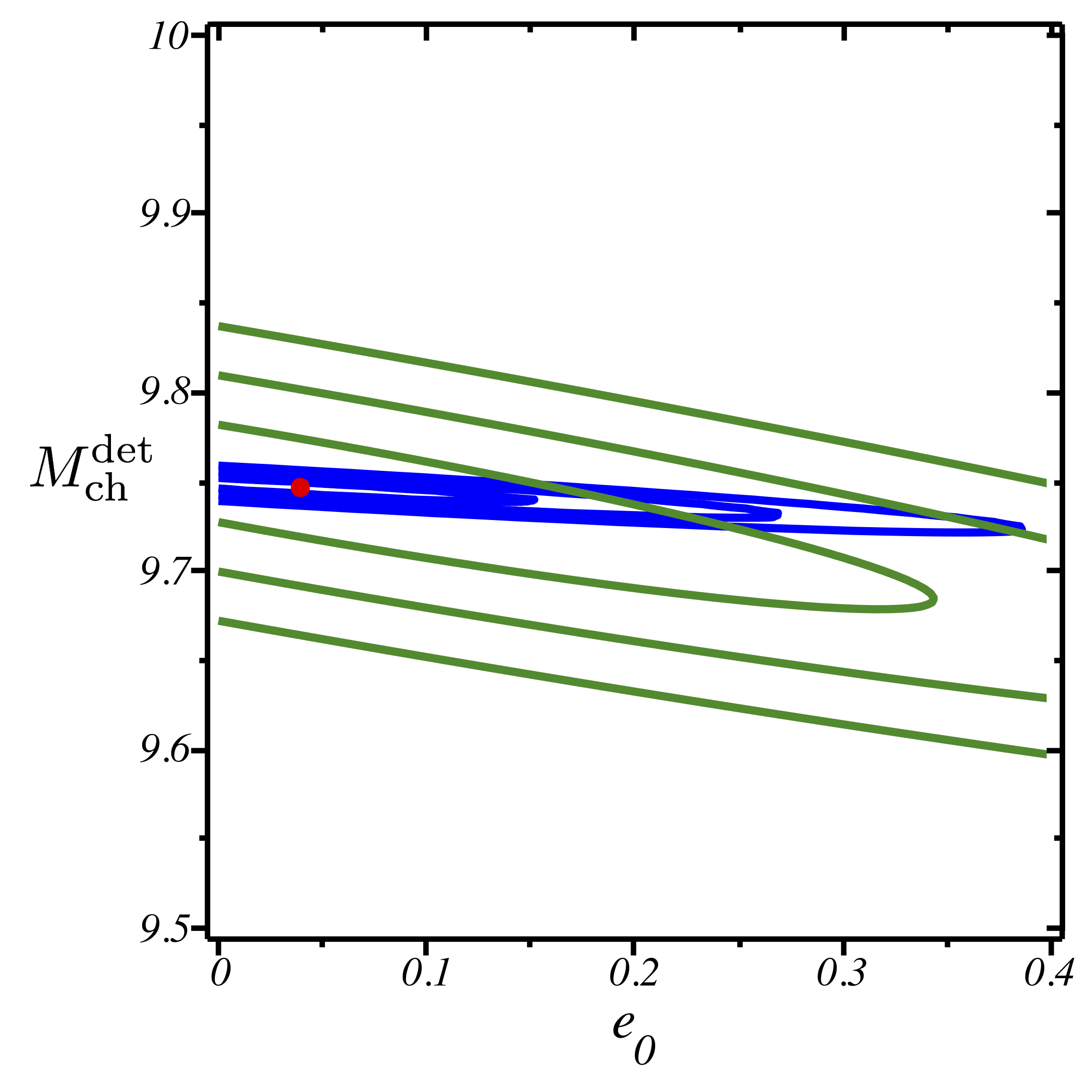} &
\includegraphics[angle=0, width=0.32\textwidth]{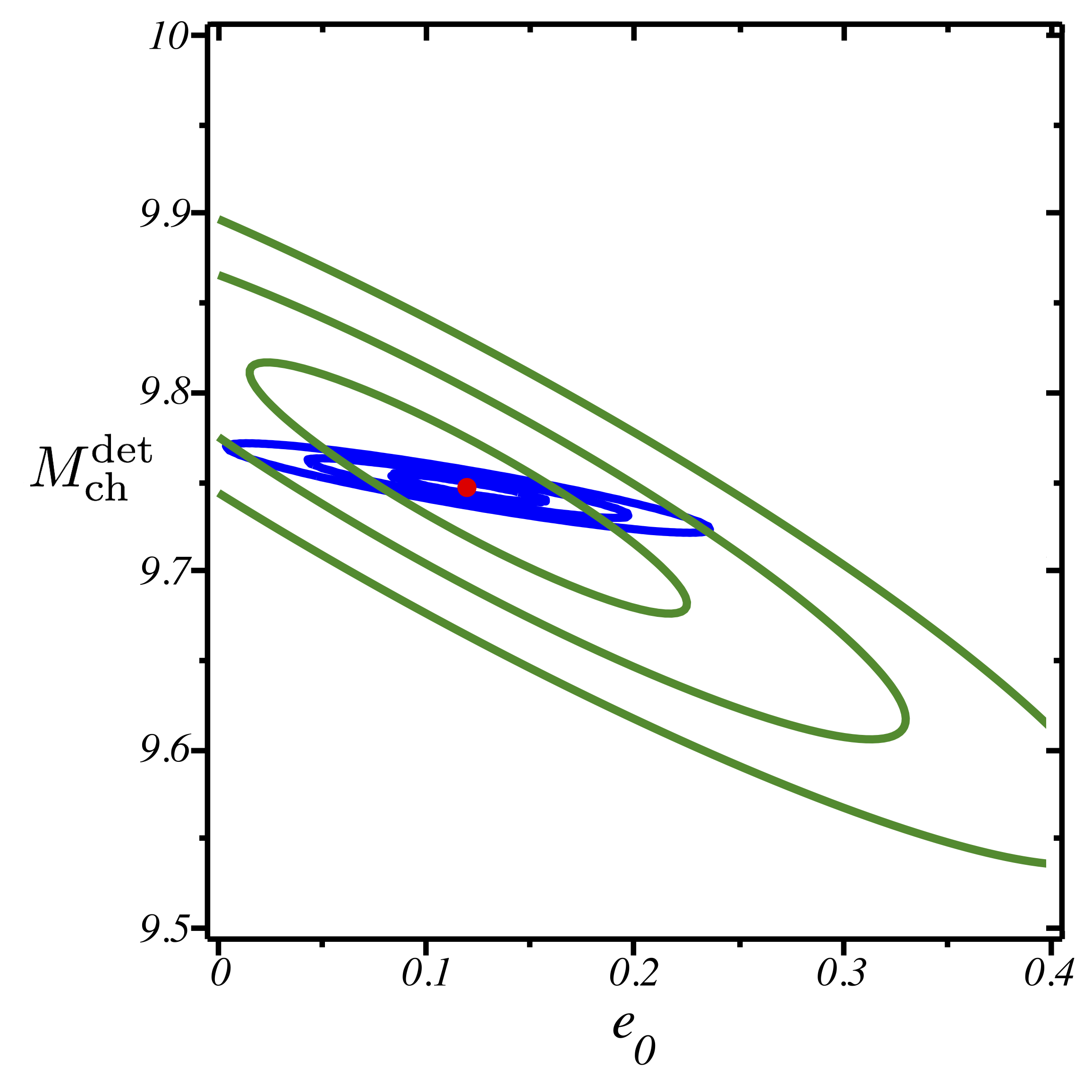} &
\includegraphics[angle=0, width=0.32\textwidth]{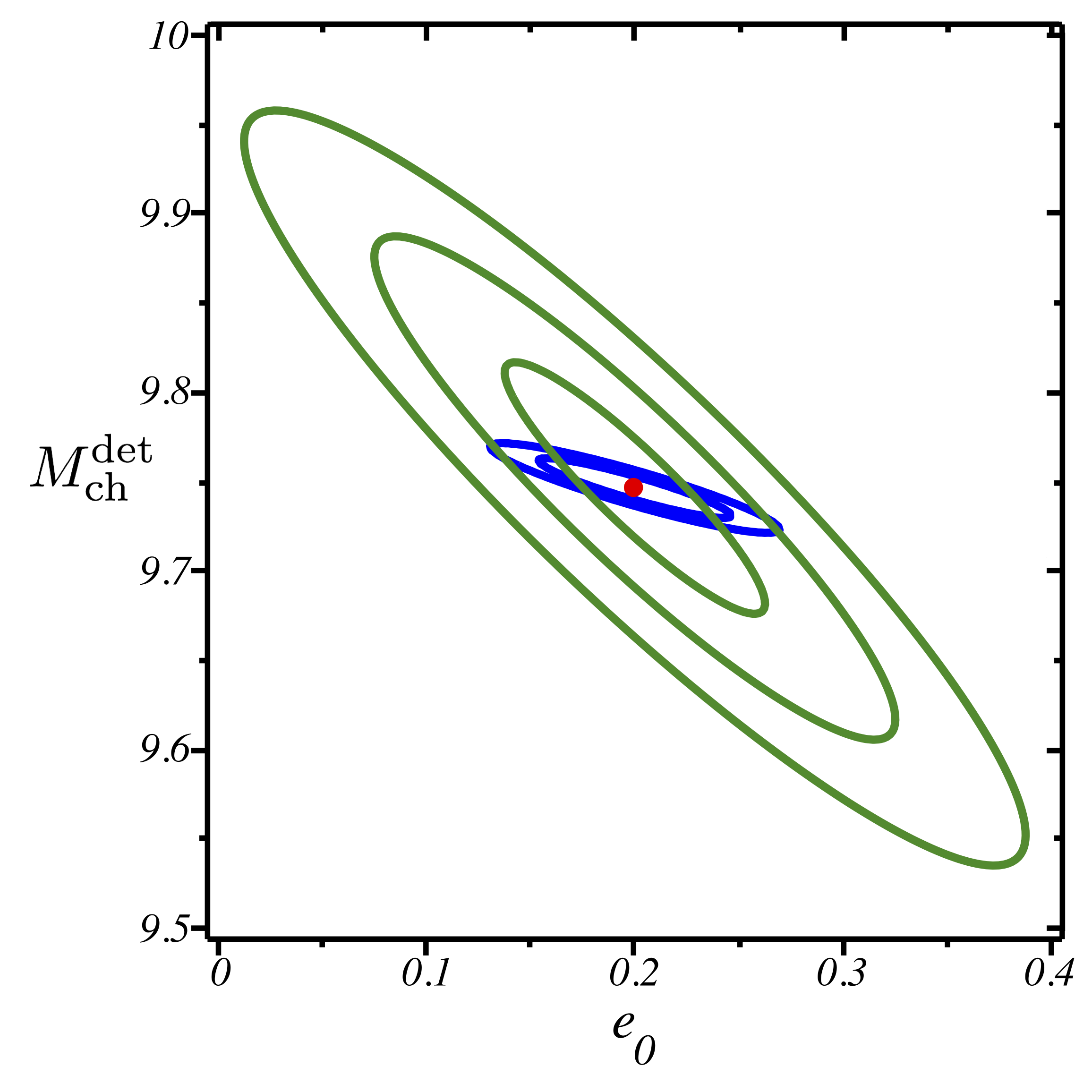}
\end{array}
$
\caption{\label{fig:e0Mchellipse}Error ellipses in the $e_0$-$M_{\rm ch}^{\rm det}$ plane computed via the Fisher matrix approach. The contours show confidence regions representing $n=1, 2,$ and $3$ standard deviations. The blue contours are computed via a 2D slice through the 5D error ellipsoid. The green contours are the 2D confidence regions after marginalizing over the parameters $t_c$, $\phi_c$, and $\eta$. The $y$-axis is in units of $M_{\odot}$. The red dot indicates the true parameter values: $M_{\rm ch, inj}^{\rm det}=9.746 M_{\odot}$ and $e_0^{\rm inj}=0.04,\,0.12, \,0.2$ from left to right, analogous to Fig.~\ref{fig:banana}. 
}
\end{figure*}
%
\subsection{MCMC results: Systematic errors}
In Fig.~\ref{fig:ecccircMCMCpdf} we show the systematic bias that results when eccentric signals are injected but recovered using circular templates (i.e., using the {\tt TaylorF2} waveform, which is identical to {\tt TaylorF2Ecc} but with $e_0$ set to zero). We focus on the resulting bias in $\eta$ and $M_{\rm ch}^{\rm det}$. There is clear indication of a growing bias in $\eta$ (top row) and $M_{\rm ch}^{\rm det}$ (bottom row) as $e_0$ is increased from $0.04$ to $0.2$. (A similar trend is noted in \cite{OSheaKumar2021}.)

Tables \ref{tab:MCMC-systematic_eta} and \ref{tab:MCMC-systematic_chirp} further quantify our MCMC results and also compare with the bias computed via the Fisher-Cutler-Vallisneri (FCV) formalism. There we see a generally increasing bias with increasing $e_0$, growing to $5\%$ in $\eta$ and $1\%$ in $M_{\rm ch}^{\rm det}$. The middle three columns of those tables compare the bias predicted by the MCMC and FCV methods. We see that they agree quite well: generally better than $1\%$ for $\eta$ and better than $0.1\%$ for $M_{\rm ch}^{\rm det}$. This confirms the utility of the Fisher/FCV approaches for estimating statistical and systematic errors in the high SNR limit. 

\section{\label{sec:degen}Investigating the $e_0 \mbox{--} M_{\rm ch}$ degeneracy}
We noted in Sec.~\ref{subsec:MCMCstat} that the bottom row of Fig.~\ref{fig:MCMCpdf-stat} shows an offset between the injected chirp mass and the peak of the recovered distribution; the offset grows with increasing $e_0$. (This behavior was independently noted in recent work by \cite{OSheaKumar2021}.) Here we investigate this further---both numerically and analytically---and show that it arises from a degeneracy between the chirp mass and the eccentricity parameter $e_0$.  

This degeneracy is clearly seen by examining the probability contours from our MCMC calculation in the 2D $e_0$-$M_{\rm ch}$ plane (Fig.~\ref{fig:banana}). These plots show a banana-shaped region, with probability density migrating rightward (increasing $e_0$) and downward (decreasing $M_{\rm ch}$) as $e_0$ increases. We note that similar behavior is also seen in Figs.~1 and 2 of \cite{lenon-nitz-brown2020MNRAS}, Fig.~A1 of \cite{wu-cao-zhu2020}, and Figs.~7 and 8 of \cite{OSheaKumar2021}.\footnote{References \cite{OSheaKumar2021,romero-shaw-eeccGWTC2} also discuss possible correlations between the eccentricity and the effective spin parameter $\chi_{\rm eff}$. While we do not investigate this in our MCMC analysis here, Fig.~\ref{fig:corr} in Appendix \ref{app:otherstaterrors} shows a weak correlation between $e_0$ and the individual spin parameters $\chi_i$ in our Fisher matrix analysis.}

To understand this behavior, we postulate that it is analogous to the well known banana-shaped degeneracy seen in probability contour plots in the $m_1$-$m_2$ plane, originating from the dominance of the chirp mass in the waveform. In Sec.~\ref{sec:eccchirp} above, we argued that a similar degeneracy exists between $e_0$ and $M_{\rm ch}$, arising from the fact that $M_{\rm ch}^{\rm ecc}$ is the ``effective parameter'' governing the phase evolution in binaries with small eccentricity. A comparison of Fig.~\ref{fig:banana} with Fig.~\ref{fig:m1m2chirp} qualitatively illustrates this, with contours of constant $M_{\rm ch}^{\rm ecc}$ bending down and to the right in the $e_0$-$M_{\rm ch}$ plane.  

This behavior can be further analyzed analytically via a simplified 2D Fisher-matrix calculation. In particular, we will show below that the increasing rotation of the probability contours in Fig.~\ref{fig:banana} as the injected eccentricity increases is predicted from the 2D $e_0$-$M_{\rm ch}$ Fisher error ellipse [Eq.~\eqref{eq:errorellipsoid}]. This ultimately arises from the relatively large value of the $\Gamma_{M_{\rm ch} e_0}$ component of the Fisher matrix.  

To understand this more fully, consider the 2D slice through the 5D error ellipsoid that spans the $e_0$-$M_{\rm ch}$ plane. Ignoring priors and setting $\theta^a=\hat{\theta}^a$ for $a \neq (M_{\rm ch}, e_0)$ in Eq.~\eqref{eq:errorellipsoid} yields an equation for the 2D $n$-sigma error ellipse,
\be
\label{eq:fisherellipse2d}
\frac{\Gamma_{e_0 e_0}}{n^2} \delta \theta_{e_0}^2 +\frac{\Gamma_{M_{\rm ch} M_{\rm ch}}}{n^2} \delta \theta_{M_{\rm ch}}^2 + \frac{\Gamma_{M_{\rm ch} e_0}}{n^2} \delta \theta_{e_0} \delta \theta_{M_{\rm ch}}  =1 \;,
\ee
where the Fisher matrix elements are given by Eq.~\eqref{eq:GammaAB} and $\delta \theta^a = \theta^a - \hat{\theta}^a$. [In this approach the other parameters in the Fisher matrix ($t_c, \phi_c, \eta$) are not marginalized over.] 

Consider separately the equation of an ellipse in the Cartesian $x$-$y$ plane that has semimajor axis $a$, semiminor axis $b$, and has been rotated about its center at $(x_0,y_0)$ by a counterclockwise angle $\theta$ from the positive $x$-axis. The equation for such an ellipse is\footnote{This is easily derived from the standard Cartesian form of an ellipse at the origin by applying a  counterclockwise rotation followed by a translation of the ellipse.}
\begin{multline}
    \label{eq:ellipse2d}
\left( \frac{\cos^2\theta}{a^2} + \frac{\sin^2\theta}{b^2} \right) \delta x^2 + \left( \frac{\cos^2\theta}{b^2} + \frac{\sin^2\theta}{a^2} \right) \delta y^2  \\ 
 - \left( \frac{1}{b^2} - \frac{1}{a^2} \right) \sin 2\theta \, \delta x \, \delta y =1 \,,
\end{multline}
where $\delta x= x-x_0$ and  $\delta y= y-y_0$. In terms of the eigenvalues of the $2 \times 2$ Fisher matrix 
\begin{multline}
\lambda_{\pm} = \frac{1}{2} \Big[ \Gamma_{M_{\rm ch} M_{\rm ch}} + \Gamma_{e_0 e_0} \\
\pm \sqrt{(\Gamma_{M_{\rm ch} M_{\rm ch}} -\Gamma_{e_0 e_0})^2 + (2\Gamma_{M_{\rm ch} e_0})^2}  \Big] ,
\end{multline} 
the error ellipse semimajor and semiminor axes are
\be
a=n/\sqrt{\lambda_{-}}, \;\;\; \text{and} \;\;\; b=n/\sqrt{\lambda_{+}} \;\;\;.
\ee
This is found by identifying $\delta x \equiv \delta \theta_{e_0}$ and $\delta y \equiv \delta \theta_{M_{\rm ch}}$, equating the coefficients of Eqs.~\eqref{eq:fisherellipse2d} and \eqref{eq:ellipse2d}, and solving for $a$, $b$, and $\theta$. 
The rotation angle of the ellipse is 
\be
\label{eq:ellipsetheta}
\theta = -\frac{1}{2} \arctan\left(\frac{2 \Gamma_{M_{\rm ch} e_0}}{\Gamma_{M_{\rm ch}M_{\rm ch}} -\Gamma_{e_0 e_0}} \right) \;.
\ee
From the form of Eq.~\eqref{eq:GammaAB} and assuming a 0PN model for $\Psi$ as in Eq.~\eqref{eq:Psiecc0PN}, we can see that the Fisher matrix elements will have the form:
\bs
\begin{align}
\Gamma_{M_{\rm ch} M_{\rm ch}} = &\; \alpha \frac{\rho^2}{M_{\rm ch}^2} \frac{1}{(\pi M_{\rm ch} f_c)^{10/3}} \left[1-O(e_0^2)\right] \,,\\
\Gamma_{e_0 e_0} = &\;  \beta \rho^2 \frac{1}{(\pi M_{\rm ch} f_c)^{10/3}} e_0^2 \left(\frac{f_0}{f_c} \right)^{38/9} \,, \; \text{and} \\
\Gamma_{M_{\rm ch} e_0} = &\; \gamma \frac{\rho^2}{M_{\rm ch}} \frac{1}{(\pi M_{\rm ch} f_c)^{10/3}}  e_0 \left(\frac{f_0}{f_c} \right)^{19/9} \nonumber \\ 
&\times \left[1-O(e_0^2)\right] ,
\end{align} 
\es
where $(\alpha, \beta, \gamma)$ are positive dimensionless constants that, along with $f_c$, serve as placeholders for the numerical integration in Eq.~\eqref{eq:GammaAB}. 
For the cases of interest here, $\Gamma_{M_{\rm ch} M_{\rm ch}} \gg \Gamma_{M_{\rm ch} e_0} \gg \Gamma_{e_0 e_0}$.

In the limit of small $e_0$, $\theta$ can be simplified to
\be
\theta \approx -\frac{\Gamma_{M_{\rm ch} e_0}}{\Gamma_{M_{\rm ch}M_{\rm ch}}} \approx -\frac{\gamma}{\alpha} M_{\rm ch} e_0 \left(\frac{f_0}{f_c} \right)^{19/9} \sim -e_0 M_{\rm ch} \;.
\ee
This shows that the semimajor axis of the error ellipse in the $e_0$-$M_{\rm ch}$ plane will be nearly horizontal for negligible $e_0$, but will slope downward as $e_0$ increases ($\theta<0$ for $e_0>0$, corresponding to a clockwise rotation of the ellipse in the plane). We confirm this analytic expectation by directly computing the error ellipses via Eq.~\eqref{eq:fisherellipse2d}, using the same parameters and Fisher matrix code used in Sec.~\ref{sec:mcmc} (e.g., a 5-D Fisher matrix using 3PN order nonspinning waveforms, and replacing $M_{\rm ch}$ with the detector-frame chirp mass). 

The result is shown in Fig.~\ref{fig:e0Mchellipse}, where the blue error contours show the expected behavior described by our analytical analysis and qualitatively match the behavior of Fig.~\ref{fig:banana}. More properly, we also compute the marginalized error ellipse, which is shown via the green contours in Fig.~\ref{fig:e0Mchellipse}. These are computed by constructing a new 2D Fisher matrix $\tilde{\Gamma}_{jk}$ by removing the $\theta_a=(t_c,\phi_c,\eta)$ rows and columns from the 5D covariance matrix $\Sigma_{AB}$, and then inverting the resulting $2 \times 2$ matrix \cite{coe-fishermatrix2009,schurrefPRD}. The marginalized error ellipse is then constructed via $\Gamma_{AB} \rightarrow \tilde{\Gamma}_{jk}$ in Eq.~\eqref{eq:fisherellipse2d}. These marginalized error ellipses (green curves in Fig.~\ref{fig:e0Mchellipse}) more appropriately match the MCMC contours in Fig.~\ref{fig:banana}. 

\section{\label{sec:concl}Conclusions}
Our goal was to explore the capability of a LIGO-type detector to constrain eccentricity for a range of inspiralling compact binary systems. We made use of a combination of Fisher matrix and Bayesian inference techniques, generally finding good agreement between the two approaches. Unlike many prior studies, we include spin interactions when applying the Fisher matrix formalism, but we work in the small-eccentricity limit. The latter is astrophysically realistic as large binary eccentricity is very rapidly reduced. Even if binaries with eccentricities as small as $\sim 0.01$ are detected, this implies very large eccentricities ($\sim 0.99$) only $\sim 10^2$ to $10^4$ years before detection. 

In practice, eccentricity as small as $\approx 0.01$ will be quite difficult to detect with LIGO (except perhaps for binary neutron star systems). Eccentricities $e_0 \gtrsim 0.02 \mbox{--} 0.2$ are more likely to be constrained, provided that LIGO reaches its design sensitivity at $10$ Hz (see, e.g., Fig.~\ref{fig:PEe0results}). 

Both binary neutron stars and neutron-star/black hole systems are promising candidates for constraining eccentricity, due to the large number of GW cycles that such systems have in the detector's frequency band. Considering GW151226 as a prototypical example of the binary black hole systems with several inspiral cycles seen by LIGO/Virgo, our Fisher and MCMC analyses indicate that a source with $e_0\gtrsim 0.1\mbox{--}0.2$ is needed to place a modest constraint on $e_0$. This is consistent with limits on the eccentricity of GW151226 placed in recent works \cite{wu-cao-zhu2020,OSheaKumar2021}.

As most GW signal templates do not include eccentricity, we also studied the resulting systematic bias on the other intrinsic parameters when eccentricity is neglected. This bias becomes significant when $e_0 \gtrsim 0.01$ to $0.1$. Binary black holes with parameters and SNRs similar to GW150914 and GW151226 are unlikely to be biased by eccentricity unless $e_0 \gtrsim 0.1$ to $0.2$. Systematic bias was studied via the Fisher-Cutler-Vallisneri formalism and the LALInference MCMC approach, which agree well with each other. The inclusion of eccentric waveform effects is important to reduce systematic bias in detected GW events.

Separate from the systematic error bias, we also discovered an intrinsic bias in the recovery of the chirp mass parameter which grows with increasing eccentricity. This was ultimately explained by introducing the \emph{eccentric chirp mass}, which acts as an effective mass parameter in the waveforms for low-eccentricity binaries. Because the eccentric chirp mass is the best measured mass parameter, this introduces a degeneracy between the eccentricity parameter $e_0$ and the standard chirp mass $M_{\rm ch}$. This degeneracy is manifested as a tilted error ellipsoid in the $e_0$-$M_{\rm ch}$ plane, with the tilt angle proportional to $e_0$. 

Combined with information on the component spins, measurements of binary eccentricity from GW observations can inform us about the formation pathways of compact object binaries. It will be interesting to investigate the possibility of nonzero eccentricity in GW events from the third and fourth observing runs of LIGO/Virgo. As demonstrated here, improvements in the detector sensitivity at low frequencies ($\approx 10 \mbox{--} 30$ Hz) is essential to increasing the odds of measuring or constraining binary eccentricity. 

\begin{acknowledgments}
We thank Shaon Ghosh for helpful feedback on the manuscript and Edwin Son for plotting assistance. M.F.~was supported by NSF (National Science Foundation) Grants No.~PHY-1308527 and No.~PHY-1653374, and a grant from the Simons Foundation (554674, MF). C.K.~was supported by National Research Foundation (NRF) Grants (No.~2018R1D1A1B07047677, No.~2021R1F1A1062969). K.G.A.~is partially supported by the Swarnajayanti Fellowship Grant No.~DST/SJF/PSA-01/2017-18, Grant No.~EMR/2016/005594, MATRICS grant (Mathematical Research Impact Centric Support) MTR/2020/000177 of the Science and Engineering Research Board (SERB), and a grant from the Infosys Foundation. H.W.L.~and J.K.~were supported by National Research Foundation (NRF) Grant (No.~2018R1D1A1B0549338).  The authors are also grateful to KISTI (Korea Institute of Science and Technology Information) for providing computing resources through the GSDC (Global Science experimental Data hub Center). This manuscript is assigned LIGO DCC number P2100284.
\end{acknowledgments}

\appendix
%
\section{\label{app:eccevolve}\uppercase{Evolution equations for binary eccentricity}}
In this Appendix we briefly discuss the evolution equations for binary eccentricity used to generate the data in Table \ref{tab:eccevol}. When we treat the compact objects as point particles and ignore their spins, then the shape of the orbital ellipse (averaged over an orbital timescale) can be described via two parameters: a time-averaged eccentricity parameter (here chosen to be the time-eccentricity $e_t$) and a second ``orbital size'' parameter (which we will denote here as $\tilde{ R}$). A number of possible choices for $\tilde{R}$ are possible, including the semimajor axes of the instantaneous ellipse, the semiminor axis, the periastron distance, the apastron distance, or any of a number of variables relating to the orbital frequency or speed. For whatever choice is convenient, it is well known that the time evolution of the parameters $(e_t, \tilde{R})$ are governed by a coupled set of ordinary differential equations of the form:
\begin{subequations}
\label{eq:eccevolve}
\begin{align}
\frac{d e_t}{dt} &= F(e_t, \tilde{R}) \;,\\    
\frac{d \tilde{R}}{dt} &= G(e_t, \tilde{R}) \;,
\end{align}
\end{subequations}
where $F$ and $G$ denote particular functions of the dynamical variables that are computable to 3PN order and depend on the system masses. Explicit solutions of time $[e_t(t), \tilde{R}(t)]$ require a numerical solution of these ODEs. However, for some choices of $\tilde{R}$, an analytic solution of the form $\tilde{R} = H[e_t; \tilde{R}_i, e_{i}]$ is possible (where $H$ is an unspecified function and $\tilde{R}_i, e_i$ are values of the dynamical variables at a particular reference time $t_i$. 

For simplicity, we evolve versions of Eq.~\eqref{eq:eccevolve} at leading post-Newtonian order (0PN order) choosing $\tilde{R}$ to be the Newtonian periastron separation $r_p$. Making use of Eqs.~(2.20), (2.26), (2.29), (2.39), and (2.40) in \cite{favata-eccentricmemory}, we numerically integrate the system
\begin{subequations}
\label{eq:eccevolve2}
\begin{align}
\frac{d e_t}{dt} &= -\frac{\eta}{15M} \left(\frac{M}{r_p}\right)^4 \frac{e_t (1-e_t^2)^{3/2}}{(1+e_t)^{4}} \left( 304+121 e_t^2 \right)    \;,\\
\frac{d r_p}{dt} &= -\frac{\eta}{15} \left(\frac{M}{r_p}\right)^3 \frac{(1-e_t)^{3/2}}{(1+e_t)^{7/2}} \nonumber \\
& \;\;\;\;\;\; \times \left( 192-112 e_t + 168 e_t^2 + 47 e_t^3 \right)  \;,    
\end{align}
\end{subequations}
where $M=m_1+m_2$, $\eta = m_1 m_2/M^2$, and Kepler's third law provides a relationship between $r_p$ and the GW frequency $f_{\rm gw}$ (denoted by $f$ in the main text),
\be
\frac{r_p}{M} = \frac{(1-e_t)}{(\pi M f_{\rm gw})^{2/3}} \; .
\ee
Given initial conditions $e_t(0)=e_0$, $f_{\rm gw}(0)=10$ Hz, and switching the sign of the time variable ($t \rightarrow -t$) to easily allow for a ``backward'' in time integration, the above equations can be numerically solved. Combined with a root-solving procedure, the time $\Delta T$ to reach a specified value of eccentricity [e.g., $e_t(\Delta T)=0.999$] is then computed to generate the values in Table \ref{tab:eccevol}.

\section{\label{app:cosmo}\uppercase{Cosmological effects}}
When considering sources at significant distances, we need to incorporate cosmological effects. This procedure is well known, but often not made explicit. Here we provide a brief explanation for clarity and pedagogical purposes. Consider a general GW signal that is observed outside (but near to) the wave zone of the source: $h_{\rm src}(t_{\rm src})\equiv H(t_{\rm src})/D$. Here $t_{\rm src}$ is the time measured in a frame comoving with the source and $H$ is the signal with the distance $D$ factored out. This distance is  naturally the proper distance between the source and the nearby ``local'' observer \cite{maggiore-GWvol1}. In terms of the Friedman-Robertson-Walker (FRW) metric, if the source and local observer are assigned comoving radial coordinates $r_1$ and $r_2$ respectively, then they are separated by a proper distance $D\approx a(t_{\rm src}) r$, where $r=r_2-r_1$, and $a(t)$ is the scale factor entering the FRW metric. (This assumes we are near the source such that the spatial curvature and expansion of the universe are negligible.) Far from the source, the scale factor is related to the redshift $z$ via $a(t_{\rm src})/a(t_{\rm obs})= 1/(1+z)$; the times at the source $t_{\rm src}$ and the distant observer $t_{\rm obs}$ are related via $dt_{\rm src} = dt_{\rm obs}/(1+z)$; and the frequencies in the two frames are related via $f_{\rm src} = (1+z) f_{\rm obs}$. In a flat universe the GW signal observed at Earth is then $h_{\rm obs}(t_{\rm obs}) = H(t_{\rm obs})/[a(t_{\rm obs}) r]$, where $r$ is constant and $a(t_{\rm obs}) r$ refers to the much larger distance to the Earth-based observer. (See Ch.~4.1.4 of \cite{maggiore-GWvol1} for a full justification.) It is conventional to replace the proper distance between the source and distant observer $a(t_{\rm obs}) r$ with the luminosity distance $d_L= (1+z) a(t_{\rm obs}) r$. [The luminosity distance is the quantity that naturally arises in the relationship between observed flux ${\mathcal F}$ and luminosity at the source ${\mathcal L}$, ${\mathcal F} ={\mathcal L}/(4\pi d_L^2)$.] Ignoring the irrelevant constant time shift between $t_{\rm src}$ and $t_{\rm obs}$, the observed waveform is thus related to the source-frame waveform via
\begin{align}
h_{\rm obs}(t_{\rm obs}) &= (1+z) \frac{H[t_{\rm src} \rightarrow t_{\rm obs}/(1+z)]}{d_L} \,, \nonumber \\
&=h_{\rm src} \left[ t_{\rm src} \rightarrow \frac{t_{\rm obs}}{(1+z)}, D\rightarrow \frac{d_L}{(1+z)} \right]\,.
\end{align}

To compute the FT of the signal [Eq.~\eqref{eq:fourier2}] in the distant observer's frame and relate it to the FT in the source frame, we substitute the above relation between the two time coordinates:
\begin{align}
\tilde{h}_{\rm obs} (f_{\rm obs}) &= \int_{-\infty}^{+\infty} h_{\rm obs}(t_{\rm obs}) e^{2\pi i f_{\rm obs} t_{\rm obs}} dt_{\rm obs} \,, \nonumber  \\
&= (1+z)  \int_{-\infty}^{+\infty} h_{\rm src}(t_{\rm src}) e^{2\pi i f_{\rm src} t_{\rm src}} dt_{\rm src} \,, \nonumber \\
&= (1+z) \tilde{h}_{\rm src}(f_{\rm src})  \,, \nonumber \\
& = (1+z) \tilde{h}_{\rm src}\left[f_{\rm src} \rightarrow (1+z) f_{\rm obs} \right],
\end{align}
where in the second line we have ignored an arbitrary phase offset of the form $e^{2\pi i f_{\rm obs}(1+z) C}$ for constant $C$ that arises from the relation between $t_{\rm obs}$ and $t_{\rm src}$; this can be absorbed into the coalescence time $t_c$. In the last line we must again replace $D \rightarrow d_L/(1+z)$ to express our result in terms of the luminosity distance. 

Taking the above transformations into account and setting henceforth $f\equiv f_{\rm obs}$ (i.e., dropping the ``obs'' and ``src'' labels), our waveform in the observer frame becomes
\begin{align}
\label{eq:waveform-z}
\tilde{h}(f) =& \;{\mathcal A}_z e^{i \Psi(f)} = \hat{{\mathcal A}}_z f^{-7/6} e^{i \Psi(f)}\,, \;\;\;\; \text{where} \\
\label{eq:Az}
{\mathcal A}_z =& -M \sqrt{\frac{5\pi}{96}} (1+z)^{5/6} \left(\frac{M}{d_L}\right) \sqrt{\eta} (\pi M f)^{-7/6} \nonumber \\
& \times \left[(1+C^2)^2 F_{+}^2 + 4 C^2 F_{\times}^2 \right]^{1/2} \,,
\end{align}
$\hat{\mathcal A}_z \equiv {\mathcal A}_z f^{7/6}$, and $\Psi(f)$ above is given by replacing the frequency in Eq.~\eqref{eq:Psiterms} via $\Psi[f\rightarrow (1+z) f]$ . In all equations used here, masses refer to their source-frame values. The above scheme for incorporating cosmological redshifts is equivalent to starting from Eq.~\eqref{eq:spawaveform} but replacing $D \rightarrow d_L$ and $M \rightarrow M_{\rm obs}=(1+z) M$, with $M$ referring to the ``source frame'' total mass and $M_{\rm obs}$ the ``detector-frame'' or ``observer-frame'' total mass.  
In our calculations we make use of the luminosity-distance/redshift relation for a flat universe \cite{hogg},
\be
\label{eq:dLz}
d_L(z) = \frac{c}{H_0} (1+z) \int_0^z \frac{dz'}{\sqrt{\Omega_M (1+z')^3 + \Omega_{\Lambda}}},
\ee
where
$H_0=100 h \, {\rm (km/s)/Mpc}$. We use the cosmological parameters given in Table 4, column 3 of \cite{planck2015-cosmoparam}: $h=0.6790$, $\Omega_M=0.3065$, and $\Omega_{\Lambda}=0.6935$. Throughout this work we make use of the unit conversions in Eqs.~(7.3) \mbox{--} (7.6) and footnote 21 of \cite{moore-etal-PRD2016}, as well as the relationship between Mpc and seconds:\footnote{This follows from the definitions of the parsec $(1 {\rm pc} \equiv 648 000/\pi \,{\rm AU})$ and the AU $(1 {\rm AU} \equiv 149\,597\,870\,700 \,{\rm m})$ \cite{IAU2012,IAU2015}.}
\be
1\, {\rm Mpc} = 1.029271250 \times 10^{14} \, {\rm sec} \,.
\ee

\section{\label{app:iscoeqs}\uppercase{Equations for the ISCO, final mass, and final spin}}
In this Appendix we specify the formulas needed to construct the redshifted ISCO (inner-most stable circular orbit) frequency in Eq.~\eqref{eq:fisco}, which is used to set the high-frequency cutoff of our SNR and Fisher-matrix integrals. That formula depends on the ISCO of a Kerr black hole with mass and spin parameters determined by the final BH merger remnant formed following a BBH or NS/BH collisions (we do not distinguish between these two types of binaries for the purpose of setting a termination frequency for our inspiral waveforms).

The dimensionless Kerr ISCO angular frequency $\hat{\Omega}_{\rm isco}(\chi)$ for a Kerr BH with mass $M_{\rm kerr}$ and spin parameter $\chi \in [-1,1]$ is given by \cite{bptkerr}: 
\bs
\begin{align}
& \hat{\Omega}_{\rm isco}(\chi) \equiv M_{\rm kerr} \Omega_{\rm isco} = \frac{1}{\hat{r}_{\rm isco}^{3/2}(\chi) + \chi} \,, \\
& \hat{r}_{\rm isco}(\chi) \equiv \frac{r_{\rm isco}}{M_{\rm kerr}} \\
& \;\;\;\;\;\;\;= 3+Z_2-\frac{\chi}{|\chi|} \sqrt{(3-Z_1) (3+Z_1+2 Z_2)} \,, \\
& Z_2 = \sqrt{3 \chi^2 + Z_1^2} \,, \\
& Z_1 = 1+(1-\chi^2)^{1/3}  \left[ (1+\chi)^{1/3}+ (1-\chi)^{1/3} \right] \,. 
\end{align}
\es
The final BH mass $M_f$ appearing in Eq.~\eqref{eq:fisco} is given by Eqs.~(3.7) and (3.8) of \cite{husa-khan-etalPRD2016} in terms of the radiated GW energy $E_{\rm rad}$:
\bs
\begin{align}
M_f =&\; M (1-E_{\rm rad}/M)\; ,  \;\;\;\; \text{where} \\
\frac{E_{\rm rad}}{M} =&\; (0.0559745 \eta + 0.580951 \eta^2  \\
&-0.960673 \eta^3 + 3.35241 \eta^4) \nonumber  \\
& \times \left[  \frac{1+\hat{S} (-0.00303023 - 2.00661 \eta + 7.70506 \eta^2)}{1+\hat{S} (-0.67144 - 1.47569 \eta + 7.30468 \eta^2)} \right] \,, \nonumber \\
& \text{and} \;\;\;\; \hat{S} = \frac{(\chi_1 m_1^2 + \chi_2 m_2^2)/M^2}{1-2\eta} \,.
\end{align}
\es
(Recall that $M=m_1+m_2$ is the sum of the source-frame component masses.)

The dimensionless spin parameter of the final BH is given by Eq.~(7) of \cite{hofmann-barausse-rezzola-finalspinApJL2016}:
\begin{align}
\label{eq:chifinal}
\chi_{\rm f} =& \;  a_{\rm tot} + \eta \big\{ \hat{L}_{\rm isco} (a_{\rm eff}) - 2 a_{\rm tot} [\hat{E}_{\rm isco} (a_{\rm eff}) - 1] \big\} \nonumber \\
& +(k_{00} + k_{01} a_{\rm eff} + k_{02} a_{\rm eff}^2) \eta^2 \nonumber \\
& + (k_{10} + k_{11}  a_{\rm eff} +k_{12} a_{\rm eff}^2) \eta^3 ,
\end{align}
where the dimensionless orbital energy and angular momentum are
\bs
\begin{align}
\hat{E}_{\rm isco}(\chi) &= \sqrt{1-\frac{2}{3 \hat{r}_{\rm isco}(\chi)}} \,, \\
\hat{L}_{\rm isco}(\chi) &= \frac{2}{3\sqrt{3}} \left[ 1+ 2\sqrt{3 \hat{r}_{\rm isco}(\chi)-2} \right] \,,
\end{align}
\es
and where
\bs
\begin{align}
a_{\rm eff} &= a_{\rm tot} + \xi \eta (\chi_1+\chi_2) \,, \\
a_{\rm tot} &= \frac{\chi_1 + \chi_2 (m_2/m_1)^2}{(1+m_2/m_1)^2} \,, \;\;\; \text{with}
\end{align}
\es
\begin{align}
k_{01} &=-1.2019\,, \;\;\; k_{02} =-1.20764\,, \;\;\; k_{10} =3.79245  \,, \nonumber \\
k_{11} &=1.18385 \,, \;\;\;  k_{12} =4.90494 \,,\;\;\; \xi =0.41616 \,, \nonumber \\
k_{00} &=-3.821158961 \,.
\end{align}
Here $k_{00}$ follows from the constraint in Eq.~(11) of \cite{hofmann-barausse-rezzola-finalspinApJL2016}. 
Throughout our calculations we also make use of the following relationship between the mass parameters $(m_1,m_2)$ and $(M,\eta)$:
\bs
\begin{align}
m_1 &= \frac{M}{2} \left( 1 + \sqrt{1-4\eta}  \right) \,, \\
m_2 &= \frac{M}{2} \left( 1 - \sqrt{1-4\eta}  \right)\,.
\end{align}
\es
\\

\section{\label{app:otherstaterrors}\uppercase{Additional Fisher matrix statistical errors and parameter correlations}}
\begin{figure*}[t]
$
\begin{array}{cc}
\includegraphics[angle=0, width=0.48\textwidth]{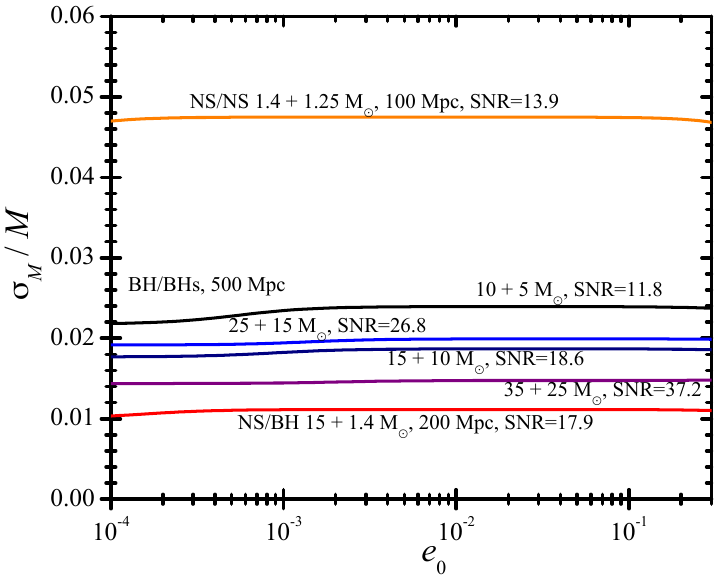} &
\includegraphics[angle=0, width=0.49\textwidth]{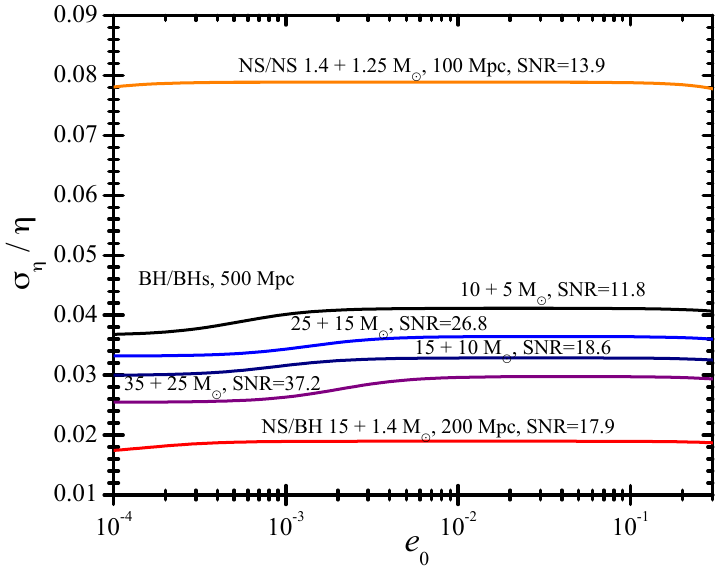} \\
\includegraphics[angle=0, width=0.48\textwidth]{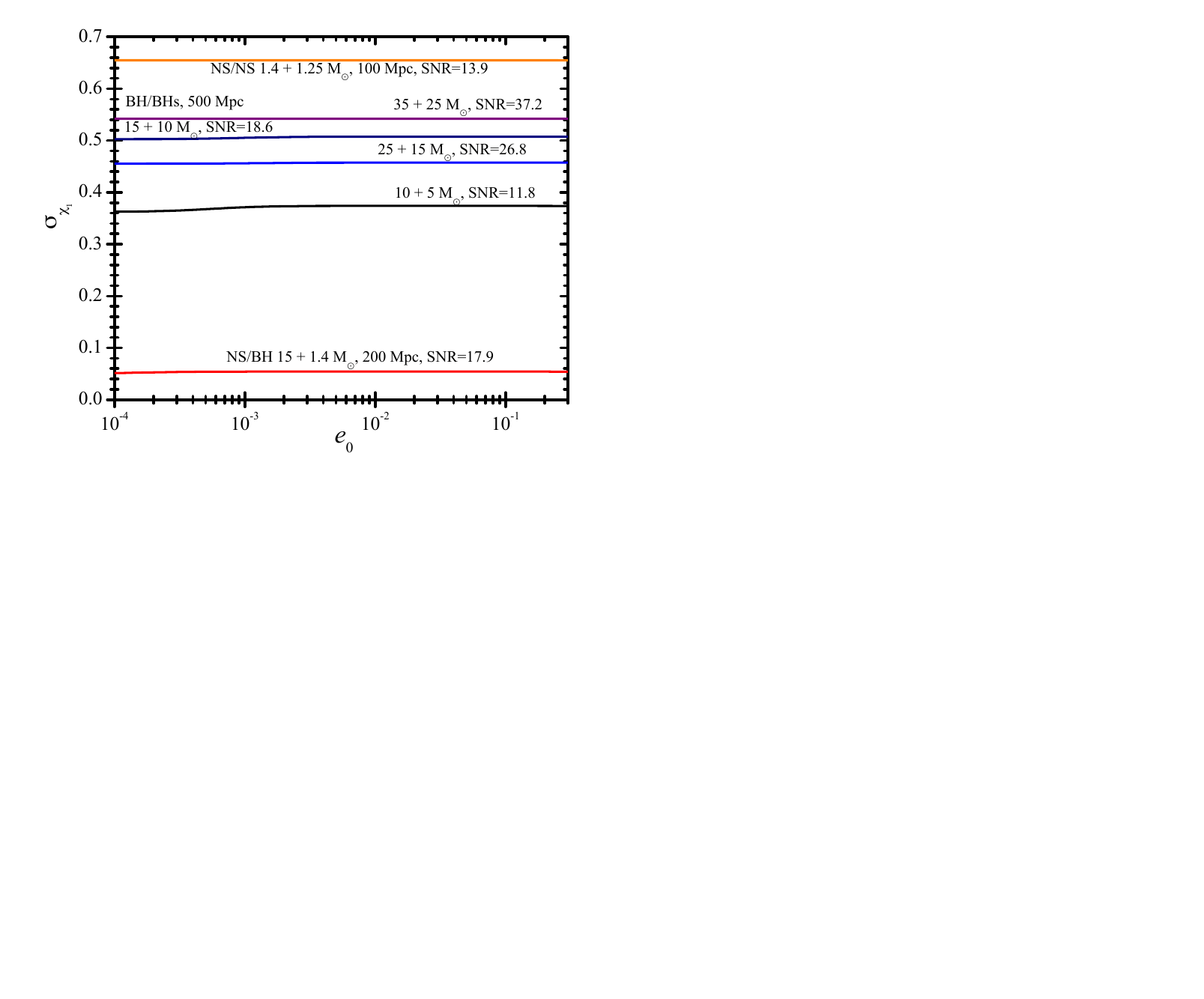} &
\includegraphics[angle=0, width=0.48\textwidth]{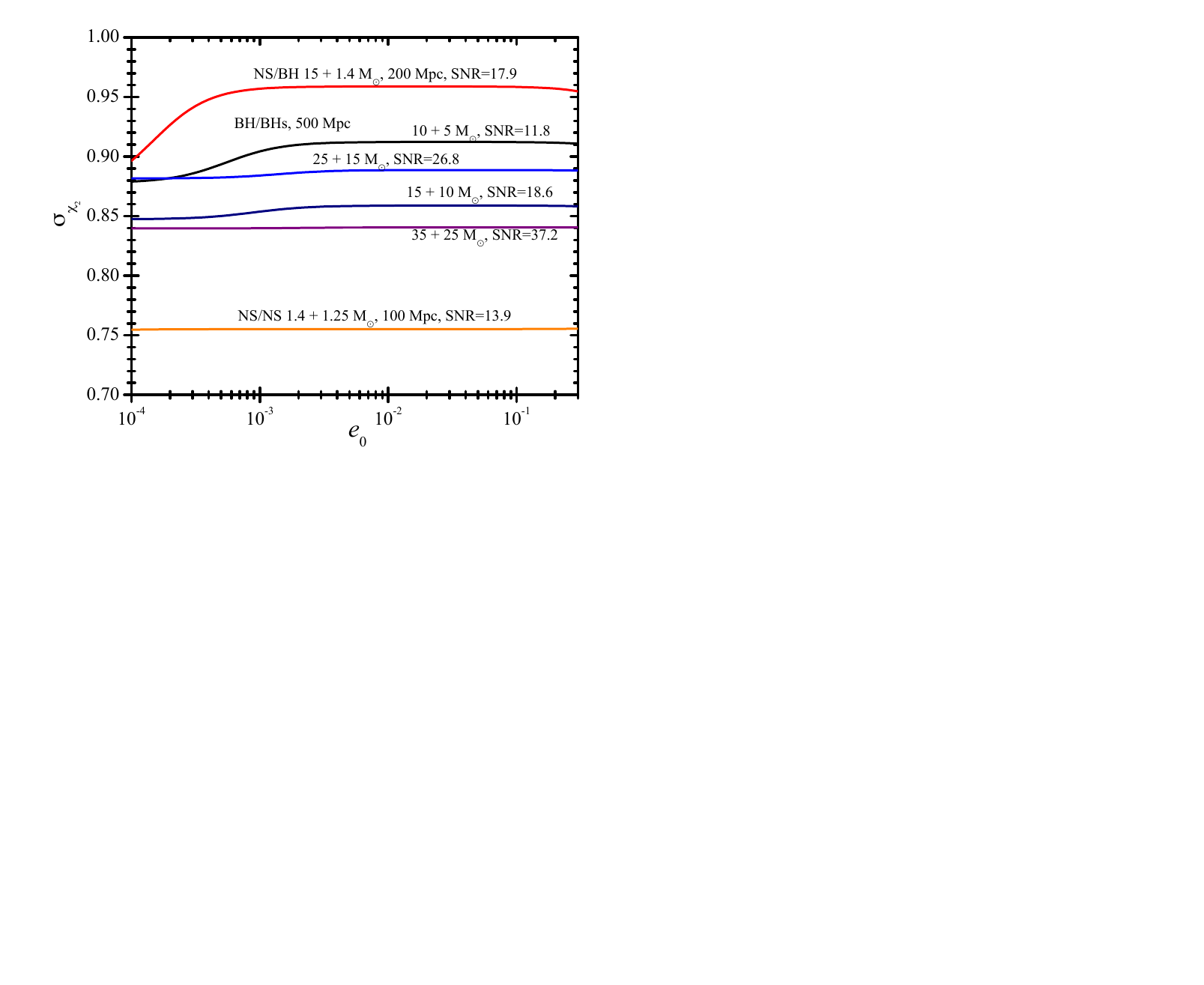}
\end{array}
$
\caption{\label{fig:PEotherresults}One-sigma parameter estimation errors for the binary total mass $M$, reduced mass ratio $\eta$, and dimensionless spin parameters $\chi_{1,2}$ for the same systems shown in the left panel of Figure \ref{fig:PEe0results}. The parameter errors are nearly independent of eccentricity, except for the secondary's spin parameter $\chi_2$ in unequal mass systems.}
\end{figure*}
\begin{figure*}[t]
$
\begin{array}{cc}
\includegraphics[angle=0, width=0.48\textwidth]{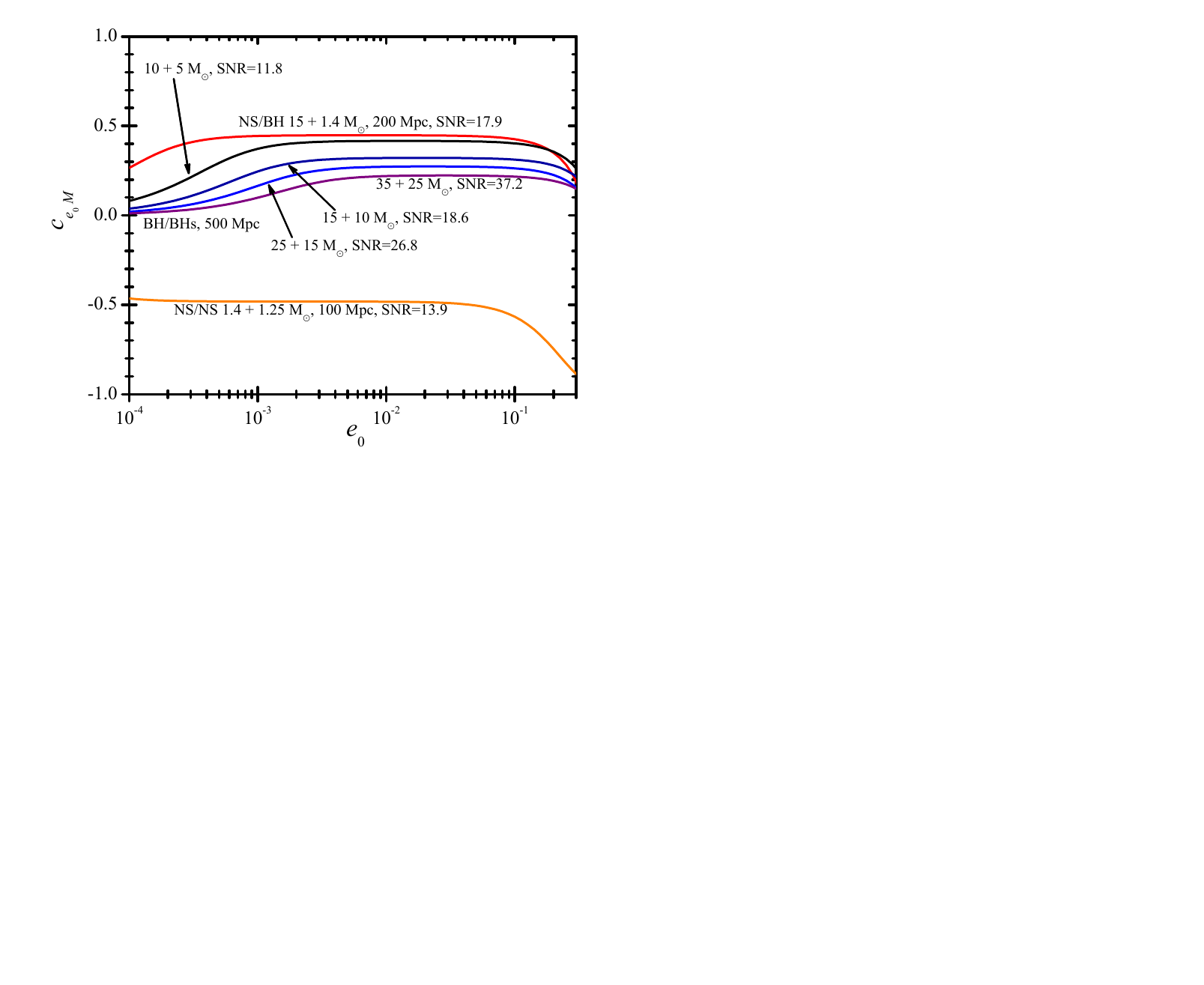} &
\includegraphics[angle=0, width=0.48\textwidth]{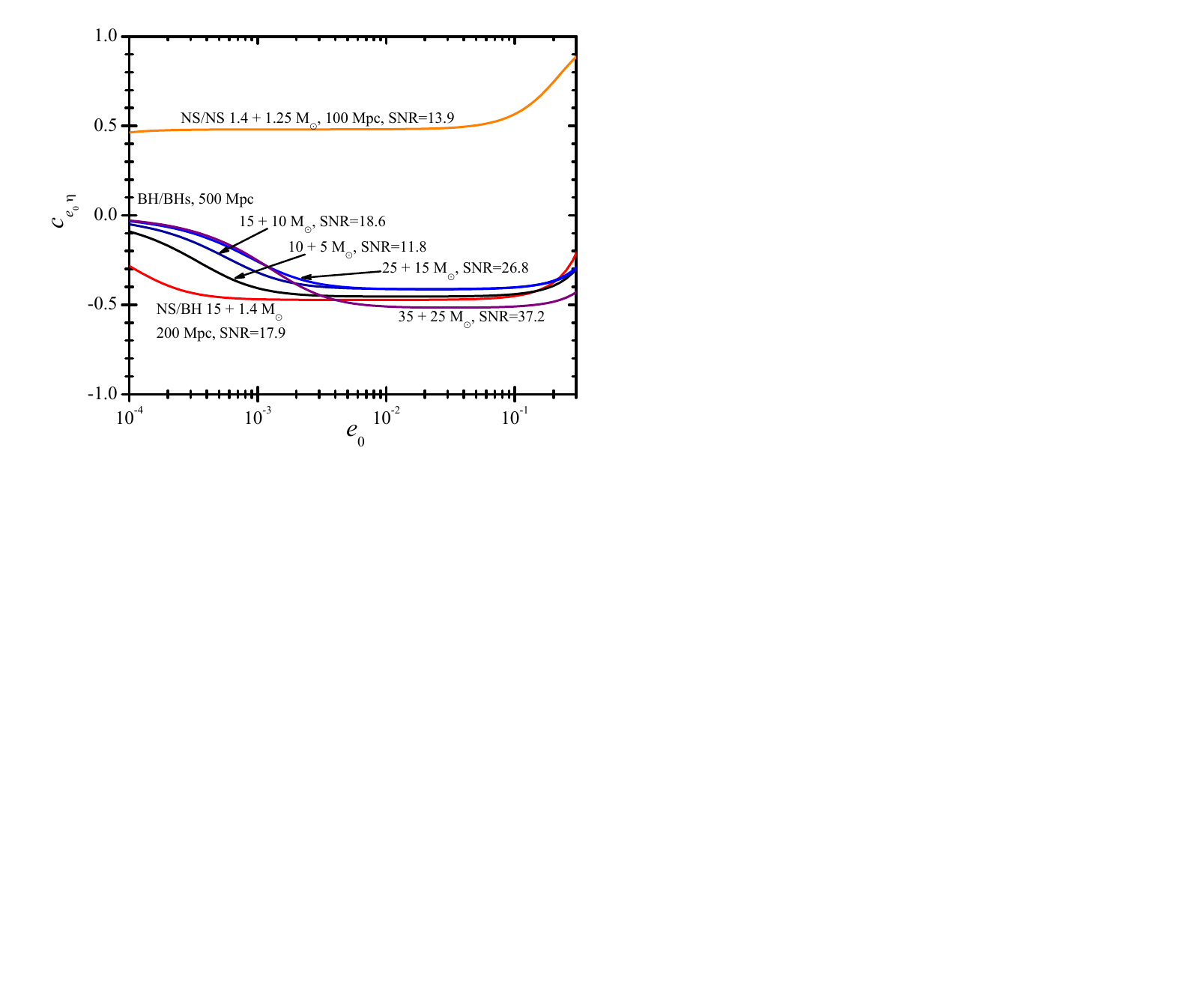} \\
\includegraphics[angle=0, width=0.48\textwidth]{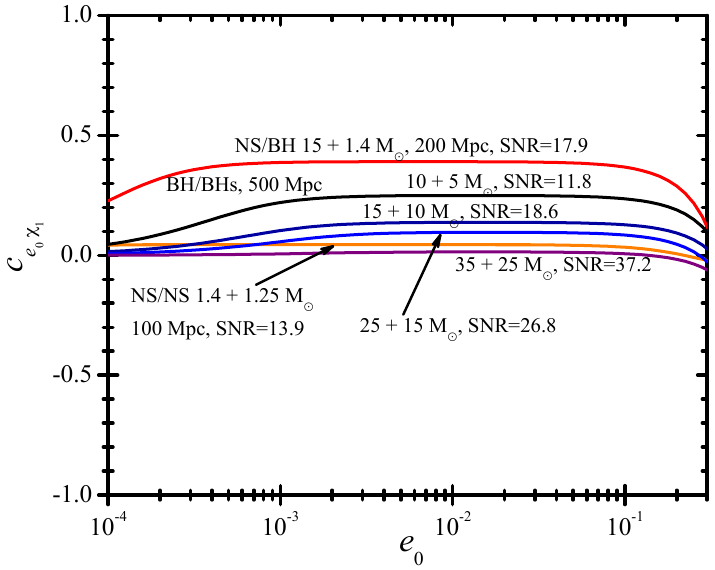} &
\includegraphics[angle=0, width=0.48\textwidth]{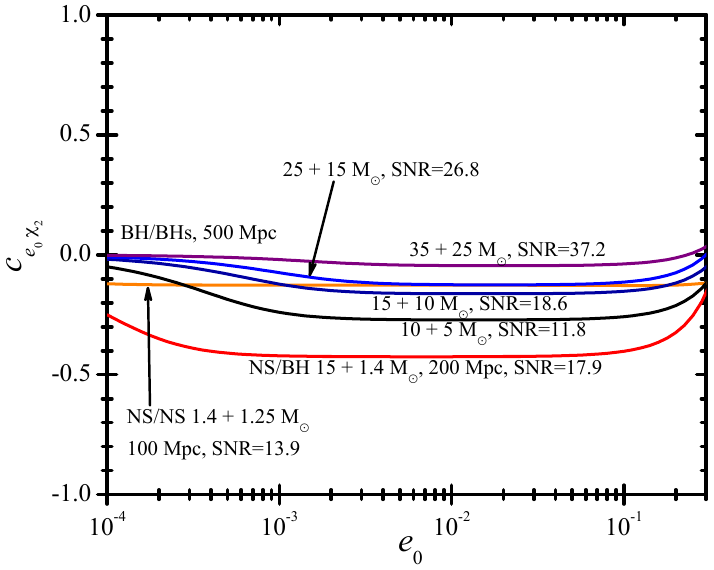}
\end{array}
$
\caption{\label{fig:corr}Correlations with eccentricity $e_0$. Each panel shows a component of the correlation matrix $c_{AB}$ for $A=e_0$ and $B=(M,\eta,\chi_1,\chi_2)$. Each curve corresponds to the same systems as in Figure \ref{fig:PEotherresults} and the left panel of Figure \ref{fig:PEe0results}. }
\end{figure*}
This Appendix shows additional Fisher-matrix results not presented in the main text. In the left panel of Fig.~\ref{fig:PEe0results} of the main text, we considered the statistical error in the eccentricity as a function of eccentricity for a variety of ``fiducial'' compact binary systems. Figure \ref{fig:PEotherresults} here shows similar results for the other intrinsic parameters of those systems, $(M, \eta, \chi_1, \chi_2)$, again as a function of $e_0$. The parameters $M$ and $\eta$ are measured with very good precision ($\sim 1\%$ to $8\%$). The spin parameters are measured with poor precision, except for the BH spin $\chi_1$ in the NS/BH system ($\sigma_{\chi_1} \sim 0.06$).  We also see that the statistical parameter errors on these four parameters are nearly constant with $e_0$. However, there is a small variation in the parameter errors with $e_0$ which is more prominent for systems with more unequal mass ratios.  

Figure \ref{fig:corr} shows the correlations of these same parameters $(M, \eta, \chi_1, \chi_2)$ with the eccentricity $e_0$. Specifically, we plot the $c_{e_0 \theta_A}$ coefficients of the correlation matrix [Eq.~\eqref{eq:cAB}] as a function of $e_0$. We see that $|c_{e_0 \theta_A}| \lesssim 0.5$ in almost all cases. Only for $c_{e_0 M}$ and $c_{e_0 \eta}$ for $e_0 \gtrsim 0.1$ in the NS/NS binary does the coefficient significantly exceed $0.5$; however, note that higher-order in $e_0$ corrections to the {\tt TaylorF2Ecc} waveform may become important for BNS systems with $e_0\gtrsim 0.1$. Correlations between $e_0$ and the spin parameters $\chi_{1,2}$ are generally weaker than those between $e_0$ and the mass parameters, with the largest values $|c_{e_0 \chi_{1,2}}| \lesssim 0.4$ for the NS/BH system.
\bibliography{masterbibdatabase_ECC}

\begin{thebibliography}{150}%
\makeatletter
\providecommand \@ifxundefined [1]{%
 \@ifx{#1\undefined}
}%
\providecommand \@ifnum [1]{%
 \ifnum #1\expandafter \@firstoftwo
 \else \expandafter \@secondoftwo
 \fi
}%
\providecommand \@ifx [1]{%
 \ifx #1\expandafter \@firstoftwo
 \else \expandafter \@secondoftwo
 \fi
}%
\providecommand \natexlab [1]{#1}%
\providecommand \enquote  [1]{``#1''}%
\providecommand \bibnamefont  [1]{#1}%
\providecommand \bibfnamefont [1]{#1}%
\providecommand \citenamefont [1]{#1}%
\providecommand \href@noop [0]{\@secondoftwo}%
\providecommand \href [0]{\begingroup \@sanitize@url \@href}%
\providecommand \@href[1]{\@@startlink{#1}\@@href}%
\providecommand \@@href[1]{\endgroup#1\@@endlink}%
\providecommand \@sanitize@url [0]{\catcode `\\12\catcode `\$12\catcode
  `\&12\catcode `\#12\catcode `\^12\catcode `\_12\catcode `\%12\relax}%
\providecommand \@@startlink[1]{}%
\providecommand \@@endlink[0]{}%
\providecommand \url  [0]{\begingroup\@sanitize@url \@url }%
\providecommand \@url [1]{\endgroup\@href {#1}{\urlprefix }}%
\providecommand \urlprefix  [0]{URL }%
\providecommand \Eprint [0]{\href }%
\providecommand \doibase [0]{http://dx.doi.org/}%
\providecommand \selectlanguage [0]{\@gobble}%
\providecommand \bibinfo  [0]{\@secondoftwo}%
\providecommand \bibfield  [0]{\@secondoftwo}%
\providecommand \translation [1]{[#1]}%
\providecommand \BibitemOpen [0]{}%
\providecommand \bibitemStop [0]{}%
\providecommand \bibitemNoStop [0]{.\EOS\space}%
\providecommand \EOS [0]{\spacefactor3000\relax}%
\providecommand \BibitemShut  [1]{\csname bibitem#1\endcsname}%
\let\auto@bib@innerbib\@empty
\bibitem [{\citenamefont {Aasi}\ \emph {et~al.}(2015)\citenamefont {Aasi} \emph
  {et~al.}}]{LIGO-stdrd-ref1}%
  \BibitemOpen
  \bibfield  {author} {\bibinfo {author} {\bibfnamefont {J.}~\bibnamefont
  {Aasi}} \emph {et~al.} (\bibinfo {collaboration} {LIGO Scientific
  Collaboration}),\ }\href {\doibase 10.1088/0264-9381/32/7/074001} {\bibfield
  {journal} {\bibinfo  {journal} {Classical Quantum Gravity}\ }\textbf
  {\bibinfo {volume} {32}},\ \bibinfo {pages} {074001} (\bibinfo {year}
  {2015})},\ \Eprint {http://arxiv.org/abs/1411.4547} {arXiv:1411.4547 [gr-qc]}
  \BibitemShut {NoStop}%
\bibitem [{\citenamefont {Acernese}\ \emph {et~al.}(2015)\citenamefont
  {Acernese} \emph {et~al.}}]{aVirgo-stdref}%
  \BibitemOpen
  \bibfield  {author} {\bibinfo {author} {\bibfnamefont {F.}~\bibnamefont
  {Acernese}} \emph {et~al.} (\bibinfo {collaboration} {Virgo Collaboration}),\
  }\href {\doibase 10.1088/0264-9381/32/2/024001} {\bibfield  {journal}
  {\bibinfo  {journal} {Classical Quantum Gravity}\ }\textbf {\bibinfo {volume}
  {32}},\ \bibinfo {pages} {024001} (\bibinfo {year} {2015})},\ \Eprint
  {http://arxiv.org/abs/1408.3978} {arXiv:1408.3978 [gr-qc]} \BibitemShut
  {NoStop}%
\bibitem [{\citenamefont {Abbott}\ \emph
  {et~al.}(2016{\natexlab{a}})\citenamefont {Abbott} \emph
  {et~al.}}]{detectionPRL2016}%
  \BibitemOpen
  \bibfield  {author} {\bibinfo {author} {\bibfnamefont {B.~P.}\ \bibnamefont
  {Abbott}} \emph {et~al.} (\bibinfo {collaboration} {LIGO Scientific and Virgo
  Collaborations}),\ }\href {\doibase 10.1103/PhysRevLett.116.061102}
  {\bibfield  {journal} {\bibinfo  {journal} {\prl}\ }\textbf {\bibinfo
  {volume} {116}},\ \bibinfo {eid} {061102} (\bibinfo {year}
  {2016}{\natexlab{a}})},\ \Eprint {http://arxiv.org/abs/1602.03837}
  {arXiv:1602.03837 [gr-qc]} \BibitemShut {NoStop}%
\bibitem [{\citenamefont {Abbott}\ \emph
  {et~al.}(2016{\natexlab{b}})\citenamefont {Abbott} \emph
  {et~al.}}]{gw151226-PRL2016}%
  \BibitemOpen
  \bibfield  {author} {\bibinfo {author} {\bibfnamefont {B.~P.}\ \bibnamefont
  {Abbott}} \emph {et~al.} (\bibinfo {collaboration} {LIGO Scientific and Virgo
  Collaborations}),\ }\href {\doibase 10.1103/PhysRevLett.116.241103}
  {\bibfield  {journal} {\bibinfo  {journal} {Phys.~Rev.~Lett.}\ }\textbf
  {\bibinfo {volume} {116}},\ \bibinfo {eid} {241103} (\bibinfo {year}
  {2016}{\natexlab{b}})},\ \Eprint {http://arxiv.org/abs/1606.04855}
  {arXiv:1606.04855} \BibitemShut {NoStop}%
\bibitem [{\citenamefont {Abbott}\ \emph
  {et~al.}(2017{\natexlab{a}})\citenamefont {Abbott} \emph
  {et~al.}}]{GW170104}%
  \BibitemOpen
  \bibfield  {author} {\bibinfo {author} {\bibfnamefont {B.~P.}\ \bibnamefont
  {Abbott}} \emph {et~al.} (\bibinfo {collaboration} {LIGO Scientific and Virgo
  Collaborations}),\ }\href {\doibase 10.1103/PhysRevLett.118.221101}
  {\bibfield  {journal} {\bibinfo  {journal} {Phys. Rev. Lett.}\ }\textbf
  {\bibinfo {volume} {118}},\ \bibinfo {pages} {221101} (\bibinfo {year}
  {2017}{\natexlab{a}})},\ \Eprint {http://arxiv.org/abs/1706.01812}
  {arXiv:1706.01812 [gr-qc]} \BibitemShut {NoStop}%
\bibitem [{\citenamefont {Abbott}\ \emph
  {et~al.}(2017{\natexlab{b}})\citenamefont {Abbott} \emph
  {et~al.}}]{GW170608}%
  \BibitemOpen
  \bibfield  {author} {\bibinfo {author} {\bibfnamefont {B.~P.}\ \bibnamefont
  {Abbott}} \emph {et~al.} (\bibinfo {collaboration} {LIGO Scientific and Virgo
  Collaborations}),\ }\href {\doibase 10.3847/2041-8213/aa9f0c} {\bibfield
  {journal} {\bibinfo  {journal} {Astrophys. J.}\ }\textbf {\bibinfo {volume}
  {851}},\ \bibinfo {pages} {L35} (\bibinfo {year} {2017}{\natexlab{b}})},\
  \Eprint {http://arxiv.org/abs/1711.05578} {arXiv:1711.05578 [astro-ph.HE]}
  \BibitemShut {NoStop}%
\bibitem [{\citenamefont {Abbott}\ \emph
  {et~al.}(2017{\natexlab{c}})\citenamefont {Abbott} \emph
  {et~al.}}]{GW170814}%
  \BibitemOpen
  \bibfield  {author} {\bibinfo {author} {\bibfnamefont {B.~P.}\ \bibnamefont
  {Abbott}} \emph {et~al.} (\bibinfo {collaboration} {LIGO Scientific and Virgo
  Collaborations}),\ }\href {\doibase 10.1103/PhysRevLett.119.141101}
  {\bibfield  {journal} {\bibinfo  {journal} {Phys. Rev. Lett.}\ }\textbf
  {\bibinfo {volume} {119}},\ \bibinfo {pages} {141101} (\bibinfo {year}
  {2017}{\natexlab{c}})},\ \Eprint {http://arxiv.org/abs/1709.09660}
  {arXiv:1709.09660 [gr-qc]} \BibitemShut {NoStop}%
\bibitem [{\citenamefont {Abbott}\ \emph
  {et~al.}(2019{\natexlab{a}})\citenamefont {Abbott} \emph
  {et~al.}}]{GWTC1-LVC2018}%
  \BibitemOpen
  \bibfield  {author} {\bibinfo {author} {\bibfnamefont {B.~P.}\ \bibnamefont
  {Abbott}} \emph {et~al.} (\bibinfo {collaboration} {LIGO Scientific and Virgo
  Collaborations}),\ }\href {\doibase 10.1103/PhysRevX.9.031040} {\bibfield
  {journal} {\bibinfo  {journal} {Phys. Rev. X}\ }\textbf {\bibinfo {volume}
  {9}},\ \bibinfo {pages} {031040} (\bibinfo {year} {2019}{\natexlab{a}})},\
  \Eprint {http://arxiv.org/abs/1811.12907} {arXiv:1811.12907 [astro-ph.HE]}
  \BibitemShut {NoStop}%
\bibitem [{\citenamefont {Abbott}\ \emph
  {et~al.}(2020{\natexlab{a}})\citenamefont {Abbott} \emph
  {et~al.}}]{GW190412}%
  \BibitemOpen
  \bibfield  {author} {\bibinfo {author} {\bibfnamefont {R.}~\bibnamefont
  {Abbott}} \emph {et~al.} (\bibinfo {collaboration} {LIGO Scientific and Virgo
  Collaborations}),\ }\href {\doibase 10.1103/PhysRevD.102.043015} {\bibfield
  {journal} {\bibinfo  {journal} {Phys. Rev. D}\ }\textbf {\bibinfo {volume}
  {102}},\ \bibinfo {pages} {043015} (\bibinfo {year} {2020}{\natexlab{a}})},\
  \Eprint {http://arxiv.org/abs/2004.08342} {arXiv:2004.08342 [astro-ph.HE]}
  \BibitemShut {NoStop}%
\bibitem [{\citenamefont {Abbott}\ \emph
  {et~al.}(2020{\natexlab{b}})\citenamefont {Abbott} \emph
  {et~al.}}]{GW190814}%
  \BibitemOpen
  \bibfield  {author} {\bibinfo {author} {\bibfnamefont {R.}~\bibnamefont
  {Abbott}} \emph {et~al.} (\bibinfo {collaboration} {LIGO Scientific and Virgo
  Collaborations}),\ }\href {\doibase 10.3847/2041-8213/ab960f} {\bibfield
  {journal} {\bibinfo  {journal} {Astrophys. J.}\ }\textbf {\bibinfo {volume}
  {896}},\ \bibinfo {pages} {L44} (\bibinfo {year} {2020}{\natexlab{b}})},\
  \Eprint {http://arxiv.org/abs/2006.12611} {arXiv:2006.12611 [astro-ph.HE]}
  \BibitemShut {NoStop}%
\bibitem [{\citenamefont {Abbott}\ \emph
  {et~al.}(2020{\natexlab{c}})\citenamefont {Abbott} \emph
  {et~al.}}]{GW190521-PRL2021}%
  \BibitemOpen
  \bibfield  {author} {\bibinfo {author} {\bibfnamefont {R.}~\bibnamefont
  {Abbott}} \emph {et~al.} (\bibinfo {collaboration} {LIGO Scientific and Virgo
  Collaborations}),\ }\href {\doibase 10.1103/PhysRevLett.125.101102}
  {\bibfield  {journal} {\bibinfo  {journal} {Phys.~Rev.~Lett.}\ }\textbf
  {\bibinfo {volume} {125}},\ \bibinfo {pages} {101102} (\bibinfo {year}
  {2020}{\natexlab{c}})}\BibitemShut {NoStop}%
\bibitem [{\citenamefont {Abbott}\ \emph
  {et~al.}(2021{\natexlab{a}})\citenamefont {Abbott} \emph
  {et~al.}}]{GWTC-2-PRX2021}%
  \BibitemOpen
  \bibfield  {author} {\bibinfo {author} {\bibfnamefont {R.}~\bibnamefont
  {Abbott}} \emph {et~al.} (\bibinfo {collaboration} {LIGO Scientific and Virgo
  Collaborations}),\ }\href {\doibase 10.1103/PhysRevX.11.021053} {\bibfield
  {journal} {\bibinfo  {journal} {Phys.~Rev.~X}\ }\textbf {\bibinfo {volume}
  {11}},\ \bibinfo {eid} {021053} (\bibinfo {year} {2021}{\natexlab{a}})},\
  \Eprint {http://arxiv.org/abs/2010.14527} {arXiv:2010.14527 [gr-qc]}
  \BibitemShut {NoStop}%
\bibitem [{\citenamefont {Abbott}\ \emph
  {et~al.}(2017{\natexlab{d}})\citenamefont {Abbott} \emph
  {et~al.}}]{GW170817}%
  \BibitemOpen
  \bibfield  {author} {\bibinfo {author} {\bibfnamefont {B.~P.}\ \bibnamefont
  {Abbott}} \emph {et~al.} (\bibinfo {collaboration} {LIGO Scientific and Virgo
  Collaborations}),\ }\href {\doibase 10.1103/PhysRevLett.119.161101}
  {\bibfield  {journal} {\bibinfo  {journal} {Phys. Rev. Lett.}\ }\textbf
  {\bibinfo {volume} {119}},\ \bibinfo {pages} {161101} (\bibinfo {year}
  {2017}{\natexlab{d}})},\ \Eprint {http://arxiv.org/abs/1710.05832}
  {arXiv:1710.05832 [gr-qc]} \BibitemShut {NoStop}%
\bibitem [{\citenamefont {Abbott}\ \emph
  {et~al.}(2020{\natexlab{d}})\citenamefont {Abbott} \emph
  {et~al.}}]{GW190425}%
  \BibitemOpen
  \bibfield  {author} {\bibinfo {author} {\bibfnamefont {B.~P.}\ \bibnamefont
  {Abbott}} \emph {et~al.} (\bibinfo {collaboration} {LIGO Scientific and Virgo
  Collaborations}),\ }\href {\doibase 10.3847/2041-8213/ab75f5} {\bibfield
  {journal} {\bibinfo  {journal} {Astrophys. J. Lett.}\ }\textbf {\bibinfo
  {volume} {892}},\ \bibinfo {pages} {L3} (\bibinfo {year}
  {2020}{\natexlab{d}})},\ \Eprint {http://arxiv.org/abs/2001.01761}
  {arXiv:2001.01761 [astro-ph.HE]} \BibitemShut {NoStop}%
\bibitem [{\citenamefont {Abbott}\ \emph
  {et~al.}(2021{\natexlab{b}})\citenamefont {Abbott} \emph
  {et~al.}}]{NSBH-ApJL2021}%
  \BibitemOpen
  \bibfield  {author} {\bibinfo {author} {\bibfnamefont {R.}~\bibnamefont
  {Abbott}} \emph {et~al.} (\bibinfo {collaboration} {LIGO Scientific, Virgo,
  and KAGRA Collaborations}),\ }\href {\doibase 10.3847/2041-8213/ac082e}
  {\bibfield  {journal} {\bibinfo  {journal} {Astrophys.~J.~Lett.}\ }\textbf
  {\bibinfo {volume} {915}},\ \bibinfo {eid} {L5} (\bibinfo {year}
  {2021}{\natexlab{b}})},\ \Eprint {http://arxiv.org/abs/2106.15163}
  {arXiv:2106.15163 [astro-ph.HE]} \BibitemShut {NoStop}%
\bibitem [{\citenamefont {Abbott}\ \emph
  {et~al.}(2019{\natexlab{b}})\citenamefont {Abbott} \emph
  {et~al.}}]{LIGOScientific:2018jsj}%
  \BibitemOpen
  \bibfield  {author} {\bibinfo {author} {\bibfnamefont {B.~P.}\ \bibnamefont
  {Abbott}} \emph {et~al.} (\bibinfo {collaboration} {LIGO Scientific and Virgo
  Collaborations}),\ }\href {\doibase 10.3847/2041-8213/ab3800} {\bibfield
  {journal} {\bibinfo  {journal} {Astrophys.~J.}\ }\textbf {\bibinfo {volume}
  {882}},\ \bibinfo {pages} {L24} (\bibinfo {year} {2019}{\natexlab{b}})},\
  \Eprint {http://arxiv.org/abs/1811.12940} {arXiv:1811.12940 [astro-ph.HE]}
  \BibitemShut {NoStop}%
\bibitem [{\citenamefont {Benacquista}\ and\ \citenamefont
  {Downing}(2013)}]{Benacquista:2011kv}%
  \BibitemOpen
  \bibfield  {author} {\bibinfo {author} {\bibfnamefont {M.~J.}\ \bibnamefont
  {Benacquista}}\ and\ \bibinfo {author} {\bibfnamefont {J.~M.~B.}\
  \bibnamefont {Downing}},\ }\href {\doibase 10.12942/lrr-2013-4} {\bibfield
  {journal} {\bibinfo  {journal} {Living Rev.~Relativity}\ }\textbf {\bibinfo
  {volume} {16}},\ \bibinfo {pages} {4} (\bibinfo {year} {2013})},\ \Eprint
  {http://arxiv.org/abs/1110.4423} {arXiv:1110.4423 [astro-ph.SR]} \BibitemShut
  {NoStop}%
\bibitem [{\citenamefont {{Bae}}\ \emph {et~al.}(2014)\citenamefont {{Bae}},
  \citenamefont {{Kim}},\ and\ \citenamefont {{Lee}}}]{bae-etal-MNRAS2013}%
  \BibitemOpen
  \bibfield  {author} {\bibinfo {author} {\bibfnamefont {Y.-B.}\ \bibnamefont
  {{Bae}}}, \bibinfo {author} {\bibfnamefont {C.}~\bibnamefont {{Kim}}}, \ and\
  \bibinfo {author} {\bibfnamefont {H.~M.}\ \bibnamefont {{Lee}}},\ }\href
  {\doibase 10.1093/mnras/stu381} {\bibfield  {journal} {\bibinfo  {journal}
  {Mon.~Not.~R.~Astron.~Soc.}\ }\textbf {\bibinfo {volume} {440}},\ \bibinfo
  {pages} {2714} (\bibinfo {year} {2014})},\ \Eprint
  {http://arxiv.org/abs/1308.1641} {arXiv:1308.1641 [astro-ph.HE]} \BibitemShut
  {NoStop}%
\bibitem [{\citenamefont {Rodriguez}\ \emph {et~al.}(2018)\citenamefont
  {Rodriguez}, \citenamefont {Amaro-Seoane}, \citenamefont {Chatterjee},\ and\
  \citenamefont {Rasio}}]{Rodriguez:2017pec}%
  \BibitemOpen
  \bibfield  {author} {\bibinfo {author} {\bibfnamefont {C.~L.}\ \bibnamefont
  {Rodriguez}}, \bibinfo {author} {\bibfnamefont {P.}~\bibnamefont
  {Amaro-Seoane}}, \bibinfo {author} {\bibfnamefont {S.}~\bibnamefont
  {Chatterjee}}, \ and\ \bibinfo {author} {\bibfnamefont {F.~A.}\ \bibnamefont
  {Rasio}},\ }\href {\doibase 10.1103/PhysRevLett.120.151101} {\bibfield
  {journal} {\bibinfo  {journal} {Phys. Rev. Lett.}\ }\textbf {\bibinfo
  {volume} {120}},\ \bibinfo {pages} {151101} (\bibinfo {year} {2018})},\
  \Eprint {http://arxiv.org/abs/1712.04937} {arXiv:1712.04937 [astro-ph.HE]}
  \BibitemShut {NoStop}%
\bibitem [{\citenamefont {Samsing}(2018)}]{Samsing:2017xmd}%
  \BibitemOpen
  \bibfield  {author} {\bibinfo {author} {\bibfnamefont {J.}~\bibnamefont
  {Samsing}},\ }\href {\doibase 10.1103/PhysRevD.97.103014} {\bibfield
  {journal} {\bibinfo  {journal} {Phys.~Rev.~D}\ }\textbf {\bibinfo {volume}
  {97}},\ \bibinfo {pages} {103014} (\bibinfo {year} {2018})},\ \Eprint
  {http://arxiv.org/abs/1711.07452} {arXiv:1711.07452 [astro-ph.HE]}
  \BibitemShut {NoStop}%
\bibitem [{\citenamefont {Damour}\ \emph {et~al.}(2004)\citenamefont {Damour},
  \citenamefont {Gopakumar},\ and\ \citenamefont {Iyer}}]{DGI}%
  \BibitemOpen
  \bibfield  {author} {\bibinfo {author} {\bibfnamefont {T.}~\bibnamefont
  {Damour}}, \bibinfo {author} {\bibfnamefont {A.}~\bibnamefont {Gopakumar}}, \
  and\ \bibinfo {author} {\bibfnamefont {B.~R.}\ \bibnamefont {Iyer}},\ }\href
  {\doibase 10.1103/PhysRevD.70.064028} {\bibfield  {journal} {\bibinfo
  {journal} {Phys.~Rev.~D}\ }\textbf {\bibinfo {volume} {70}},\ \bibinfo
  {pages} {064028} (\bibinfo {year} {2004})},\ \Eprint
  {http://arxiv.org/abs/gr-qc/0404128} {arXiv:gr-qc/0404128} \BibitemShut
  {NoStop}%
\bibitem [{\citenamefont {{Moore}}\ \emph {et~al.}(2016)\citenamefont
  {{Moore}}, \citenamefont {{Favata}}, \citenamefont {{Arun}},\ and\
  \citenamefont {{Mishra}}}]{moore-etal-PRD2016}%
  \BibitemOpen
  \bibfield  {author} {\bibinfo {author} {\bibfnamefont {B.}~\bibnamefont
  {{Moore}}}, \bibinfo {author} {\bibfnamefont {M.}~\bibnamefont {{Favata}}},
  \bibinfo {author} {\bibfnamefont {K.~G.}\ \bibnamefont {{Arun}}}, \ and\
  \bibinfo {author} {\bibfnamefont {C.~K.}\ \bibnamefont {{Mishra}}},\ }\href
  {\doibase 10.1103/PhysRevD.93.124061} {\bibfield  {journal} {\bibinfo
  {journal} {\prd}\ }\textbf {\bibinfo {volume} {93}},\ \bibinfo {eid} {124061}
  (\bibinfo {year} {2016})},\ \Eprint {http://arxiv.org/abs/1605.00304}
  {arXiv:1605.00304 [gr-qc]} \BibitemShut {NoStop}%
\bibitem [{\citenamefont {{Peters}}\ and\ \citenamefont
  {{Mathews}}(1963)}]{petersmathews}%
  \BibitemOpen
  \bibfield  {author} {\bibinfo {author} {\bibfnamefont {P.~C.}\ \bibnamefont
  {{Peters}}}\ and\ \bibinfo {author} {\bibfnamefont {J.}~\bibnamefont
  {{Mathews}}},\ }\href {\doibase 10.1103/PhysRev.131.435} {\bibfield
  {journal} {\bibinfo  {journal} {Phys.~Rev.}\ }\textbf {\bibinfo {volume}
  {131}},\ \bibinfo {pages} {435} (\bibinfo {year} {1963})}\BibitemShut
  {NoStop}%
\bibitem [{\citenamefont {{Peters}}(1964)}]{peters}%
  \BibitemOpen
  \bibfield  {author} {\bibinfo {author} {\bibfnamefont {P.~C.}\ \bibnamefont
  {{Peters}}},\ }\href {\doibase 10.1103/PhysRev.136.B1224} {\bibfield
  {journal} {\bibinfo  {journal} {Phys.~Rev.}\ }\textbf {\bibinfo {volume}
  {136}},\ \bibinfo {pages} {B1224} (\bibinfo {year} {1964})}\BibitemShut
  {NoStop}%
\bibitem [{\citenamefont {{Favata}}(2014)}]{favata-PRL2014}%
  \BibitemOpen
  \bibfield  {author} {\bibinfo {author} {\bibfnamefont {M.}~\bibnamefont
  {{Favata}}},\ }\href {\doibase 10.1103/PhysRevLett.112.101101} {\bibfield
  {journal} {\bibinfo  {journal} {Phys.~Rev.~Lett.}\ }\textbf {\bibinfo
  {volume} {112}},\ \bibinfo {eid} {101101} (\bibinfo {year} {2014})},\ \Eprint
  {http://arxiv.org/abs/1310.8288} {arXiv:1310.8288 [gr-qc]} \BibitemShut
  {NoStop}%
\bibitem [{\citenamefont {{Kowalska}}\ \emph {et~al.}(2011)\citenamefont
  {{Kowalska}}, \citenamefont {{Bulik}}, \citenamefont {{Belczynski}},
  \citenamefont {{Dominik}},\ and\ \citenamefont
  {{Gondek-Rosinska}}}]{kowalska-etal-eccentricity-distribution-AA2011}%
  \BibitemOpen
  \bibfield  {author} {\bibinfo {author} {\bibfnamefont {I.}~\bibnamefont
  {{Kowalska}}}, \bibinfo {author} {\bibfnamefont {T.}~\bibnamefont {{Bulik}}},
  \bibinfo {author} {\bibfnamefont {K.}~\bibnamefont {{Belczynski}}}, \bibinfo
  {author} {\bibfnamefont {M.}~\bibnamefont {{Dominik}}}, \ and\ \bibinfo
  {author} {\bibfnamefont {D.}~\bibnamefont {{Gondek-Rosinska}}},\ }\href
  {\doibase 10.1051/0004-6361/201015777} {\bibfield  {journal} {\bibinfo
  {journal} {Astron.~Astrophys.}\ }\textbf {\bibinfo {volume} {527}},\ \bibinfo
  {eid} {A70} (\bibinfo {year} {2011})},\ \Eprint
  {http://arxiv.org/abs/1010.0511} {arXiv:1010.0511} \BibitemShut {NoStop}%
\bibitem [{\citenamefont {{Silsbee}}\ and\ \citenamefont
  {{Tremaine}}(2017)}]{silsbee-tremaine2016}%
  \BibitemOpen
  \bibfield  {author} {\bibinfo {author} {\bibfnamefont {K.}~\bibnamefont
  {{Silsbee}}}\ and\ \bibinfo {author} {\bibfnamefont {S.}~\bibnamefont
  {{Tremaine}}},\ }\href {\doibase 10.3847/1538-4357/aa5729} {\bibfield
  {journal} {\bibinfo  {journal} {Astrophys.~J.}\ }\textbf {\bibinfo {volume}
  {836}},\ \bibinfo {eid} {39} (\bibinfo {year} {2017})},\ \Eprint
  {http://arxiv.org/abs/1608.07642} {arXiv:1608.07642 [astro-ph.HE]}
  \BibitemShut {NoStop}%
\bibitem [{\citenamefont {{Antonini}}\ \emph {et~al.}(2016)\citenamefont
  {{Antonini}}, \citenamefont {{Chatterjee}}, \citenamefont {{Rodriguez}},
  \citenamefont {{Morscher}}, \citenamefont {{Pattabiraman}}, \citenamefont
  {{Kalogera}},\ and\ \citenamefont {{Rasio}}}]{antonini-etalApJ2015}%
  \BibitemOpen
  \bibfield  {author} {\bibinfo {author} {\bibfnamefont {F.}~\bibnamefont
  {{Antonini}}}, \bibinfo {author} {\bibfnamefont {S.}~\bibnamefont
  {{Chatterjee}}}, \bibinfo {author} {\bibfnamefont {C.~L.}\ \bibnamefont
  {{Rodriguez}}}, \bibinfo {author} {\bibfnamefont {M.}~\bibnamefont
  {{Morscher}}}, \bibinfo {author} {\bibfnamefont {B.}~\bibnamefont
  {{Pattabiraman}}}, \bibinfo {author} {\bibfnamefont {V.}~\bibnamefont
  {{Kalogera}}}, \ and\ \bibinfo {author} {\bibfnamefont {F.~A.}\ \bibnamefont
  {{Rasio}}},\ }\href {\doibase 10.3847/0004-637X/816/2/65} {\bibfield
  {journal} {\bibinfo  {journal} {\apj}\ }\textbf {\bibinfo {volume} {816}},\
  \bibinfo {eid} {65} (\bibinfo {year} {2016})},\ \Eprint
  {http://arxiv.org/abs/1509.05080} {arXiv:1509.05080} \BibitemShut {NoStop}%
\bibitem [{\citenamefont {{Rodriguez}}\ \emph {et~al.}(2016)\citenamefont
  {{Rodriguez}}, \citenamefont {{Chatterjee}},\ and\ \citenamefont
  {{Rasio}}}]{rodriguez-chatterjee-rasio2016}%
  \BibitemOpen
  \bibfield  {author} {\bibinfo {author} {\bibfnamefont {C.~L.}\ \bibnamefont
  {{Rodriguez}}}, \bibinfo {author} {\bibfnamefont {S.}~\bibnamefont
  {{Chatterjee}}}, \ and\ \bibinfo {author} {\bibfnamefont {F.~A.}\
  \bibnamefont {{Rasio}}},\ }\href {\doibase 10.1103/PhysRevD.93.084029}
  {\bibfield  {journal} {\bibinfo  {journal} {\prd}\ }\textbf {\bibinfo
  {volume} {93}},\ \bibinfo {eid} {084029} (\bibinfo {year} {2016})},\ \Eprint
  {http://arxiv.org/abs/1602.02444} {arXiv:1602.02444 [astro-ph.HE]}
  \BibitemShut {NoStop}%
\bibitem [{\citenamefont {{Wen}}(2003)}]{wen-eccentricity-ApJ2003}%
  \BibitemOpen
  \bibfield  {author} {\bibinfo {author} {\bibfnamefont {L.}~\bibnamefont
  {{Wen}}},\ }\href {\doibase 10.1086/378794} {\bibfield  {journal} {\bibinfo
  {journal} {Astrophys.~J.}\ }\textbf {\bibinfo {volume} {598}},\ \bibinfo
  {pages} {419} (\bibinfo {year} {2003})},\ \Eprint
  {http://arxiv.org/abs/arXiv:astro-ph/0211492} {arXiv:astro-ph/0211492}
  \BibitemShut {NoStop}%
\bibitem [{\citenamefont {{Antonini}}\ \emph {et~al.}(2014)\citenamefont
  {{Antonini}}, \citenamefont {{Murray}},\ and\ \citenamefont
  {{Mikkola}}}]{antonini-murray-mikkola-ApJ2014}%
  \BibitemOpen
  \bibfield  {author} {\bibinfo {author} {\bibfnamefont {F.}~\bibnamefont
  {{Antonini}}}, \bibinfo {author} {\bibfnamefont {N.}~\bibnamefont
  {{Murray}}}, \ and\ \bibinfo {author} {\bibfnamefont {S.}~\bibnamefont
  {{Mikkola}}},\ }\href {\doibase 10.1088/0004-637X/781/1/45} {\bibfield
  {journal} {\bibinfo  {journal} {\apj}\ }\textbf {\bibinfo {volume} {781}},\
  \bibinfo {eid} {45} (\bibinfo {year} {2014})},\ \Eprint
  {http://arxiv.org/abs/1308.3674} {arXiv:1308.3674 [astro-ph.HE]} \BibitemShut
  {NoStop}%
\bibitem [{\citenamefont {{Antognini}}\ \emph {et~al.}(2014)\citenamefont
  {{Antognini}}, \citenamefont {{Shappee}}, \citenamefont {{Thompson}},\ and\
  \citenamefont {{Amaro-Seoane}}}]{antognini-etalMNRAS2014}%
  \BibitemOpen
  \bibfield  {author} {\bibinfo {author} {\bibfnamefont {J.~M.}\ \bibnamefont
  {{Antognini}}}, \bibinfo {author} {\bibfnamefont {B.~J.}\ \bibnamefont
  {{Shappee}}}, \bibinfo {author} {\bibfnamefont {T.~A.}\ \bibnamefont
  {{Thompson}}}, \ and\ \bibinfo {author} {\bibfnamefont {P.}~\bibnamefont
  {{Amaro-Seoane}}},\ }\href {\doibase 10.1093/mnras/stu039} {\bibfield
  {journal} {\bibinfo  {journal} {Mon.~Not.~R.~Astron.~Soc.}\ }\textbf
  {\bibinfo {volume} {439}},\ \bibinfo {pages} {1079} (\bibinfo {year}
  {2014})},\ \Eprint {http://arxiv.org/abs/1308.5682} {arXiv:1308.5682
  [astro-ph.HE]} \BibitemShut {NoStop}%
\bibitem [{\citenamefont {{G{\"u}ltekin}}\ \emph {et~al.}(2004)\citenamefont
  {{G{\"u}ltekin}}, \citenamefont {{Miller}},\ and\ \citenamefont
  {{Hamilton}}}]{gultekin-miller-hamilton2004ApJ}%
  \BibitemOpen
  \bibfield  {author} {\bibinfo {author} {\bibfnamefont {K.}~\bibnamefont
  {{G{\"u}ltekin}}}, \bibinfo {author} {\bibfnamefont {M.~C.}\ \bibnamefont
  {{Miller}}}, \ and\ \bibinfo {author} {\bibfnamefont {D.~P.}\ \bibnamefont
  {{Hamilton}}},\ }\href {\doibase 10.1086/424809} {\bibfield  {journal}
  {\bibinfo  {journal} {\apj}\ }\textbf {\bibinfo {volume} {616}},\ \bibinfo
  {pages} {221} (\bibinfo {year} {2004})},\ \Eprint
  {http://arxiv.org/abs/astro-ph/0402532} {astro-ph/0402532} \BibitemShut
  {NoStop}%
\bibitem [{\citenamefont {{O'Leary}}\ \emph {et~al.}(2006)\citenamefont
  {{O'Leary}}, \citenamefont {{Rasio}}, \citenamefont {{Fregeau}},
  \citenamefont {{Ivanova}},\ and\ \citenamefont
  {{O'Shaughnessy}}}]{oleary-etal-BHmergersGC-ApJ2006}%
  \BibitemOpen
  \bibfield  {author} {\bibinfo {author} {\bibfnamefont {R.~M.}\ \bibnamefont
  {{O'Leary}}}, \bibinfo {author} {\bibfnamefont {F.~A.}\ \bibnamefont
  {{Rasio}}}, \bibinfo {author} {\bibfnamefont {J.~M.}\ \bibnamefont
  {{Fregeau}}}, \bibinfo {author} {\bibfnamefont {N.}~\bibnamefont
  {{Ivanova}}}, \ and\ \bibinfo {author} {\bibfnamefont {R.}~\bibnamefont
  {{O'Shaughnessy}}},\ }\href {\doibase 10.1086/498446} {\bibfield  {journal}
  {\bibinfo  {journal} {Astrophys.~J.}\ }\textbf {\bibinfo {volume} {637}},\
  \bibinfo {pages} {937} (\bibinfo {year} {2006})},\ \Eprint
  {http://arxiv.org/abs/arXiv:astro-ph/0508224} {arXiv:astro-ph/0508224}
  \BibitemShut {NoStop}%
\bibitem [{\citenamefont {{Haster}}\ \emph {et~al.}(2016)\citenamefont
  {{Haster}}, \citenamefont {{Antonini}}, \citenamefont {{Kalogera}},\ and\
  \citenamefont {{Mandel}}}]{haster-etal2016}%
  \BibitemOpen
  \bibfield  {author} {\bibinfo {author} {\bibfnamefont {C.-J.}\ \bibnamefont
  {{Haster}}}, \bibinfo {author} {\bibfnamefont {F.}~\bibnamefont
  {{Antonini}}}, \bibinfo {author} {\bibfnamefont {V.}~\bibnamefont
  {{Kalogera}}}, \ and\ \bibinfo {author} {\bibfnamefont {I.}~\bibnamefont
  {{Mandel}}},\ }\href {\doibase 10.3847/0004-637X/832/2/192} {\bibfield
  {journal} {\bibinfo  {journal} {Astrophys.~J.}\ }\textbf {\bibinfo {volume}
  {832}},\ \bibinfo {eid} {192} (\bibinfo {year} {2016})},\ \Eprint
  {http://arxiv.org/abs/1606.07097} {arXiv:1606.07097 [astro-ph.HE]}
  \BibitemShut {NoStop}%
\bibitem [{\citenamefont {{D'Orazio}}\ and\ \citenamefont
  {{Samsing}}(2018)}]{dorazio-samsingMNRAS2018}%
  \BibitemOpen
  \bibfield  {author} {\bibinfo {author} {\bibfnamefont {D.~J.}\ \bibnamefont
  {{D'Orazio}}}\ and\ \bibinfo {author} {\bibfnamefont {J.}~\bibnamefont
  {{Samsing}}},\ }\href {\doibase 10.1093/mnras/sty2568} {\bibfield  {journal}
  {\bibinfo  {journal} {Mon.~Not.~R.~Astron.~Soc.}\ }\textbf {\bibinfo {volume}
  {481}},\ \bibinfo {pages} {4775} (\bibinfo {year} {2018})},\ \Eprint
  {http://arxiv.org/abs/1805.06194} {arXiv:1805.06194 [astro-ph.HE]}
  \BibitemShut {NoStop}%
\bibitem [{\citenamefont {{Fragione}}\ and\ \citenamefont
  {{Loeb}}(2019)}]{fragione-loebMNRAS2019}%
  \BibitemOpen
  \bibfield  {author} {\bibinfo {author} {\bibfnamefont {G.}~\bibnamefont
  {{Fragione}}}\ and\ \bibinfo {author} {\bibfnamefont {A.}~\bibnamefont
  {{Loeb}}},\ }\href {\doibase 10.1093/mnras/stz1131} {\bibfield  {journal}
  {\bibinfo  {journal} {Mon.~Not.~R.~Astron.~Soc.}\ }\textbf {\bibinfo {volume}
  {486}},\ \bibinfo {pages} {4443} (\bibinfo {year} {2019})},\ \Eprint
  {http://arxiv.org/abs/1903.10511} {arXiv:1903.10511 [astro-ph.GA]}
  \BibitemShut {NoStop}%
\bibitem [{\citenamefont {{Zevin}}\ \emph {et~al.}(2021)\citenamefont
  {{Zevin}}, \citenamefont {{Romero-Shaw}}, \citenamefont {{Kremer}},
  \citenamefont {{Thrane}},\ and\ \citenamefont {{Lasky}}}]{zevin-etal2021}%
  \BibitemOpen
  \bibfield  {author} {\bibinfo {author} {\bibfnamefont {M.}~\bibnamefont
  {{Zevin}}}, \bibinfo {author} {\bibfnamefont {I.~M.}\ \bibnamefont
  {{Romero-Shaw}}}, \bibinfo {author} {\bibfnamefont {K.}~\bibnamefont
  {{Kremer}}}, \bibinfo {author} {\bibfnamefont {E.}~\bibnamefont {{Thrane}}},
  \ and\ \bibinfo {author} {\bibfnamefont {P.~D.}\ \bibnamefont {{Lasky}}},\
  }\href {\doibase 10.3847/2041-8213/ac32dc} {\bibfield  {journal} {\bibinfo
  {journal} {Astrophys.~J.~Lett.}\ }\textbf {\bibinfo {volume} {921}},\
  \bibinfo {eid} {L43} (\bibinfo {year} {2021})},\ \Eprint
  {http://arxiv.org/abs/2106.09042} {arXiv:2106.09042 [astro-ph.HE]}
  \BibitemShut {NoStop}%
\bibitem [{\citenamefont {{Breivik}}\ \emph {et~al.}(2016)\citenamefont
  {{Breivik}}, \citenamefont {{Rodriguez}}, \citenamefont {{Larson}},
  \citenamefont {{Kalogera}},\ and\ \citenamefont
  {{Rasio}}}]{breivik-etal2016}%
  \BibitemOpen
  \bibfield  {author} {\bibinfo {author} {\bibfnamefont {K.}~\bibnamefont
  {{Breivik}}}, \bibinfo {author} {\bibfnamefont {C.~L.}\ \bibnamefont
  {{Rodriguez}}}, \bibinfo {author} {\bibfnamefont {S.~L.}\ \bibnamefont
  {{Larson}}}, \bibinfo {author} {\bibfnamefont {V.}~\bibnamefont
  {{Kalogera}}}, \ and\ \bibinfo {author} {\bibfnamefont {F.~A.}\ \bibnamefont
  {{Rasio}}},\ }\href {\doibase 10.3847/2041-8205/830/1/L18} {\bibfield
  {journal} {\bibinfo  {journal} {Astrophys.~J.~Lett.}\ }\textbf {\bibinfo
  {volume} {830}},\ \bibinfo {eid} {L18} (\bibinfo {year} {2016})},\ \Eprint
  {http://arxiv.org/abs/1606.09558} {arXiv:1606.09558 [astro-ph.GA]}
  \BibitemShut {NoStop}%
\bibitem [{\citenamefont {{O'Leary}}\ \emph {et~al.}(2009)\citenamefont
  {{O'Leary}}, \citenamefont {{Kocsis}},\ and\ \citenamefont
  {{Loeb}}}]{oleary-kocsis-loeb-MNRAS2009}%
  \BibitemOpen
  \bibfield  {author} {\bibinfo {author} {\bibfnamefont {R.~M.}\ \bibnamefont
  {{O'Leary}}}, \bibinfo {author} {\bibfnamefont {B.}~\bibnamefont {{Kocsis}}},
  \ and\ \bibinfo {author} {\bibfnamefont {A.}~\bibnamefont {{Loeb}}},\ }\href
  {\doibase 10.1111/j.1365-2966.2009.14653.x} {\bibfield  {journal} {\bibinfo
  {journal} {Mon.~Not.~R.~Astron.~Soc.}\ }\textbf {\bibinfo {volume} {395}},\
  \bibinfo {pages} {2127} (\bibinfo {year} {2009})},\ \Eprint
  {http://arxiv.org/abs/arXiv:0807.2638 [astro-ph]} {arXiv:0807.2638
  [astro-ph]} \BibitemShut {NoStop}%
\bibitem [{\citenamefont {{Antonini}}\ and\ \citenamefont
  {{Perets}}(2012)}]{antonini-peretsApJ2012}%
  \BibitemOpen
  \bibfield  {author} {\bibinfo {author} {\bibfnamefont {F.}~\bibnamefont
  {{Antonini}}}\ and\ \bibinfo {author} {\bibfnamefont {H.~B.}\ \bibnamefont
  {{Perets}}},\ }\href {\doibase 10.1088/0004-637X/757/1/27} {\bibfield
  {journal} {\bibinfo  {journal} {\apj}\ }\textbf {\bibinfo {volume} {757}},\
  \bibinfo {eid} {27} (\bibinfo {year} {2012})},\ \Eprint
  {http://arxiv.org/abs/1203.2938} {arXiv:1203.2938} \BibitemShut {NoStop}%
\bibitem [{\citenamefont {{Hong}}\ and\ \citenamefont
  {{Lee}}(2015)}]{hong-lee_MNRAS2015}%
  \BibitemOpen
  \bibfield  {author} {\bibinfo {author} {\bibfnamefont {J.}~\bibnamefont
  {{Hong}}}\ and\ \bibinfo {author} {\bibfnamefont {H.~M.}\ \bibnamefont
  {{Lee}}},\ }\href {\doibase 10.1093/mnras/stv035} {\bibfield  {journal}
  {\bibinfo  {journal} {Mon.~Not.~R.~Astron.~Soc.}\ }\textbf {\bibinfo {volume}
  {448}},\ \bibinfo {pages} {754} (\bibinfo {year} {2015})},\ \Eprint
  {http://arxiv.org/abs/1501.02717} {arXiv:1501.02717} \BibitemShut {NoStop}%
\bibitem [{\citenamefont {{Gond{\'a}n}}\ \emph {et~al.}(2018)\citenamefont
  {{Gond{\'a}n}}, \citenamefont {{Kocsis}}, \citenamefont {{Raffai}},\ and\
  \citenamefont {{Frei}}}]{laszlo-etalApJ2018}%
  \BibitemOpen
  \bibfield  {author} {\bibinfo {author} {\bibfnamefont {L.}~\bibnamefont
  {{Gond{\'a}n}}}, \bibinfo {author} {\bibfnamefont {B.}~\bibnamefont
  {{Kocsis}}}, \bibinfo {author} {\bibfnamefont {P.}~\bibnamefont {{Raffai}}},
  \ and\ \bibinfo {author} {\bibfnamefont {Z.}~\bibnamefont {{Frei}}},\ }\href
  {\doibase 10.3847/1538-4357/aabfee} {\bibfield  {journal} {\bibinfo
  {journal} {Astrophys.~J.}\ }\textbf {\bibinfo {volume} {860}},\ \bibinfo
  {eid} {5} (\bibinfo {year} {2018})},\ \Eprint
  {http://arxiv.org/abs/1711.09989} {arXiv:1711.09989 [astro-ph.HE]}
  \BibitemShut {NoStop}%
\bibitem [{\citenamefont {{Gond{\'a}n}}\ and\ \citenamefont
  {{Kocsis}}(2021)}]{gondan-bence-MNRAS2021}%
  \BibitemOpen
  \bibfield  {author} {\bibinfo {author} {\bibfnamefont {L.}~\bibnamefont
  {{Gond{\'a}n}}}\ and\ \bibinfo {author} {\bibfnamefont {B.}~\bibnamefont
  {{Kocsis}}},\ }\href {\doibase 10.1093/mnras/stab1722} {\bibfield  {journal}
  {\bibinfo  {journal} {Mon.~Not.~R.~Astron.~Soc.~}\ }\textbf {\bibinfo
  {volume} {506}},\ \bibinfo {pages} {1665} (\bibinfo {year} {2021})},\ \Eprint
  {http://arxiv.org/abs/2011.02507} {arXiv:2011.02507 [astro-ph.HE]}
  \BibitemShut {NoStop}%
\bibitem [{\citenamefont {{Tak{\'a}tsy}}\ \emph {et~al.}(2019)\citenamefont
  {{Tak{\'a}tsy}}, \citenamefont {{B{\'e}csy}},\ and\ \citenamefont
  {{Raffai}}}]{takatsy-etalMNRAS2019}%
  \BibitemOpen
  \bibfield  {author} {\bibinfo {author} {\bibfnamefont {J.}~\bibnamefont
  {{Tak{\'a}tsy}}}, \bibinfo {author} {\bibfnamefont {B.}~\bibnamefont
  {{B{\'e}csy}}}, \ and\ \bibinfo {author} {\bibfnamefont {P.}~\bibnamefont
  {{Raffai}}},\ }\href {\doibase 10.1093/mnras/stz820} {\bibfield  {journal}
  {\bibinfo  {journal} {Mon.~Not.~R.~Astron.~Soc.}\ }\textbf {\bibinfo {volume}
  {486}},\ \bibinfo {pages} {570} (\bibinfo {year} {2019})},\ \Eprint
  {http://arxiv.org/abs/1812.04012} {arXiv:1812.04012 [astro-ph.HE]}
  \BibitemShut {NoStop}%
\bibitem [{\citenamefont {{Samsing}}\ \emph {et~al.}(2020)\citenamefont
  {{Samsing}}, \citenamefont {{Bartos}}, \citenamefont {{D'Orazio}},
  \citenamefont {{Haiman}}, \citenamefont {{Kocsis}}, \citenamefont {{Leigh}},
  \citenamefont {{Liu}}, \citenamefont {{Pessah}},\ and\ \citenamefont
  {{Tagawa}}}]{samsing-etal2020}%
  \BibitemOpen
  \bibfield  {author} {\bibinfo {author} {\bibfnamefont {J.}~\bibnamefont
  {{Samsing}}}, \bibinfo {author} {\bibfnamefont {I.}~\bibnamefont {{Bartos}}},
  \bibinfo {author} {\bibfnamefont {D.~J.}\ \bibnamefont {{D'Orazio}}},
  \bibinfo {author} {\bibfnamefont {Z.}~\bibnamefont {{Haiman}}}, \bibinfo
  {author} {\bibfnamefont {B.}~\bibnamefont {{Kocsis}}}, \bibinfo {author}
  {\bibfnamefont {N.~W.~C.}\ \bibnamefont {{Leigh}}}, \bibinfo {author}
  {\bibfnamefont {B.}~\bibnamefont {{Liu}}}, \bibinfo {author} {\bibfnamefont
  {M.~E.}\ \bibnamefont {{Pessah}}}, \ and\ \bibinfo {author} {\bibfnamefont
  {H.}~\bibnamefont {{Tagawa}}},\ }\href@noop {} {\  (\bibinfo {year}
  {2020})},\ \Eprint {http://arxiv.org/abs/2010.09765} {arXiv:2010.09765
  [astro-ph.HE]} \BibitemShut {NoStop}%
\bibitem [{\citenamefont {{Tagawa}}\ \emph {et~al.}(2021)\citenamefont
  {{Tagawa}}, \citenamefont {{Kocsis}}, \citenamefont {{Haiman}}, \citenamefont
  {{Bartos}}, \citenamefont {{Omukai}},\ and\ \citenamefont
  {{Samsing}}}]{tagawa-etalApJL2021}%
  \BibitemOpen
  \bibfield  {author} {\bibinfo {author} {\bibfnamefont {H.}~\bibnamefont
  {{Tagawa}}}, \bibinfo {author} {\bibfnamefont {B.}~\bibnamefont {{Kocsis}}},
  \bibinfo {author} {\bibfnamefont {Z.}~\bibnamefont {{Haiman}}}, \bibinfo
  {author} {\bibfnamefont {I.}~\bibnamefont {{Bartos}}}, \bibinfo {author}
  {\bibfnamefont {K.}~\bibnamefont {{Omukai}}}, \ and\ \bibinfo {author}
  {\bibfnamefont {J.}~\bibnamefont {{Samsing}}},\ }\href {\doibase
  10.3847/2041-8213/abd4d3} {\bibfield  {journal} {\bibinfo  {journal}
  {Astrophys.~J.~Lett.}\ }\textbf {\bibinfo {volume} {907}},\ \bibinfo {eid}
  {L20} (\bibinfo {year} {2021})},\ \Eprint {http://arxiv.org/abs/2010.10526}
  {arXiv:2010.10526 [astro-ph.HE]} \BibitemShut {NoStop}%
\bibitem [{\citenamefont
  {{Favata}}(2011{\natexlab{a}})}]{favata-eccentricmemory}%
  \BibitemOpen
  \bibfield  {author} {\bibinfo {author} {\bibfnamefont {M.}~\bibnamefont
  {{Favata}}},\ }\href {\doibase 10.1103/PhysRevD.84.124013} {\bibfield
  {journal} {\bibinfo  {journal} {Phys.~Rev.~D}\ }\textbf {\bibinfo {volume}
  {84}},\ \bibinfo {eid} {124013} (\bibinfo {year} {2011}{\natexlab{a}})},\
  \Eprint {http://arxiv.org/abs/arXiv:1108.3121} {arXiv:1108.3121 [gr-qc]}
  \BibitemShut {NoStop}%
\bibitem [{\citenamefont {{Mandel}}\ \emph {et~al.}(2008)\citenamefont
  {{Mandel}}, \citenamefont {{Brown}}, \citenamefont {{Gair}},\ and\
  \citenamefont {{Miller}}}]{mandel-brown}%
  \BibitemOpen
  \bibfield  {author} {\bibinfo {author} {\bibfnamefont {I.}~\bibnamefont
  {{Mandel}}}, \bibinfo {author} {\bibfnamefont {D.~A.}\ \bibnamefont
  {{Brown}}}, \bibinfo {author} {\bibfnamefont {J.~R.}\ \bibnamefont {{Gair}}},
  \ and\ \bibinfo {author} {\bibfnamefont {M.~C.}\ \bibnamefont {{Miller}}},\
  }\href {\doibase 10.1086/588246} {\bibfield  {journal} {\bibinfo  {journal}
  {\apj}\ }\textbf {\bibinfo {volume} {681}},\ \bibinfo {pages} {1431}
  (\bibinfo {year} {2008})},\ \Eprint {http://arxiv.org/abs/0705.0285}
  {arXiv:0705.0285} \BibitemShut {NoStop}%
\bibitem [{\citenamefont {{Martel}}\ and\ \citenamefont
  {{Poisson}}(1999)}]{martel-poisson-eccentric-PRD1999}%
  \BibitemOpen
  \bibfield  {author} {\bibinfo {author} {\bibfnamefont {K.}~\bibnamefont
  {{Martel}}}\ and\ \bibinfo {author} {\bibfnamefont {E.}~\bibnamefont
  {{Poisson}}},\ }\href {\doibase 10.1103/PhysRevD.60.124008} {\bibfield
  {journal} {\bibinfo  {journal} {Phys.~Rev.~D}\ }\textbf {\bibinfo {volume}
  {60}},\ \bibinfo {pages} {124008} (\bibinfo {year} {1999})},\ \Eprint
  {http://arxiv.org/abs/arXiv:gr-qc/9907006} {arXiv:gr-qc/9907006} \BibitemShut
  {NoStop}%
\bibitem [{\citenamefont {{Tessmer}}\ and\ \citenamefont
  {{Gopakumar}}(2008)}]{tessmer-gopu-PRD2008}%
  \BibitemOpen
  \bibfield  {author} {\bibinfo {author} {\bibfnamefont {M.}~\bibnamefont
  {{Tessmer}}}\ and\ \bibinfo {author} {\bibfnamefont {A.}~\bibnamefont
  {{Gopakumar}}},\ }\href {\doibase 10.1103/PhysRevD.78.084029} {\bibfield
  {journal} {\bibinfo  {journal} {Phys.~Rev.~D}\ }\textbf {\bibinfo {volume}
  {78}},\ \bibinfo {pages} {084029} (\bibinfo {year} {2008})},\ \Eprint
  {http://arxiv.org/abs/arXiv:0712.3199} {arXiv:0712.3199} \BibitemShut
  {NoStop}%
\bibitem [{\citenamefont {{Cokelaer}}\ and\ \citenamefont
  {{Pathak}}(2009)}]{cokelaer-pathak-detecteccentric-CQG2009}%
  \BibitemOpen
  \bibfield  {author} {\bibinfo {author} {\bibfnamefont {T.}~\bibnamefont
  {{Cokelaer}}}\ and\ \bibinfo {author} {\bibfnamefont {D.}~\bibnamefont
  {{Pathak}}},\ }\href {\doibase 10.1088/0264-9381/26/4/045013} {\bibfield
  {journal} {\bibinfo  {journal} {Classical Quantum Gravity}\ }\textbf
  {\bibinfo {volume} {26}},\ \bibinfo {pages} {045013} (\bibinfo {year}
  {2009})},\ \Eprint {http://arxiv.org/abs/0903.4791} {arXiv:0903.4791 [gr-qc]}
  \BibitemShut {NoStop}%
\bibitem [{\citenamefont {{Brown}}\ and\ \citenamefont
  {{Zimmerman}}(2010)}]{brown-zimmerman-eccentric-PRD2010}%
  \BibitemOpen
  \bibfield  {author} {\bibinfo {author} {\bibfnamefont {D.~A.}\ \bibnamefont
  {{Brown}}}\ and\ \bibinfo {author} {\bibfnamefont {P.~J.}\ \bibnamefont
  {{Zimmerman}}},\ }\href {\doibase 10.1103/PhysRevD.81.024007} {\bibfield
  {journal} {\bibinfo  {journal} {Phys.~Rev.~D}\ }\textbf {\bibinfo {volume}
  {81}},\ \bibinfo {eid} {024007} (\bibinfo {year} {2010})},\ \Eprint
  {http://arxiv.org/abs/0909.0066} {arXiv:0909.0066 [gr-qc]} \BibitemShut
  {NoStop}%
\bibitem [{\citenamefont {{Huerta}}\ and\ \citenamefont
  {{Brown}}(2013)}]{huerta-brown2013PRD}%
  \BibitemOpen
  \bibfield  {author} {\bibinfo {author} {\bibfnamefont {E.~A.}\ \bibnamefont
  {{Huerta}}}\ and\ \bibinfo {author} {\bibfnamefont {D.~A.}\ \bibnamefont
  {{Brown}}},\ }\href {\doibase 10.1103/PhysRevD.87.127501} {\bibfield
  {journal} {\bibinfo  {journal} {\prd}\ }\textbf {\bibinfo {volume} {87}},\
  \bibinfo {eid} {127501} (\bibinfo {year} {2013})},\ \Eprint
  {http://arxiv.org/abs/arXiv:1301.1895} {arXiv:1301.1895} \BibitemShut
  {NoStop}%
\bibitem [{\citenamefont {{Porter}}\ and\ \citenamefont
  {{Sesana}}(2010)}]{porter-sesana2010}%
  \BibitemOpen
  \bibfield  {author} {\bibinfo {author} {\bibfnamefont {E.~K.}\ \bibnamefont
  {{Porter}}}\ and\ \bibinfo {author} {\bibfnamefont {A.}~\bibnamefont
  {{Sesana}}},\ }\href@noop {} {\  (\bibinfo {year} {2010})},\ \Eprint
  {http://arxiv.org/abs/1005.5296} {arXiv:1005.5296 [gr-qc]} \BibitemShut
  {NoStop}%
\bibitem [{\citenamefont {{Kocsis}}\ and\ \citenamefont
  {{Levin}}(2012)}]{kocsis-levinPRD2012}%
  \BibitemOpen
  \bibfield  {author} {\bibinfo {author} {\bibfnamefont {B.}~\bibnamefont
  {{Kocsis}}}\ and\ \bibinfo {author} {\bibfnamefont {J.}~\bibnamefont
  {{Levin}}},\ }\href {\doibase 10.1103/PhysRevD.85.123005} {\bibfield
  {journal} {\bibinfo  {journal} {\prd}\ }\textbf {\bibinfo {volume} {85}},\
  \bibinfo {eid} {123005} (\bibinfo {year} {2012})},\ \Eprint
  {http://arxiv.org/abs/1109.4170} {arXiv:1109.4170 [astro-ph.CO]} \BibitemShut
  {NoStop}%
\bibitem [{\citenamefont {{East}}\ \emph {et~al.}(2013)\citenamefont {{East}},
  \citenamefont {{McWilliams}}, \citenamefont {{Levin}},\ and\ \citenamefont
  {{Pretorius}}}]{east-etalPRD2013}%
  \BibitemOpen
  \bibfield  {author} {\bibinfo {author} {\bibfnamefont {W.~E.}\ \bibnamefont
  {{East}}}, \bibinfo {author} {\bibfnamefont {S.~T.}\ \bibnamefont
  {{McWilliams}}}, \bibinfo {author} {\bibfnamefont {J.}~\bibnamefont
  {{Levin}}}, \ and\ \bibinfo {author} {\bibfnamefont {F.}~\bibnamefont
  {{Pretorius}}},\ }\href {\doibase 10.1103/PhysRevD.87.043004} {\bibfield
  {journal} {\bibinfo  {journal} {\prd}\ }\textbf {\bibinfo {volume} {87}},\
  \bibinfo {eid} {043004} (\bibinfo {year} {2013})},\ \Eprint
  {http://arxiv.org/abs/1212.0837} {arXiv:1212.0837 [gr-qc]} \BibitemShut
  {NoStop}%
\bibitem [{\citenamefont {{Tai}}\ \emph {et~al.}(2014)\citenamefont {{Tai}},
  \citenamefont {{McWilliams}},\ and\ \citenamefont
  {{Pretorius}}}]{tai-mcwilliams-pretoriusPRD2014}%
  \BibitemOpen
  \bibfield  {author} {\bibinfo {author} {\bibfnamefont {K.~S.}\ \bibnamefont
  {{Tai}}}, \bibinfo {author} {\bibfnamefont {S.~T.}\ \bibnamefont
  {{McWilliams}}}, \ and\ \bibinfo {author} {\bibfnamefont {F.}~\bibnamefont
  {{Pretorius}}},\ }\href {\doibase 10.1103/PhysRevD.90.103001} {\bibfield
  {journal} {\bibinfo  {journal} {\prd}\ }\textbf {\bibinfo {volume} {90}},\
  \bibinfo {eid} {103001} (\bibinfo {year} {2014})},\ \Eprint
  {http://arxiv.org/abs/1403.7754} {arXiv:1403.7754 [gr-qc]} \BibitemShut
  {NoStop}%
\bibitem [{\citenamefont {{Coughlin}}\ \emph {et~al.}(2015)\citenamefont
  {{Coughlin}}, \citenamefont {{Meyers}}, \citenamefont {{Thrane}},
  \citenamefont {{Luo}},\ and\ \citenamefont
  {{Christensen}}}]{coughlin-etalPRD2015}%
  \BibitemOpen
  \bibfield  {author} {\bibinfo {author} {\bibfnamefont {M.}~\bibnamefont
  {{Coughlin}}}, \bibinfo {author} {\bibfnamefont {P.}~\bibnamefont
  {{Meyers}}}, \bibinfo {author} {\bibfnamefont {E.}~\bibnamefont {{Thrane}}},
  \bibinfo {author} {\bibfnamefont {J.}~\bibnamefont {{Luo}}}, \ and\ \bibinfo
  {author} {\bibfnamefont {N.}~\bibnamefont {{Christensen}}},\ }\href {\doibase
  10.1103/PhysRevD.91.063004} {\bibfield  {journal} {\bibinfo  {journal}
  {\prd}\ }\textbf {\bibinfo {volume} {91}},\ \bibinfo {eid} {063004} (\bibinfo
  {year} {2015})},\ \Eprint {http://arxiv.org/abs/1412.4665} {arXiv:1412.4665
  [gr-qc]} \BibitemShut {NoStop}%
\bibitem [{\citenamefont {{Tiwari}}\ \emph {et~al.}(2016)\citenamefont
  {{Tiwari}}, \citenamefont {{Klimenko}}, \citenamefont {{Christensen}},
  \citenamefont {{Huerta}}, \citenamefont {{Mohapatra}}, \citenamefont
  {{Gopakumar}}, \citenamefont {{Haney}}, \citenamefont {{Ajith}},
  \citenamefont {{McWilliams}}, \citenamefont {{Vedovato}}, \citenamefont
  {{Drago}}, \citenamefont {{Salemi}}, \citenamefont {{Prodi}}, \citenamefont
  {{Lazzaro}}, \citenamefont {{Tiwari}}, \citenamefont {{Mitselmakher}},\ and\
  \citenamefont {{Da Silva}}}]{tiwari-etal-eBBHsearch2016PhRvD}%
  \BibitemOpen
  \bibfield  {author} {\bibinfo {author} {\bibfnamefont {V.}~\bibnamefont
  {{Tiwari}}}, \bibinfo {author} {\bibfnamefont {S.}~\bibnamefont
  {{Klimenko}}}, \bibinfo {author} {\bibfnamefont {N.}~\bibnamefont
  {{Christensen}}}, \bibinfo {author} {\bibfnamefont {E.~A.}\ \bibnamefont
  {{Huerta}}}, \bibinfo {author} {\bibfnamefont {S.~R.~P.}\ \bibnamefont
  {{Mohapatra}}}, \bibinfo {author} {\bibfnamefont {A.}~\bibnamefont
  {{Gopakumar}}}, \bibinfo {author} {\bibfnamefont {M.}~\bibnamefont
  {{Haney}}}, \bibinfo {author} {\bibfnamefont {P.}~\bibnamefont {{Ajith}}},
  \bibinfo {author} {\bibfnamefont {S.~T.}\ \bibnamefont {{McWilliams}}},
  \bibinfo {author} {\bibfnamefont {G.}~\bibnamefont {{Vedovato}}}, \bibinfo
  {author} {\bibfnamefont {M.}~\bibnamefont {{Drago}}}, \bibinfo {author}
  {\bibfnamefont {F.}~\bibnamefont {{Salemi}}}, \bibinfo {author}
  {\bibfnamefont {G.~A.}\ \bibnamefont {{Prodi}}}, \bibinfo {author}
  {\bibfnamefont {C.}~\bibnamefont {{Lazzaro}}}, \bibinfo {author}
  {\bibfnamefont {S.}~\bibnamefont {{Tiwari}}}, \bibinfo {author}
  {\bibfnamefont {G.}~\bibnamefont {{Mitselmakher}}}, \ and\ \bibinfo {author}
  {\bibfnamefont {F.}~\bibnamefont {{Da Silva}}},\ }\href {\doibase
  10.1103/PhysRevD.93.043007} {\bibfield  {journal} {\bibinfo  {journal}
  {\prd}\ }\textbf {\bibinfo {volume} {93}},\ \bibinfo {eid} {043007} (\bibinfo
  {year} {2016})},\ \Eprint {http://arxiv.org/abs/1511.09240} {arXiv:1511.09240
  [gr-qc]} \BibitemShut {NoStop}%
\bibitem [{\citenamefont {Klimenko}\ \emph {et~al.}(2008)\citenamefont
  {Klimenko}, \citenamefont {Yakushin}, \citenamefont {Mercer},\ and\
  \citenamefont {Mitselmakher}}]{Klimenko:2008fu}%
  \BibitemOpen
  \bibfield  {author} {\bibinfo {author} {\bibfnamefont {S.}~\bibnamefont
  {Klimenko}}, \bibinfo {author} {\bibfnamefont {I.}~\bibnamefont {Yakushin}},
  \bibinfo {author} {\bibfnamefont {A.}~\bibnamefont {Mercer}}, \ and\ \bibinfo
  {author} {\bibfnamefont {G.}~\bibnamefont {Mitselmakher}},\ }\href {\doibase
  10.1088/0264-9381/25/11/114029} {\bibfield  {journal} {\bibinfo  {journal}
  {Classical Quantum Gravity}\ }\textbf {\bibinfo {volume} {25}},\ \bibinfo
  {pages} {114029} (\bibinfo {year} {2008})},\ \Eprint
  {http://arxiv.org/abs/0802.3232} {arXiv:0802.3232 [gr-qc]} \BibitemShut
  {NoStop}%
\bibitem [{\citenamefont {Abbott}\ \emph
  {et~al.}(2019{\natexlab{c}})\citenamefont {Abbott} \emph
  {et~al.}}]{Salemi:2019owp}%
  \BibitemOpen
  \bibfield  {author} {\bibinfo {author} {\bibfnamefont {B.~P.}\ \bibnamefont
  {Abbott}} \emph {et~al.} (\bibinfo {collaboration} {LIGO Scientific and Virgo
  Collaborations}),\ }\href {\doibase 10.3847/1538-4357/ab3c2d} {\bibfield
  {journal} {\bibinfo  {journal} {Astrophys.~J.}\ }\textbf {\bibinfo {volume}
  {883}},\ \bibinfo {pages} {149} (\bibinfo {year} {2019}{\natexlab{c}})},\
  \Eprint {http://arxiv.org/abs/1907.09384} {arXiv:1907.09384 [astro-ph.HE]}
  \BibitemShut {NoStop}%
\bibitem [{\citenamefont {Abbott}\ \emph
  {et~al.}(2021{\natexlab{c}})\citenamefont {Abbott} \emph
  {et~al.}}]{LVKallskyBurst2021}%
  \BibitemOpen
  \bibfield  {author} {\bibinfo {author} {\bibfnamefont {R.}~\bibnamefont
  {Abbott}} \emph {et~al.} (\bibinfo {collaboration} {LIGO Scientific, Virgo,
  and KAGRA Collaborations}),\ }\href {\doibase 10.1103/PhysRevD.104.102001}
  {\bibfield  {journal} {\bibinfo  {journal} {Phys.~Rev.~D}\ }\textbf {\bibinfo
  {volume} {104}},\ \bibinfo {pages} {102001} (\bibinfo {year}
  {2021}{\natexlab{c}})},\ \Eprint {http://arxiv.org/abs/2107.13796}
  {arXiv:2107.13796 [gr-qc]} \BibitemShut {NoStop}%
\bibitem [{\citenamefont {Ramos-Buades}\ \emph {et~al.}(2020)\citenamefont
  {Ramos-Buades}, \citenamefont {Tiwari}, \citenamefont {Haney},\ and\
  \citenamefont {Husa}}]{Ramos-Buades:2020eju}%
  \BibitemOpen
  \bibfield  {author} {\bibinfo {author} {\bibfnamefont {A.}~\bibnamefont
  {Ramos-Buades}}, \bibinfo {author} {\bibfnamefont {S.}~\bibnamefont
  {Tiwari}}, \bibinfo {author} {\bibfnamefont {M.}~\bibnamefont {Haney}}, \
  and\ \bibinfo {author} {\bibfnamefont {S.}~\bibnamefont {Husa}},\ }\href
  {\doibase 10.1103/PhysRevD.102.043005} {\bibfield  {journal} {\bibinfo
  {journal} {Phys. Rev. D}\ }\textbf {\bibinfo {volume} {102}},\ \bibinfo
  {pages} {043005} (\bibinfo {year} {2020})},\ \Eprint
  {http://arxiv.org/abs/2005.14016} {arXiv:2005.14016 [gr-qc]} \BibitemShut
  {NoStop}%
\bibitem [{\citenamefont {{Kyutoku}}\ and\ \citenamefont
  {{Seto}}(2014)}]{kyutoku-seto2014MNRAS}%
  \BibitemOpen
  \bibfield  {author} {\bibinfo {author} {\bibfnamefont {K.}~\bibnamefont
  {{Kyutoku}}}\ and\ \bibinfo {author} {\bibfnamefont {N.}~\bibnamefont
  {{Seto}}},\ }\href {\doibase 10.1093/mnras/stu698} {\bibfield  {journal}
  {\bibinfo  {journal} {Mon.~Not.~R.~Astron.~Soc.}\ }\textbf {\bibinfo {volume}
  {441}},\ \bibinfo {pages} {1934} (\bibinfo {year} {2014})},\ \Eprint
  {http://arxiv.org/abs/1312.2953} {arXiv:1312.2953 [astro-ph.HE]} \BibitemShut
  {NoStop}%
\bibitem [{\citenamefont {{Sun}}\ \emph {et~al.}(2015)\citenamefont {{Sun}},
  \citenamefont {{Cao}}, \citenamefont {{Wang}},\ and\ \citenamefont
  {{Yeh}}}]{sun-cao-etal2015PhRvD}%
  \BibitemOpen
  \bibfield  {author} {\bibinfo {author} {\bibfnamefont {B.}~\bibnamefont
  {{Sun}}}, \bibinfo {author} {\bibfnamefont {Z.}~\bibnamefont {{Cao}}},
  \bibinfo {author} {\bibfnamefont {Y.}~\bibnamefont {{Wang}}}, \ and\ \bibinfo
  {author} {\bibfnamefont {H.-C.}\ \bibnamefont {{Yeh}}},\ }\href {\doibase
  10.1103/PhysRevD.92.044034} {\bibfield  {journal} {\bibinfo  {journal}
  {\prd}\ }\textbf {\bibinfo {volume} {92}},\ \bibinfo {eid} {044034} (\bibinfo
  {year} {2015})}\BibitemShut {NoStop}%
\bibitem [{\citenamefont {Gondán}\ \emph {et~al.}(2018)\citenamefont
  {Gondán}, \citenamefont {Kocsis}, \citenamefont {Raffai},\ and\
  \citenamefont {Frei}}]{Gondan:2017hbp}%
  \BibitemOpen
  \bibfield  {author} {\bibinfo {author} {\bibfnamefont {L.}~\bibnamefont
  {Gondán}}, \bibinfo {author} {\bibfnamefont {B.}~\bibnamefont {Kocsis}},
  \bibinfo {author} {\bibfnamefont {P.}~\bibnamefont {Raffai}}, \ and\ \bibinfo
  {author} {\bibfnamefont {Z.}~\bibnamefont {Frei}},\ }\href {\doibase
  10.3847/1538-4357/aaad0e} {\bibfield  {journal} {\bibinfo  {journal}
  {Astrophys. J.}\ }\textbf {\bibinfo {volume} {855}},\ \bibinfo {pages} {34}
  (\bibinfo {year} {2018})},\ \Eprint {http://arxiv.org/abs/1705.10781}
  {arXiv:1705.10781 [astro-ph.HE]} \BibitemShut {NoStop}%
\bibitem [{\citenamefont {{Gond{\'a}n}}\ and\ \citenamefont
  {{Kocsis}}(2019)}]{Gondan:2018khr}%
  \BibitemOpen
  \bibfield  {author} {\bibinfo {author} {\bibfnamefont {L.}~\bibnamefont
  {{Gond{\'a}n}}}\ and\ \bibinfo {author} {\bibfnamefont {B.}~\bibnamefont
  {{Kocsis}}},\ }\href {\doibase 10.3847/1538-4357/aaf893} {\bibfield
  {journal} {\bibinfo  {journal} {Astrophys.~J.}\ }\textbf {\bibinfo {volume}
  {871}},\ \bibinfo {eid} {178} (\bibinfo {year} {2019})},\ \Eprint
  {http://arxiv.org/abs/1809.00672} {arXiv:1809.00672 [astro-ph.HE]}
  \BibitemShut {NoStop}%
\bibitem [{\citenamefont {{Ashton}}\ \emph {et~al.}(2019)\citenamefont
  {{Ashton}} \emph {et~al.}}]{bilby2019ApJS}%
  \BibitemOpen
  \bibfield  {author} {\bibinfo {author} {\bibfnamefont {G.}~\bibnamefont
  {{Ashton}}} \emph {et~al.},\ }\href {\doibase 10.3847/1538-4365/ab06fc}
  {\bibfield  {journal} {\bibinfo  {journal} {Astrophys.~J.~Suppl.~Ser.}\
  }\textbf {\bibinfo {volume} {241}},\ \bibinfo {eid} {27} (\bibinfo {year}
  {2019})},\ \Eprint {http://arxiv.org/abs/1811.02042} {arXiv:1811.02042
  [astro-ph.IM]} \BibitemShut {NoStop}%
\bibitem [{\citenamefont {{Lower}}\ \emph {et~al.}(2018)\citenamefont
  {{Lower}}, \citenamefont {{Thrane}}, \citenamefont {{Lasky}},\ and\
  \citenamefont {{Smith}}}]{Lower:2018seu}%
  \BibitemOpen
  \bibfield  {author} {\bibinfo {author} {\bibfnamefont {M.~E.}\ \bibnamefont
  {{Lower}}}, \bibinfo {author} {\bibfnamefont {E.}~\bibnamefont {{Thrane}}},
  \bibinfo {author} {\bibfnamefont {P.~D.}\ \bibnamefont {{Lasky}}}, \ and\
  \bibinfo {author} {\bibfnamefont {R.}~\bibnamefont {{Smith}}},\ }\href
  {\doibase 10.1103/PhysRevD.98.083028} {\bibfield  {journal} {\bibinfo
  {journal} {\prd}\ }\textbf {\bibinfo {volume} {98}},\ \bibinfo {eid} {083028}
  (\bibinfo {year} {2018})},\ \Eprint {http://arxiv.org/abs/1806.05350}
  {arXiv:1806.05350 [astro-ph.HE]} \BibitemShut {NoStop}%
\bibitem [{\citenamefont {Abbott}\ \emph
  {et~al.}(2016{\natexlab{c}})\citenamefont {Abbott} \emph
  {et~al.}}]{detection-PEpaper2016}%
  \BibitemOpen
  \bibfield  {author} {\bibinfo {author} {\bibfnamefont {B.~P.}\ \bibnamefont
  {Abbott}} \emph {et~al.} (\bibinfo {collaboration} {LIGO Scientific and Virgo
  Collaborations}),\ }\href {\doibase 10.1103/PhysRevLett.116.241102}
  {\bibfield  {journal} {\bibinfo  {journal} {\prl}\ }\textbf {\bibinfo
  {volume} {116}},\ \bibinfo {eid} {241102} (\bibinfo {year}
  {2016}{\natexlab{c}})},\ \Eprint {http://arxiv.org/abs/1602.03840}
  {arXiv:1602.03840 [gr-qc]} \BibitemShut {NoStop}%
\bibitem [{\citenamefont {Abbott}\ \emph
  {et~al.}(2016{\natexlab{d}})\citenamefont {Abbott} \emph
  {et~al.}}]{detection-Astropaper2016ApJL}%
  \BibitemOpen
  \bibfield  {author} {\bibinfo {author} {\bibfnamefont {B.~P.}\ \bibnamefont
  {Abbott}} \emph {et~al.} (\bibinfo {collaboration} {LIGO Scientific and Virgo
  Collaborations}),\ }\href {\doibase 10.3847/2041-8205/818/2/L22} {\bibfield
  {journal} {\bibinfo  {journal} {Astrophys.~J.~Lett.}\ }\textbf {\bibinfo
  {volume} {818}},\ \bibinfo {eid} {L22} (\bibinfo {year}
  {2016}{\natexlab{d}})},\ \Eprint {http://arxiv.org/abs/1602.03846}
  {arXiv:1602.03846 [astro-ph.HE]} \BibitemShut {NoStop}%
\bibitem [{\citenamefont {Abbott}\ \emph
  {et~al.}(2017{\natexlab{e}})\citenamefont {Abbott} \emph
  {et~al.}}]{Abbott:2016wiq}%
  \BibitemOpen
  \bibfield  {author} {\bibinfo {author} {\bibfnamefont {B.~P.}\ \bibnamefont
  {Abbott}} \emph {et~al.} (\bibinfo {collaboration} {LIGO Scientific and Virgo
  Collaborations}),\ }\href {\doibase 10.1088/1361-6382/aa6854} {\bibfield
  {journal} {\bibinfo  {journal} {Classical Quantum Gravity}\ }\textbf
  {\bibinfo {volume} {34}},\ \bibinfo {pages} {104002} (\bibinfo {year}
  {2017}{\natexlab{e}})},\ \Eprint {http://arxiv.org/abs/1611.07531}
  {arXiv:1611.07531 [gr-qc]} \BibitemShut {NoStop}%
\bibitem [{\citenamefont {Cao}\ and\ \citenamefont
  {Han}(2017)}]{cao-han-SEOBeccPRD2017}%
  \BibitemOpen
  \bibfield  {author} {\bibinfo {author} {\bibfnamefont {Z.}~\bibnamefont
  {Cao}}\ and\ \bibinfo {author} {\bibfnamefont {W.-B.}\ \bibnamefont {Han}},\
  }\href {\doibase 10.1103/PhysRevD.96.044028} {\bibfield  {journal} {\bibinfo
  {journal} {Phys. Rev. D}\ }\textbf {\bibinfo {volume} {96}},\ \bibinfo
  {pages} {044028} (\bibinfo {year} {2017})}\BibitemShut {NoStop}%
\bibitem [{\citenamefont {{Romero-Shaw}}\ \emph {et~al.}(2019)\citenamefont
  {{Romero-Shaw}}, \citenamefont {{Lasky}},\ and\ \citenamefont
  {{Thrane}}}]{romero-shaw2019MNRAS}%
  \BibitemOpen
  \bibfield  {author} {\bibinfo {author} {\bibfnamefont {I.~M.}\ \bibnamefont
  {{Romero-Shaw}}}, \bibinfo {author} {\bibfnamefont {P.~D.}\ \bibnamefont
  {{Lasky}}}, \ and\ \bibinfo {author} {\bibfnamefont {E.}~\bibnamefont
  {{Thrane}}},\ }\href {\doibase 10.1093/mnras/stz2996} {\bibfield  {journal}
  {\bibinfo  {journal} {Mon.~Not.~R.~Astron.~Soc.}\ }\textbf {\bibinfo {volume}
  {490}},\ \bibinfo {pages} {5210} (\bibinfo {year} {2019})},\ \Eprint
  {http://arxiv.org/abs/1909.05466} {arXiv:1909.05466 [astro-ph.HE]}
  \BibitemShut {NoStop}%
\bibitem [{\citenamefont {{Huerta}}\ \emph {et~al.}(2014)\citenamefont
  {{Huerta}}, \citenamefont {{Kumar}}, \citenamefont {{McWilliams}},
  \citenamefont {{O'Shaughnessy}},\ and\ \citenamefont
  {{Yunes}}}]{huerta-etal-PRD2014}%
  \BibitemOpen
  \bibfield  {author} {\bibinfo {author} {\bibfnamefont {E.~A.}\ \bibnamefont
  {{Huerta}}}, \bibinfo {author} {\bibfnamefont {P.}~\bibnamefont {{Kumar}}},
  \bibinfo {author} {\bibfnamefont {S.~T.}\ \bibnamefont {{McWilliams}}},
  \bibinfo {author} {\bibfnamefont {R.}~\bibnamefont {{O'Shaughnessy}}}, \ and\
  \bibinfo {author} {\bibfnamefont {N.}~\bibnamefont {{Yunes}}},\ }\href
  {\doibase 10.1103/PhysRevD.90.084016} {\bibfield  {journal} {\bibinfo
  {journal} {\prd}\ }\textbf {\bibinfo {volume} {90}},\ \bibinfo {eid} {084016}
  (\bibinfo {year} {2014})},\ \Eprint {http://arxiv.org/abs/1408.3406}
  {arXiv:1408.3406 [gr-qc]} \BibitemShut {NoStop}%
\bibitem [{\citenamefont {{Wu}}\ \emph {et~al.}(2020)\citenamefont {{Wu}},
  \citenamefont {{Cao}},\ and\ \citenamefont {{Zhu}}}]{wu-cao-zhu2020}%
  \BibitemOpen
  \bibfield  {author} {\bibinfo {author} {\bibfnamefont {S.}~\bibnamefont
  {{Wu}}}, \bibinfo {author} {\bibfnamefont {Z.}~\bibnamefont {{Cao}}}, \ and\
  \bibinfo {author} {\bibfnamefont {Z.-H.}\ \bibnamefont {{Zhu}}},\ }\href
  {\doibase 10.1093/mnras/staa1176} {\bibfield  {journal} {\bibinfo  {journal}
  {Mon.~Not.~R.~Astron.~Soc.}\ }\textbf {\bibinfo {volume} {495}},\ \bibinfo
  {pages} {466} (\bibinfo {year} {2020})},\ \Eprint
  {http://arxiv.org/abs/2002.05528} {arXiv:2002.05528 [astro-ph.IM]}
  \BibitemShut {NoStop}%
\bibitem [{\citenamefont {{Nitz}}\ \emph {et~al.}(2020)\citenamefont {{Nitz}},
  \citenamefont {{Lenon}},\ and\ \citenamefont
  {{Brown}}}]{nitz-lenon-brown2020ApJ}%
  \BibitemOpen
  \bibfield  {author} {\bibinfo {author} {\bibfnamefont {A.~H.}\ \bibnamefont
  {{Nitz}}}, \bibinfo {author} {\bibfnamefont {A.}~\bibnamefont {{Lenon}}}, \
  and\ \bibinfo {author} {\bibfnamefont {D.~A.}\ \bibnamefont {{Brown}}},\
  }\href {\doibase 10.3847/1538-4357/ab6611} {\bibfield  {journal} {\bibinfo
  {journal} {\apj}\ }\textbf {\bibinfo {volume} {890}},\ \bibinfo {eid} {1}
  (\bibinfo {year} {2020})},\ \Eprint {http://arxiv.org/abs/1912.05464}
  {arXiv:1912.05464 [astro-ph.HE]} \BibitemShut {NoStop}%
\bibitem [{\citenamefont {{Lenon}}\ \emph {et~al.}(2020)\citenamefont
  {{Lenon}}, \citenamefont {{Nitz}},\ and\ \citenamefont
  {{Brown}}}]{lenon-nitz-brown2020MNRAS}%
  \BibitemOpen
  \bibfield  {author} {\bibinfo {author} {\bibfnamefont {A.~K.}\ \bibnamefont
  {{Lenon}}}, \bibinfo {author} {\bibfnamefont {A.~H.}\ \bibnamefont {{Nitz}}},
  \ and\ \bibinfo {author} {\bibfnamefont {D.~A.}\ \bibnamefont {{Brown}}},\
  }\href {\doibase 10.1093/mnras/staa2120} {\bibfield  {journal} {\bibinfo
  {journal} {Mon.~Not.~R.~Astron.~Soc.}\ }\textbf {\bibinfo {volume} {497}},\
  \bibinfo {pages} {1966} (\bibinfo {year} {2020})},\ \Eprint
  {http://arxiv.org/abs/2005.14146} {arXiv:2005.14146 [astro-ph.HE]}
  \BibitemShut {NoStop}%
\bibitem [{\citenamefont {{Biwer}}\ \emph {et~al.}(2019)\citenamefont
  {{Biwer}}, \citenamefont {{Capano}}, \citenamefont {{De}}, \citenamefont
  {{Cabero}}, \citenamefont {{Brown}}, \citenamefont {{Nitz}},\ and\
  \citenamefont {{Raymond}}}]{PyCBCInfereceA2019PASP}%
  \BibitemOpen
  \bibfield  {author} {\bibinfo {author} {\bibfnamefont {C.~M.}\ \bibnamefont
  {{Biwer}}}, \bibinfo {author} {\bibfnamefont {C.~D.}\ \bibnamefont
  {{Capano}}}, \bibinfo {author} {\bibfnamefont {S.}~\bibnamefont {{De}}},
  \bibinfo {author} {\bibfnamefont {M.}~\bibnamefont {{Cabero}}}, \bibinfo
  {author} {\bibfnamefont {D.~A.}\ \bibnamefont {{Brown}}}, \bibinfo {author}
  {\bibfnamefont {A.~H.}\ \bibnamefont {{Nitz}}}, \ and\ \bibinfo {author}
  {\bibfnamefont {V.}~\bibnamefont {{Raymond}}},\ }\href {\doibase
  10.1088/1538-3873/aaef0b} {\bibfield  {journal} {\bibinfo  {journal}
  {Publ.~Astron.~Soc.~Pac.}\ }\textbf {\bibinfo {volume} {131}},\ \bibinfo
  {pages} {024503} (\bibinfo {year} {2019})},\ \Eprint
  {http://arxiv.org/abs/1807.10312} {arXiv:1807.10312 [astro-ph.IM]}
  \BibitemShut {NoStop}%
\bibitem [{\citenamefont {Kim}\ \emph {et~al.}(2019)\citenamefont {Kim},
  \citenamefont {Kim}, \citenamefont {Lee.}, \citenamefont {Favata},\ and\
  \citenamefont {Arun.}}]{TaylorF2EccURL}%
  \BibitemOpen
  \bibfield  {author} {\bibinfo {author} {\bibfnamefont {J.}~\bibnamefont
  {Kim}}, \bibinfo {author} {\bibfnamefont {C.~L.}\ \bibnamefont {Kim}},
  \bibinfo {author} {\bibfnamefont {H.~W.}\ \bibnamefont {Lee.}}, \bibinfo
  {author} {\bibfnamefont {M.}~\bibnamefont {Favata}}, \ and\ \bibinfo {author}
  {\bibfnamefont {K.~G.}\ \bibnamefont {Arun.}},\ }\href@noop {} {\enquote
  {\bibinfo {title} {{LALSimInspiralTaylorF2Ecc.c File Reference}},}\ }\bibinfo
  {howpublished}
  {\url{https://lscsoft.docs.ligo.org/lalsuite/lalsimulation/_l_a_l_sim_inspiral_taylor_f2_ecc_8c.html}}
  (\bibinfo {year} {2019})\BibitemShut {NoStop}%
\bibitem [{\citenamefont {{O'Shea}}\ and\ \citenamefont
  {{Kumar}}(2021)}]{OSheaKumar2021}%
  \BibitemOpen
  \bibfield  {author} {\bibinfo {author} {\bibfnamefont {E.}~\bibnamefont
  {{O'Shea}}}\ and\ \bibinfo {author} {\bibfnamefont {P.}~\bibnamefont
  {{Kumar}}},\ }\href@noop {} {\  (\bibinfo {year} {2021})},\ \Eprint
  {http://arxiv.org/abs/2107.07981} {arXiv:2107.07981 [astro-ph.HE]}
  \BibitemShut {NoStop}%
\bibitem [{\citenamefont {{Romero-Shaw}}\ \emph {et~al.}(2021)\citenamefont
  {{Romero-Shaw}}, \citenamefont {{Lasky}},\ and\ \citenamefont
  {{Thrane}}}]{romero-shaw-eeccGWTC2}%
  \BibitemOpen
  \bibfield  {author} {\bibinfo {author} {\bibfnamefont {I.}~\bibnamefont
  {{Romero-Shaw}}}, \bibinfo {author} {\bibfnamefont {P.~D.}\ \bibnamefont
  {{Lasky}}}, \ and\ \bibinfo {author} {\bibfnamefont {E.}~\bibnamefont
  {{Thrane}}},\ }\href {\doibase 10.3847/2041-8213/ac3138} {\bibfield
  {journal} {\bibinfo  {journal} {Astrophys.~J.~Lett.}\ }\textbf {\bibinfo
  {volume} {921}},\ \bibinfo {eid} {L31} (\bibinfo {year} {2021})},\ \Eprint
  {http://arxiv.org/abs/2108.01284} {arXiv:2108.01284 [astro-ph.HE]}
  \BibitemShut {NoStop}%
\bibitem [{\citenamefont {Abbott}\ \emph
  {et~al.}(2020{\natexlab{e}})\citenamefont {Abbott} \emph
  {et~al.}}]{GW190521-astroApJL}%
  \BibitemOpen
  \bibfield  {author} {\bibinfo {author} {\bibfnamefont {R.}~\bibnamefont
  {Abbott}} \emph {et~al.} (\bibinfo {collaboration} {LIGO Scientific and Virgo
  Collaborations}),\ }\href {\doibase 10.3847/2041-8213/aba493} {\bibfield
  {journal} {\bibinfo  {journal} {Astrophys.~J.~Lett.}\ }\textbf {\bibinfo
  {volume} {900}},\ \bibinfo {eid} {L13} (\bibinfo {year}
  {2020}{\natexlab{e}})},\ \Eprint {http://arxiv.org/abs/2009.01190}
  {arXiv:2009.01190 [astro-ph.HE]} \BibitemShut {NoStop}%
\bibitem [{\citenamefont {{Bustillo}}\ \emph {et~al.}(2021)\citenamefont
  {{Bustillo}}, \citenamefont {{Sanchis-Gual}}, \citenamefont
  {{Torres-Forn{\'e}}},\ and\ \citenamefont
  {{Font}}}]{Bustillo-headon190521PRL2021}%
  \BibitemOpen
  \bibfield  {author} {\bibinfo {author} {\bibfnamefont {J.~C.}\ \bibnamefont
  {{Bustillo}}}, \bibinfo {author} {\bibfnamefont {N.}~\bibnamefont
  {{Sanchis-Gual}}}, \bibinfo {author} {\bibfnamefont {A.}~\bibnamefont
  {{Torres-Forn{\'e}}}}, \ and\ \bibinfo {author} {\bibfnamefont {J.~A.}\
  \bibnamefont {{Font}}},\ }\href {\doibase 10.1103/PhysRevLett.126.201101}
  {\bibfield  {journal} {\bibinfo  {journal} {\prl}\ }\textbf {\bibinfo
  {volume} {126}},\ \bibinfo {eid} {201101} (\bibinfo {year} {2021})},\ \Eprint
  {http://arxiv.org/abs/2009.01066} {arXiv:2009.01066 [gr-qc]} \BibitemShut
  {NoStop}%
\bibitem [{\citenamefont {{Romero-Shaw}}\ \emph {et~al.}(2020)\citenamefont
  {{Romero-Shaw}}, \citenamefont {{Lasky}}, \citenamefont {{Thrane}},\ and\
  \citenamefont {{Calder{\'o}n Bustillo}}}]{romero-shaw-etal2020ApJ}%
  \BibitemOpen
  \bibfield  {author} {\bibinfo {author} {\bibfnamefont {I.}~\bibnamefont
  {{Romero-Shaw}}}, \bibinfo {author} {\bibfnamefont {P.~D.}\ \bibnamefont
  {{Lasky}}}, \bibinfo {author} {\bibfnamefont {E.}~\bibnamefont {{Thrane}}}, \
  and\ \bibinfo {author} {\bibfnamefont {J.}~\bibnamefont {{Calder{\'o}n
  Bustillo}}},\ }\href {\doibase 10.3847/2041-8213/abbe26} {\bibfield
  {journal} {\bibinfo  {journal} {Astrophys.~J.~Lett.}\ }\textbf {\bibinfo
  {volume} {903}},\ \bibinfo {eid} {L5} (\bibinfo {year} {2020})},\ \Eprint
  {http://arxiv.org/abs/2009.04771} {arXiv:2009.04771 [astro-ph.HE]}
  \BibitemShut {NoStop}%
\bibitem [{\citenamefont {{Gayathri}}\ \emph {et~al.}(2020)\citenamefont
  {{Gayathri}}, \citenamefont {{Healy}}, \citenamefont {{Lange}}, \citenamefont
  {{O'Brien}}, \citenamefont {{Szczepanczyk}}, \citenamefont {{Bartos}},
  \citenamefont {{Campanelli}}, \citenamefont {{Klimenko}}, \citenamefont
  {{Lousto}},\ and\ \citenamefont {{O'Shaughnessy}}}]{gayathri-etal2020}%
  \BibitemOpen
  \bibfield  {author} {\bibinfo {author} {\bibfnamefont {V.}~\bibnamefont
  {{Gayathri}}}, \bibinfo {author} {\bibfnamefont {J.}~\bibnamefont {{Healy}}},
  \bibinfo {author} {\bibfnamefont {J.}~\bibnamefont {{Lange}}}, \bibinfo
  {author} {\bibfnamefont {B.}~\bibnamefont {{O'Brien}}}, \bibinfo {author}
  {\bibfnamefont {M.}~\bibnamefont {{Szczepanczyk}}}, \bibinfo {author}
  {\bibfnamefont {I.}~\bibnamefont {{Bartos}}}, \bibinfo {author}
  {\bibfnamefont {M.}~\bibnamefont {{Campanelli}}}, \bibinfo {author}
  {\bibfnamefont {S.}~\bibnamefont {{Klimenko}}}, \bibinfo {author}
  {\bibfnamefont {C.}~\bibnamefont {{Lousto}}}, \ and\ \bibinfo {author}
  {\bibfnamefont {R.}~\bibnamefont {{O'Shaughnessy}}},\ }\href@noop {} {\
  (\bibinfo {year} {2020})},\ \Eprint {http://arxiv.org/abs/2009.05461}
  {arXiv:2009.05461 [astro-ph.HE]} \BibitemShut {NoStop}%
\bibitem [{\citenamefont {{Cutler}}\ and\ \citenamefont
  {{Vallisneri}}(2007)}]{cutler-vallisneri-systematicerrors-PRD2007}%
  \BibitemOpen
  \bibfield  {author} {\bibinfo {author} {\bibfnamefont {C.}~\bibnamefont
  {{Cutler}}}\ and\ \bibinfo {author} {\bibfnamefont {M.}~\bibnamefont
  {{Vallisneri}}},\ }\href {\doibase 10.1103/PhysRevD.76.104018} {\bibfield
  {journal} {\bibinfo  {journal} {Phys.~Rev.~D}\ }\textbf {\bibinfo {volume}
  {76}},\ \bibinfo {eid} {104018} (\bibinfo {year} {2007})},\ \Eprint
  {http://arxiv.org/abs/arXiv:0707.2982} {arXiv:0707.2982} \BibitemShut
  {NoStop}%
\bibitem [{\citenamefont {{R{\"o}ver}}\ \emph {et~al.}(2007)\citenamefont
  {{R{\"o}ver}}, \citenamefont {{Meyer}},\ and\ \citenamefont
  {{Christensen}}}]{rover-meyer-christensen-PRD2007}%
  \BibitemOpen
  \bibfield  {author} {\bibinfo {author} {\bibfnamefont {C.}~\bibnamefont
  {{R{\"o}ver}}}, \bibinfo {author} {\bibfnamefont {R.}~\bibnamefont
  {{Meyer}}}, \ and\ \bibinfo {author} {\bibfnamefont {N.}~\bibnamefont
  {{Christensen}}},\ }\href {\doibase 10.1103/PhysRevD.75.062004} {\bibfield
  {journal} {\bibinfo  {journal} {Phys.~Rev.~D.}\ }\textbf {\bibinfo {volume}
  {75}},\ \bibinfo {eid} {062004} (\bibinfo {year} {2007})},\ \Eprint
  {http://arxiv.org/abs/arXiv:gr-qc/0609131} {arXiv:gr-qc/0609131} \BibitemShut
  {NoStop}%
\bibitem [{\citenamefont {{van der Sluys}}\ \emph {et~al.}(2008)\citenamefont
  {{van der Sluys}}, \citenamefont {{R{\"o}ver}}, \citenamefont {{Stroeer}},
  \citenamefont {{Raymond}}, \citenamefont {{Mandel}}, \citenamefont
  {{Christensen}}, \citenamefont {{Kalogera}}, \citenamefont {{Meyer}},\ and\
  \citenamefont {{Vecchio}}}]{vandersluys-etal-ApJ2008}%
  \BibitemOpen
  \bibfield  {author} {\bibinfo {author} {\bibfnamefont {M.~V.}\ \bibnamefont
  {{van der Sluys}}}, \bibinfo {author} {\bibfnamefont {C.}~\bibnamefont
  {{R{\"o}ver}}}, \bibinfo {author} {\bibfnamefont {A.}~\bibnamefont
  {{Stroeer}}}, \bibinfo {author} {\bibfnamefont {V.}~\bibnamefont
  {{Raymond}}}, \bibinfo {author} {\bibfnamefont {I.}~\bibnamefont {{Mandel}}},
  \bibinfo {author} {\bibfnamefont {N.}~\bibnamefont {{Christensen}}}, \bibinfo
  {author} {\bibfnamefont {V.}~\bibnamefont {{Kalogera}}}, \bibinfo {author}
  {\bibfnamefont {R.}~\bibnamefont {{Meyer}}}, \ and\ \bibinfo {author}
  {\bibfnamefont {A.}~\bibnamefont {{Vecchio}}},\ }\href {\doibase
  10.1086/595279} {\bibfield  {journal} {\bibinfo  {journal}
  {Astrophys.~J.~Lett.}\ }\textbf {\bibinfo {volume} {688}},\ \bibinfo {pages}
  {L61} (\bibinfo {year} {2008})},\ \Eprint {http://arxiv.org/abs/0710.1897}
  {arXiv:0710.1897} \BibitemShut {NoStop}%
\bibitem [{\citenamefont {{Veitch}}\ \emph {et~al.}(2015)\citenamefont
  {{Veitch}} \emph {et~al.}}]{veitch-etalPEpaper-PRD2015}%
  \BibitemOpen
  \bibfield  {author} {\bibinfo {author} {\bibfnamefont {J.}~\bibnamefont
  {{Veitch}}} \emph {et~al.},\ }\href {\doibase 10.1103/PhysRevD.91.042003}
  {\bibfield  {journal} {\bibinfo  {journal} {\prd}\ }\textbf {\bibinfo
  {volume} {91}},\ \bibinfo {eid} {042003} (\bibinfo {year} {2015})},\ \Eprint
  {http://arxiv.org/abs/1409.7215} {arXiv:1409.7215 [gr-qc]} \BibitemShut
  {NoStop}%
\bibitem [{\citenamefont {{LIGO Scientific Collaboration}}(2018)}]{LALSuite}%
  \BibitemOpen
  \bibfield  {author} {\bibinfo {author} {\bibnamefont {{LIGO Scientific
  Collaboration}}},\ }\href {\doibase 10.7935/GT1W-FZ16} {\enquote {\bibinfo
  {title} {{LIGO Algorithm Library}},}\ }\bibinfo {howpublished}
  {\url{https://git.ligo.org/lscsoft/lalsuite}} (\bibinfo {year}
  {2018})\BibitemShut {NoStop}%
\bibitem [{\citenamefont {{Buonanno}}\ \emph {et~al.}(2009)\citenamefont
  {{Buonanno}}, \citenamefont {{Iyer}}, \citenamefont {{Ochsner}},
  \citenamefont {{Pan}},\ and\ \citenamefont
  {{Sathyaprakash}}}]{buonanno-iyer-oshsner-pan-sathya-templatecomparison-PRD2009}%
  \BibitemOpen
  \bibfield  {author} {\bibinfo {author} {\bibfnamefont {A.}~\bibnamefont
  {{Buonanno}}}, \bibinfo {author} {\bibfnamefont {B.~R.}\ \bibnamefont
  {{Iyer}}}, \bibinfo {author} {\bibfnamefont {E.}~\bibnamefont {{Ochsner}}},
  \bibinfo {author} {\bibfnamefont {Y.}~\bibnamefont {{Pan}}}, \ and\ \bibinfo
  {author} {\bibfnamefont {B.~S.}\ \bibnamefont {{Sathyaprakash}}},\ }\href
  {\doibase 10.1103/PhysRevD.80.084043} {\bibfield  {journal} {\bibinfo
  {journal} {Phys.~Rev.~D}\ }\textbf {\bibinfo {volume} {80}},\ \bibinfo
  {pages} {084043} (\bibinfo {year} {2009})},\ \Eprint
  {http://arxiv.org/abs/arXiv:0907.0700 [gr-qc]} {arXiv:0907.0700 [gr-qc]}
  \BibitemShut {NoStop}%
\bibitem [{\citenamefont {{Kidder}}\ \emph {et~al.}(1993)\citenamefont
  {{Kidder}}, \citenamefont {{Will}},\ and\ \citenamefont
  {{Wiseman}}}]{kidderwillwiseman-spineffects}%
  \BibitemOpen
  \bibfield  {author} {\bibinfo {author} {\bibfnamefont {L.~E.}\ \bibnamefont
  {{Kidder}}}, \bibinfo {author} {\bibfnamefont {C.~M.}\ \bibnamefont
  {{Will}}}, \ and\ \bibinfo {author} {\bibfnamefont {A.~G.}\ \bibnamefont
  {{Wiseman}}},\ }\href@noop {} {\bibfield  {journal} {\bibinfo  {journal}
  {Phys.~Rev.~D}\ }\textbf {\bibinfo {volume} {47}},\ \bibinfo {pages} {R4183}
  (\bibinfo {year} {1993})},\ \Eprint
  {http://arxiv.org/abs/arXiv:gr-qc/9211025} {arXiv:gr-qc/9211025} \BibitemShut
  {NoStop}%
\bibitem [{\citenamefont {{Poisson}}(1993)}]{poisson-BHpertIV-slowrotPRD1993}%
  \BibitemOpen
  \bibfield  {author} {\bibinfo {author} {\bibfnamefont {E.}~\bibnamefont
  {{Poisson}}},\ }\href {\doibase 10.1103/PhysRevD.48.1860} {\bibfield
  {journal} {\bibinfo  {journal} {Phys.~Rev.~D}\ }\textbf {\bibinfo {volume}
  {48}},\ \bibinfo {pages} {1860} (\bibinfo {year} {1993})}\BibitemShut
  {NoStop}%
\bibitem [{\citenamefont {{Kidder}}(1995)}]{kidder-spineffects}%
  \BibitemOpen
  \bibfield  {author} {\bibinfo {author} {\bibfnamefont {L.~E.}\ \bibnamefont
  {{Kidder}}},\ }\href {\doibase 10.1103/PhysRevD.52.821} {\bibfield  {journal}
  {\bibinfo  {journal} {Phys.~Rev.~D}\ }\textbf {\bibinfo {volume} {52}},\
  \bibinfo {pages} {821} (\bibinfo {year} {1995})},\ \Eprint
  {http://arxiv.org/abs/arXiv:gr-qc/9506022} {arXiv:gr-qc/9506022} \BibitemShut
  {NoStop}%
\bibitem [{\citenamefont {{Mik{\'o}czi}}\ \emph {et~al.}(2005)\citenamefont
  {{Mik{\'o}czi}}, \citenamefont {{Vas{\'u}th}},\ and\ \citenamefont
  {{Gergely}}}]{gergely-selfspinPRD05}%
  \BibitemOpen
  \bibfield  {author} {\bibinfo {author} {\bibfnamefont {B.}~\bibnamefont
  {{Mik{\'o}czi}}}, \bibinfo {author} {\bibfnamefont {M.}~\bibnamefont
  {{Vas{\'u}th}}}, \ and\ \bibinfo {author} {\bibfnamefont {L.~{\'A}.}\
  \bibnamefont {{Gergely}}},\ }\href {\doibase 10.1103/PhysRevD.71.124043}
  {\bibfield  {journal} {\bibinfo  {journal} {Phys.~Rev.~D}\ }\textbf {\bibinfo
  {volume} {71}},\ \bibinfo {pages} {124043} (\bibinfo {year} {2005})},\
  \Eprint {http://arxiv.org/abs/arXiv:astro-ph/0504538}
  {arXiv:astro-ph/0504538} \BibitemShut {NoStop}%
\bibitem [{\citenamefont
  {{Poisson}}(1998)}]{poisson-quadrupolemonopoleterm-PRD1998}%
  \BibitemOpen
  \bibfield  {author} {\bibinfo {author} {\bibfnamefont {E.}~\bibnamefont
  {{Poisson}}},\ }\href {\doibase 10.1103/PhysRevD.57.5287} {\bibfield
  {journal} {\bibinfo  {journal} {Phys.~Rev.~D}\ }\textbf {\bibinfo {volume}
  {57}},\ \bibinfo {pages} {5287} (\bibinfo {year} {1998})},\ \Eprint
  {http://arxiv.org/abs/arXiv:gr-qc/9709032} {arXiv:gr-qc/9709032} \BibitemShut
  {NoStop}%
\bibitem [{\citenamefont {{Laarakkers}}\ and\ \citenamefont
  {{Poisson}}(1999)}]{laarakker-poisson-quadmomentApJ1999}%
  \BibitemOpen
  \bibfield  {author} {\bibinfo {author} {\bibfnamefont {W.~G.}\ \bibnamefont
  {{Laarakkers}}}\ and\ \bibinfo {author} {\bibfnamefont {E.}~\bibnamefont
  {{Poisson}}},\ }\href {\doibase 10.1086/306732} {\bibfield  {journal}
  {\bibinfo  {journal} {Astrophys.~J.}\ }\textbf {\bibinfo {volume} {512}},\
  \bibinfo {pages} {282} (\bibinfo {year} {1999})}\BibitemShut {NoStop}%
\bibitem [{\citenamefont {{Gergely}}(1999)}]{gergely-selfspin-PRD2000}%
  \BibitemOpen
  \bibfield  {author} {\bibinfo {author} {\bibfnamefont {L.~{\'A}.}\
  \bibnamefont {{Gergely}}},\ }\href {\doibase 10.1103/PhysRevD.61.024035}
  {\bibfield  {journal} {\bibinfo  {journal} {Phys.~Rev.~D}\ }\textbf {\bibinfo
  {volume} {61}},\ \bibinfo {pages} {024035} (\bibinfo {year} {1999})},\
  \Eprint {http://arxiv.org/abs/arXiv:gr-qc/9911082} {arXiv:gr-qc/9911082}
  \BibitemShut {NoStop}%
\bibitem [{\citenamefont {{Faye}}\ \emph {et~al.}(2006)\citenamefont {{Faye}},
  \citenamefont {{Blanchet}},\ and\ \citenamefont
  {{Buonanno}}}]{faye-buonanno-luc-higherorderspinI}%
  \BibitemOpen
  \bibfield  {author} {\bibinfo {author} {\bibfnamefont {G.}~\bibnamefont
  {{Faye}}}, \bibinfo {author} {\bibfnamefont {L.}~\bibnamefont {{Blanchet}}},
  \ and\ \bibinfo {author} {\bibfnamefont {A.}~\bibnamefont {{Buonanno}}},\
  }\href {\doibase 10.1103/PhysRevD.74.104033} {\bibfield  {journal} {\bibinfo
  {journal} {Phys.~Rev.~D}\ }\textbf {\bibinfo {volume} {74}},\ \bibinfo
  {pages} {104033} (\bibinfo {year} {2006})},\ \Eprint
  {http://arxiv.org/abs/arXiv:gr-qc/0605139} {arXiv:gr-qc/0605139} \BibitemShut
  {NoStop}%
\bibitem [{\citenamefont {{Blanchet}}\ \emph {et~al.}(2006)\citenamefont
  {{Blanchet}}, \citenamefont {{Buonanno}},\ and\ \citenamefont
  {{Faye}}}]{faye-buonanno-luc-higherorderspinII}%
  \BibitemOpen
  \bibfield  {author} {\bibinfo {author} {\bibfnamefont {L.}~\bibnamefont
  {{Blanchet}}}, \bibinfo {author} {\bibfnamefont {A.}~\bibnamefont
  {{Buonanno}}}, \ and\ \bibinfo {author} {\bibfnamefont {G.}~\bibnamefont
  {{Faye}}},\ }\href {\doibase 10.1103/PhysRevD.74.104034} {\bibfield
  {journal} {\bibinfo  {journal} {Phys.~Rev.~D}\ }\textbf {\bibinfo {volume}
  {74}},\ \bibinfo {pages} {104034} (\bibinfo {year} {2006})},\ \Eprint
  {http://arxiv.org/abs/arXiv:gr-qc/0605140} {arXiv:gr-qc/0605140} \BibitemShut
  {NoStop}%
\bibitem [{\citenamefont {{Blanchet}}\ \emph {et~al.}(2007)\citenamefont
  {{Blanchet}}, \citenamefont {{Buonanno}},\ and\ \citenamefont
  {{Faye}}}]{faye-buonanno-luc-higherorderspinIIerratum}%
  \BibitemOpen
  \bibfield  {author} {\bibinfo {author} {\bibfnamefont {L.}~\bibnamefont
  {{Blanchet}}}, \bibinfo {author} {\bibfnamefont {A.}~\bibnamefont
  {{Buonanno}}}, \ and\ \bibinfo {author} {\bibfnamefont {G.}~\bibnamefont
  {{Faye}}},\ }\href {\doibase 10.1103/PhysRevD.75.049903} {\bibfield
  {journal} {\bibinfo  {journal} {Phys.~Rev.~D}\ }\textbf {\bibinfo {volume}
  {75}},\ \bibinfo {pages} {049903(E)} (\bibinfo {year} {2007})}\BibitemShut
  {NoStop}%
\bibitem [{\citenamefont {{Blanchet}}\ \emph {et~al.}(2010)\citenamefont
  {{Blanchet}}, \citenamefont {{Buonanno}},\ and\ \citenamefont
  {{Faye}}}]{faye-buonanno-luc-higherorderspinIIerratum2}%
  \BibitemOpen
  \bibfield  {author} {\bibinfo {author} {\bibfnamefont {L.}~\bibnamefont
  {{Blanchet}}}, \bibinfo {author} {\bibfnamefont {A.}~\bibnamefont
  {{Buonanno}}}, \ and\ \bibinfo {author} {\bibfnamefont {G.}~\bibnamefont
  {{Faye}}},\ }\href {\doibase 10.1103/PhysRevD.81.089901} {\bibfield
  {journal} {\bibinfo  {journal} {Phys.~Rev.~D}\ }\textbf {\bibinfo {volume}
  {81}},\ \bibinfo {pages} {089901(E)} (\bibinfo {year} {2010})}\BibitemShut
  {NoStop}%
\bibitem [{\citenamefont {{Favata}}(2011{\natexlab{b}})}]{favata-PNspinisco}%
  \BibitemOpen
  \bibfield  {author} {\bibinfo {author} {\bibfnamefont {M.}~\bibnamefont
  {{Favata}}},\ }\href {\doibase 10.1103/PhysRevD.83.024028} {\bibfield
  {journal} {\bibinfo  {journal} {Phys.~Rev.~D}\ }\textbf {\bibinfo {volume}
  {83}},\ \bibinfo {pages} {024028} (\bibinfo {year} {2011}{\natexlab{b}})},\
  \Eprint {http://arxiv.org/abs/arXiv:1010.2553} {arXiv:1010.2553 [gr-qc]}
  \BibitemShut {NoStop}%
\bibitem [{\citenamefont {{Alvi}}(2001)}]{alvi-BHabsorptionPRD2001}%
  \BibitemOpen
  \bibfield  {author} {\bibinfo {author} {\bibfnamefont {K.}~\bibnamefont
  {{Alvi}}},\ }\href {\doibase 10.1103/PhysRevD.64.104020} {\bibfield
  {journal} {\bibinfo  {journal} {Phys.~Rev.~D}\ }\textbf {\bibinfo {volume}
  {64}},\ \bibinfo {pages} {104020} (\bibinfo {year} {2001})},\ \Eprint
  {http://arxiv.org/abs/arXiv:gr-qc/0107080} {arXiv:gr-qc/0107080} \BibitemShut
  {NoStop}%
\bibitem [{\citenamefont {{Mishra}}\ \emph {et~al.}(2016)\citenamefont
  {{Mishra}}, \citenamefont {{Kela}}, \citenamefont {{Arun}},\ and\
  \citenamefont {{Faye}}}]{mishra-etal-spinsPRD2016}%
  \BibitemOpen
  \bibfield  {author} {\bibinfo {author} {\bibfnamefont {C.~K.}\ \bibnamefont
  {{Mishra}}}, \bibinfo {author} {\bibfnamefont {A.}~\bibnamefont {{Kela}}},
  \bibinfo {author} {\bibfnamefont {K.~G.}\ \bibnamefont {{Arun}}}, \ and\
  \bibinfo {author} {\bibfnamefont {G.}~\bibnamefont {{Faye}}},\ }\href
  {\doibase 10.1103/PhysRevD.93.084054} {\bibfield  {journal} {\bibinfo
  {journal} {\prd}\ }\textbf {\bibinfo {volume} {93}},\ \bibinfo {eid} {084054}
  (\bibinfo {year} {2016})},\ \Eprint {http://arxiv.org/abs/1601.05588}
  {arXiv:1601.05588 [gr-qc]} \BibitemShut {NoStop}%
\bibitem [{\citenamefont {Blanchet}\ \emph {et~al.}(2011)\citenamefont
  {Blanchet}, \citenamefont {Buonanno},\ and\ \citenamefont {Faye}}]{BBF2011}%
  \BibitemOpen
  \bibfield  {author} {\bibinfo {author} {\bibfnamefont {L.}~\bibnamefont
  {Blanchet}}, \bibinfo {author} {\bibfnamefont {A.}~\bibnamefont {Buonanno}},
  \ and\ \bibinfo {author} {\bibfnamefont {G.}~\bibnamefont {Faye}},\ }\href
  {\doibase 10.1103/PhysRevD.84.064041} {\bibfield  {journal} {\bibinfo
  {journal} {Phys. Rev. D}\ }\textbf {\bibinfo {volume} {84}},\ \bibinfo
  {pages} {064041} (\bibinfo {year} {2011})},\ \Eprint
  {http://arxiv.org/abs/1104.5659} {arXiv:1104.5659 [gr-qc]} \BibitemShut
  {NoStop}%
\bibitem [{\citenamefont {Boh\'e}\ \emph {et~al.}(2015)\citenamefont {Boh\'e},
  \citenamefont {Faye}, \citenamefont {Marsat},\ and\ \citenamefont
  {Porter}}]{BFMP2015}%
  \BibitemOpen
  \bibfield  {author} {\bibinfo {author} {\bibfnamefont {A.}~\bibnamefont
  {Boh\'e}}, \bibinfo {author} {\bibfnamefont {G.}~\bibnamefont {Faye}},
  \bibinfo {author} {\bibfnamefont {S.}~\bibnamefont {Marsat}}, \ and\ \bibinfo
  {author} {\bibfnamefont {E.~K.}\ \bibnamefont {Porter}},\ }\href {\doibase
  10.1088/0264-9381/32/19/195010} {\bibfield  {journal} {\bibinfo  {journal}
  {Classical Quantum Gravity}\ }\textbf {\bibinfo {volume} {32}},\ \bibinfo
  {pages} {195010} (\bibinfo {year} {2015})},\ \Eprint
  {http://arxiv.org/abs/1501.01529} {arXiv:1501.01529 [gr-qc]} \BibitemShut
  {NoStop}%
\bibitem [{\citenamefont {Boh\'e}\ \emph {et~al.}(2013)\citenamefont {Boh\'e},
  \citenamefont {Marsat},\ and\ \citenamefont {Blanchet}}]{BMB2013}%
  \BibitemOpen
  \bibfield  {author} {\bibinfo {author} {\bibfnamefont {A.}~\bibnamefont
  {Boh\'e}}, \bibinfo {author} {\bibfnamefont {S.}~\bibnamefont {Marsat}}, \
  and\ \bibinfo {author} {\bibfnamefont {L.}~\bibnamefont {Blanchet}},\ }\href
  {\doibase 10.1088/0264-9381/30/13/135009} {\bibfield  {journal} {\bibinfo
  {journal} {Classical Quantum Gravity}\ }\textbf {\bibinfo {volume} {30}},\
  \bibinfo {pages} {135009} (\bibinfo {year} {2013})},\ \Eprint
  {http://arxiv.org/abs/1303.7412} {arXiv:1303.7412 [gr-qc]} \BibitemShut
  {NoStop}%
\bibitem [{\citenamefont {Marsat}(2015)}]{MarsatCubic}%
  \BibitemOpen
  \bibfield  {author} {\bibinfo {author} {\bibfnamefont {S.}~\bibnamefont
  {Marsat}},\ }\href {\doibase 10.1088/0264-9381/32/8/085008} {\bibfield
  {journal} {\bibinfo  {journal} {Classical Quantum Gravity}\ }\textbf
  {\bibinfo {volume} {32}},\ \bibinfo {pages} {085008} (\bibinfo {year}
  {2015})},\ \Eprint {http://arxiv.org/abs/1411.4118} {arXiv:1411.4118 [gr-qc]}
  \BibitemShut {NoStop}%
\bibitem [{\citenamefont {Marsat}\ \emph {et~al.}(2014)\citenamefont {Marsat},
  \citenamefont {Boh\'e}, \citenamefont {Blanchet},\ and\ \citenamefont
  {Buonanno}}]{M3B}%
  \BibitemOpen
  \bibfield  {author} {\bibinfo {author} {\bibfnamefont {S.}~\bibnamefont
  {Marsat}}, \bibinfo {author} {\bibfnamefont {A.}~\bibnamefont {Boh\'e}},
  \bibinfo {author} {\bibfnamefont {L.}~\bibnamefont {Blanchet}}, \ and\
  \bibinfo {author} {\bibfnamefont {A.}~\bibnamefont {Buonanno}},\ }\href
  {\doibase 10.1088/0264-9381/31/2/025023} {\bibfield  {journal} {\bibinfo
  {journal} {Classical Quantum Gravity}\ }\textbf {\bibinfo {volume} {31}},\
  \bibinfo {pages} {025023} (\bibinfo {year} {2014})},\ \Eprint
  {http://arxiv.org/abs/1307.6793} {arXiv:1307.6793 [gr-qc]} \BibitemShut
  {NoStop}%
\bibitem [{\citenamefont {Porto}(2016)}]{PortoEFTrev}%
  \BibitemOpen
  \bibfield  {author} {\bibinfo {author} {\bibfnamefont {R.~A.}\ \bibnamefont
  {Porto}},\ }\href {\doibase 10.1016/j.physrep.2016.04.003} {\bibfield
  {journal} {\bibinfo  {journal} {Phys.~Rep.}\ }\textbf {\bibinfo {volume}
  {633}},\ \bibinfo {pages} {1} (\bibinfo {year} {2016})},\ \Eprint
  {http://arxiv.org/abs/1601.04914} {arXiv:1601.04914 [hep-th]} \BibitemShut
  {NoStop}%
\bibitem [{\citenamefont {Maia}\ \emph {et~al.}(2017)\citenamefont {Maia},
  \citenamefont {Galley}, \citenamefont {Leibovich},\ and\ \citenamefont
  {Porto}}]{EFTSpinOrbit}%
  \BibitemOpen
  \bibfield  {author} {\bibinfo {author} {\bibfnamefont {N.~T.}\ \bibnamefont
  {Maia}}, \bibinfo {author} {\bibfnamefont {C.~R.}\ \bibnamefont {Galley}},
  \bibinfo {author} {\bibfnamefont {A.~K.}\ \bibnamefont {Leibovich}}, \ and\
  \bibinfo {author} {\bibfnamefont {R.~A.}\ \bibnamefont {Porto}},\ }\href
  {\doibase 10.1103/PhysRevD.96.084064} {\bibfield  {journal} {\bibinfo
  {journal} {Phys. Rev. D}\ }\textbf {\bibinfo {volume} {96}},\ \bibinfo
  {pages} {084064} (\bibinfo {year} {2017})},\ \Eprint
  {http://arxiv.org/abs/1705.07934} {arXiv:1705.07934 [gr-qc]} \BibitemShut
  {NoStop}%
\bibitem [{\citenamefont {Cho}\ \emph {et~al.}(2021)\citenamefont {Cho},
  \citenamefont {Pardo},\ and\ \citenamefont {Porto}}]{EFTSpinSpin}%
  \BibitemOpen
  \bibfield  {author} {\bibinfo {author} {\bibfnamefont {G.}~\bibnamefont
  {Cho}}, \bibinfo {author} {\bibfnamefont {B.}~\bibnamefont {Pardo}}, \ and\
  \bibinfo {author} {\bibfnamefont {R.~A.}\ \bibnamefont {Porto}},\ }\href
  {\doibase 10.1103/PhysRevD.104.024037} {\bibfield  {journal} {\bibinfo
  {journal} {Phys. Rev. D}\ }\textbf {\bibinfo {volume} {104}},\ \bibinfo
  {pages} {024037} (\bibinfo {year} {2021})},\ \Eprint
  {http://arxiv.org/abs/2103.14612} {arXiv:2103.14612 [gr-qc]} \BibitemShut
  {NoStop}%
\bibitem [{\citenamefont {{Junker}}\ and\ \citenamefont
  {{Sch\"{a}fer}}(1992)}]{junker-schafer}%
  \BibitemOpen
  \bibfield  {author} {\bibinfo {author} {\bibfnamefont {W.}~\bibnamefont
  {{Junker}}}\ and\ \bibinfo {author} {\bibfnamefont {G.}~\bibnamefont
  {{Sch\"{a}fer}}},\ }\href@noop {} {\bibfield  {journal} {\bibinfo  {journal}
  {Mon.~Not.~R.~Astron.~Soc.}\ }\textbf {\bibinfo {volume} {254}},\ \bibinfo
  {pages} {146} (\bibinfo {year} {1992})}\BibitemShut {NoStop}%
\bibitem [{\citenamefont {{Gopakumar}}\ \emph {et~al.}(1997)\citenamefont
  {{Gopakumar}}, \citenamefont {{Iyer}},\ and\ \citenamefont
  {{Iyer}}}]{gopakumar-iyer-iyer-PRD1997}%
  \BibitemOpen
  \bibfield  {author} {\bibinfo {author} {\bibfnamefont {A.}~\bibnamefont
  {{Gopakumar}}}, \bibinfo {author} {\bibfnamefont {B.~R.}\ \bibnamefont
  {{Iyer}}}, \ and\ \bibinfo {author} {\bibfnamefont {S.}~\bibnamefont
  {{Iyer}}},\ }\href {\doibase 10.1103/PhysRevD.55.6030} {\bibfield  {journal}
  {\bibinfo  {journal} {Phys.~Rev.~D}\ }\textbf {\bibinfo {volume} {55}},\
  \bibinfo {pages} {6030} (\bibinfo {year} {1997})},\ \Eprint
  {http://arxiv.org/abs/arXiv:gr-qc/9703075} {arXiv:gr-qc/9703075} \BibitemShut
  {NoStop}%
\bibitem [{\citenamefont {{Arun}}\ \emph
  {et~al.}(2008{\natexlab{a}})\citenamefont {{Arun}}, \citenamefont
  {{Blanchet}}, \citenamefont {{Iyer}},\ and\ \citenamefont
  {{Qusailah}}}]{arun-eccentrictailsEflux}%
  \BibitemOpen
  \bibfield  {author} {\bibinfo {author} {\bibfnamefont {K.~G.}\ \bibnamefont
  {{Arun}}}, \bibinfo {author} {\bibfnamefont {L.}~\bibnamefont {{Blanchet}}},
  \bibinfo {author} {\bibfnamefont {B.~R.}\ \bibnamefont {{Iyer}}}, \ and\
  \bibinfo {author} {\bibfnamefont {M.~S.~S.}\ \bibnamefont {{Qusailah}}},\
  }\href {\doibase 10.1103/PhysRevD.77.064034} {\bibfield  {journal} {\bibinfo
  {journal} {Phys.~Rev.~D}\ }\textbf {\bibinfo {volume} {77}},\ \bibinfo
  {pages} {064034} (\bibinfo {year} {2008}{\natexlab{a}})},\ \Eprint
  {http://arxiv.org/abs/arXiv:0711.0250 [gr-qc]} {arXiv:0711.0250 [gr-qc]}
  \BibitemShut {NoStop}%
\bibitem [{\citenamefont {{Arun}}\ \emph
  {et~al.}(2008{\natexlab{b}})\citenamefont {{Arun}}, \citenamefont
  {{Blanchet}}, \citenamefont {{Iyer}},\ and\ \citenamefont
  {{Qusailah}}}]{arun-eccentricEflux3PN}%
  \BibitemOpen
  \bibfield  {author} {\bibinfo {author} {\bibfnamefont {K.~G.}\ \bibnamefont
  {{Arun}}}, \bibinfo {author} {\bibfnamefont {L.}~\bibnamefont {{Blanchet}}},
  \bibinfo {author} {\bibfnamefont {B.~R.}\ \bibnamefont {{Iyer}}}, \ and\
  \bibinfo {author} {\bibfnamefont {M.~S.~S.}\ \bibnamefont {{Qusailah}}},\
  }\href {\doibase 10.1103/PhysRevD.77.064035} {\bibfield  {journal} {\bibinfo
  {journal} {Phys.~Rev.~D}\ }\textbf {\bibinfo {volume} {77}},\ \bibinfo
  {pages} {064035} (\bibinfo {year} {2008}{\natexlab{b}})},\ \Eprint
  {http://arxiv.org/abs/arXiv:0711.0302 [gr-qc]} {arXiv:0711.0302 [gr-qc]}
  \BibitemShut {NoStop}%
\bibitem [{\citenamefont {{Arun}}\ \emph {et~al.}(2009)\citenamefont {{Arun}},
  \citenamefont {{Blanchet}}, \citenamefont {{Iyer}},\ and\ \citenamefont
  {{Sinha}}}]{arun-etal-eccentric-orbitalelements-PRD2009}%
  \BibitemOpen
  \bibfield  {author} {\bibinfo {author} {\bibfnamefont {K.~G.}\ \bibnamefont
  {{Arun}}}, \bibinfo {author} {\bibfnamefont {L.}~\bibnamefont {{Blanchet}}},
  \bibinfo {author} {\bibfnamefont {B.~R.}\ \bibnamefont {{Iyer}}}, \ and\
  \bibinfo {author} {\bibfnamefont {S.}~\bibnamefont {{Sinha}}},\ }\href
  {\doibase 10.1103/PhysRevD.80.124018} {\bibfield  {journal} {\bibinfo
  {journal} {Phys.~Rev.~D}\ }\textbf {\bibinfo {volume} {80}},\ \bibinfo
  {pages} {124018} (\bibinfo {year} {2009})},\ \Eprint
  {http://arxiv.org/abs/0908.3854} {arXiv:0908.3854 [gr-qc]} \BibitemShut
  {NoStop}%
\bibitem [{\citenamefont {{Khalil}}\ \emph {et~al.}(2021)\citenamefont
  {{Khalil}}, \citenamefont {{Buonanno}}, \citenamefont {{Steinhoff}},\ and\
  \citenamefont {{Vines}}}]{khalil-etalEOBeccPRD2021}%
  \BibitemOpen
  \bibfield  {author} {\bibinfo {author} {\bibfnamefont {M.}~\bibnamefont
  {{Khalil}}}, \bibinfo {author} {\bibfnamefont {A.}~\bibnamefont
  {{Buonanno}}}, \bibinfo {author} {\bibfnamefont {J.}~\bibnamefont
  {{Steinhoff}}}, \ and\ \bibinfo {author} {\bibfnamefont {J.}~\bibnamefont
  {{Vines}}},\ }\href {\doibase 10.1103/PhysRevD.104.024046} {\bibfield
  {journal} {\bibinfo  {journal} {\prd}\ }\textbf {\bibinfo {volume} {104}},\
  \bibinfo {eid} {024046} (\bibinfo {year} {2021})},\ \Eprint
  {http://arxiv.org/abs/2104.11705} {arXiv:2104.11705 [gr-qc]} \BibitemShut
  {NoStop}%
\bibitem [{\citenamefont {{Bernuzzi}}\ \emph {et~al.}(2012)\citenamefont
  {{Bernuzzi}}, \citenamefont {{Nagar}}, \citenamefont {{Thierfelder}},\ and\
  \citenamefont {{Br{\"u}gmann}}}]{bernuzzi-nagar-brugmann-PRD2012-tidalNS}%
  \BibitemOpen
  \bibfield  {author} {\bibinfo {author} {\bibfnamefont {S.}~\bibnamefont
  {{Bernuzzi}}}, \bibinfo {author} {\bibfnamefont {A.}~\bibnamefont {{Nagar}}},
  \bibinfo {author} {\bibfnamefont {M.}~\bibnamefont {{Thierfelder}}}, \ and\
  \bibinfo {author} {\bibfnamefont {B.}~\bibnamefont {{Br{\"u}gmann}}},\ }\href
  {\doibase 10.1103/PhysRevD.86.044030} {\bibfield  {journal} {\bibinfo
  {journal} {Phys.~Rev.~D}\ }\textbf {\bibinfo {volume} {86}},\ \bibinfo {eid}
  {044030} (\bibinfo {year} {2012})},\ \Eprint
  {http://arxiv.org/abs/arXiv:1205.3403} {arXiv:1205.3403} \BibitemShut
  {NoStop}%
\bibitem [{\citenamefont {Bardeen}\ \emph {et~al.}(1972)\citenamefont
  {Bardeen}, \citenamefont {Press},\ and\ \citenamefont {Teukolsky}}]{bptkerr}%
  \BibitemOpen
  \bibfield  {author} {\bibinfo {author} {\bibfnamefont {J.~M.}\ \bibnamefont
  {Bardeen}}, \bibinfo {author} {\bibfnamefont {W.~H.}\ \bibnamefont {Press}},
  \ and\ \bibinfo {author} {\bibfnamefont {S.~A.}\ \bibnamefont {Teukolsky}},\
  }\href@noop {} {\bibfield  {journal} {\bibinfo  {journal} {Astrophys. J.}\
  }\textbf {\bibinfo {volume} {178}},\ \bibinfo {pages} {347} (\bibinfo {year}
  {1972})}\BibitemShut {NoStop}%
\bibitem [{\citenamefont {{Yunes}}\ \emph {et~al.}(2009)\citenamefont
  {{Yunes}}, \citenamefont {{Arun}}, \citenamefont {{Berti}},\ and\
  \citenamefont {{Will}}}]{yunes-arun-berti-will-eccentric-PRD2009}%
  \BibitemOpen
  \bibfield  {author} {\bibinfo {author} {\bibfnamefont {N.}~\bibnamefont
  {{Yunes}}}, \bibinfo {author} {\bibfnamefont {K.~G.}\ \bibnamefont {{Arun}}},
  \bibinfo {author} {\bibfnamefont {E.}~\bibnamefont {{Berti}}}, \ and\
  \bibinfo {author} {\bibfnamefont {C.~M.}\ \bibnamefont {{Will}}},\ }\href
  {\doibase 10.1103/PhysRevD.80.084001} {\bibfield  {journal} {\bibinfo
  {journal} {Phys.~Rev.~D}\ }\textbf {\bibinfo {volume} {80}},\ \bibinfo
  {pages} {084001} (\bibinfo {year} {2009})},\ \Eprint
  {http://arxiv.org/abs/0906.0313} {arXiv:0906.0313 [gr-qc]} \BibitemShut
  {NoStop}%
\bibitem [{\citenamefont {{Mik{\'o}czi}}\ \emph {et~al.}(2012)\citenamefont
  {{Mik{\'o}czi}}, \citenamefont {{Kocsis}}, \citenamefont {{Forg{\'a}cs}},\
  and\ \citenamefont {{Vas{\'u}th}}}]{mikoczi-etal2012PhRvD}%
  \BibitemOpen
  \bibfield  {author} {\bibinfo {author} {\bibfnamefont {B.}~\bibnamefont
  {{Mik{\'o}czi}}}, \bibinfo {author} {\bibfnamefont {B.}~\bibnamefont
  {{Kocsis}}}, \bibinfo {author} {\bibfnamefont {P.}~\bibnamefont
  {{Forg{\'a}cs}}}, \ and\ \bibinfo {author} {\bibfnamefont {M.}~\bibnamefont
  {{Vas{\'u}th}}},\ }\href {\doibase 10.1103/PhysRevD.86.104027} {\bibfield
  {journal} {\bibinfo  {journal} {\prd}\ }\textbf {\bibinfo {volume} {86}},\
  \bibinfo {eid} {104027} (\bibinfo {year} {2012})},\ \Eprint
  {http://arxiv.org/abs/1206.5786} {arXiv:1206.5786 [gr-qc]} \BibitemShut
  {NoStop}%
\bibitem [{\citenamefont {{Bose}}\ and\ \citenamefont
  {{Pai}}(2021)}]{EccMcBurst}%
  \BibitemOpen
  \bibfield  {author} {\bibinfo {author} {\bibfnamefont {N.}~\bibnamefont
  {{Bose}}}\ and\ \bibinfo {author} {\bibfnamefont {A.}~\bibnamefont {{Pai}}},\
  }\href@noop {} {\  (\bibinfo {year} {2021})},\ \Eprint
  {http://arxiv.org/abs/2107.14736} {arXiv:2107.14736 [gr-qc]} \BibitemShut
  {NoStop}%
\bibitem [{\citenamefont {{Finn}}(1992)}]{finn-PRD1992}%
  \BibitemOpen
  \bibfield  {author} {\bibinfo {author} {\bibfnamefont {L.~S.}\ \bibnamefont
  {{Finn}}},\ }\href@noop {} {\bibfield  {journal} {\bibinfo  {journal}
  {Phys.~Rev.~D}\ }\textbf {\bibinfo {volume} {46}},\ \bibinfo {pages} {5236}
  (\bibinfo {year} {1992})}\BibitemShut {NoStop}%
\bibitem [{\citenamefont {{Finn}}\ and\ \citenamefont
  {{Chernoff}}(1993)}]{finn-chernoff-PRD1993}%
  \BibitemOpen
  \bibfield  {author} {\bibinfo {author} {\bibfnamefont {L.~S.}\ \bibnamefont
  {{Finn}}}\ and\ \bibinfo {author} {\bibfnamefont {D.~F.}\ \bibnamefont
  {{Chernoff}}},\ }\href@noop {} {\bibfield  {journal} {\bibinfo  {journal}
  {Phys.~Rev.~D}\ }\textbf {\bibinfo {volume} {47}},\ \bibinfo {pages} {2198}
  (\bibinfo {year} {1993})}\BibitemShut {NoStop}%
\bibitem [{\citenamefont {{Cutler}}\ and\ \citenamefont
  {{Flanagan}}(1994)}]{flanagancutler}%
  \BibitemOpen
  \bibfield  {author} {\bibinfo {author} {\bibfnamefont {C.}~\bibnamefont
  {{Cutler}}}\ and\ \bibinfo {author} {\bibfnamefont {{\'E}.~E.}\ \bibnamefont
  {{Flanagan}}},\ }\href {\doibase 10.1103/PhysRevD.49.2658} {\bibfield
  {journal} {\bibinfo  {journal} {\prd}\ }\textbf {\bibinfo {volume} {49}},\
  \bibinfo {pages} {2658} (\bibinfo {year} {1994})},\ \Eprint
  {http://arxiv.org/abs/gr-qc/9402014} {gr-qc/9402014} \BibitemShut {NoStop}%
\bibitem [{\citenamefont {{Poisson}}\ and\ \citenamefont
  {{Will}}(1995)}]{poisson-will-2PNparameterestimate}%
  \BibitemOpen
  \bibfield  {author} {\bibinfo {author} {\bibfnamefont {E.}~\bibnamefont
  {{Poisson}}}\ and\ \bibinfo {author} {\bibfnamefont {C.~M.}\ \bibnamefont
  {{Will}}},\ }\href@noop {} {\bibfield  {journal} {\bibinfo  {journal}
  {Phys.~Rev.~D}\ }\textbf {\bibinfo {volume} {52}},\ \bibinfo {pages} {848}
  (\bibinfo {year} {1995})},\ \Eprint
  {http://arxiv.org/abs/arXiv:gr-qc/9502040} {arXiv:gr-qc/9502040} \BibitemShut
  {NoStop}%
\bibitem [{\citenamefont {{Arun}}\ \emph
  {et~al.}(2005{\natexlab{a}})\citenamefont {{Arun}}, \citenamefont {{Iyer}},
  \citenamefont {{Sathyaprakash}},\ and\ \citenamefont
  {{Sundararajan}}}]{arun-etal-PRD2005-35PNparameterestimation}%
  \BibitemOpen
  \bibfield  {author} {\bibinfo {author} {\bibfnamefont {K.~G.}\ \bibnamefont
  {{Arun}}}, \bibinfo {author} {\bibfnamefont {B.~R.}\ \bibnamefont {{Iyer}}},
  \bibinfo {author} {\bibfnamefont {B.~S.}\ \bibnamefont {{Sathyaprakash}}}, \
  and\ \bibinfo {author} {\bibfnamefont {P.~A.}\ \bibnamefont
  {{Sundararajan}}},\ }\href {\doibase 10.1103/PhysRevD.71.084008} {\bibfield
  {journal} {\bibinfo  {journal} {Phys.~Rev.~D}\ }\textbf {\bibinfo {volume}
  {71}},\ \bibinfo {pages} {084008} (\bibinfo {year}
  {2005}{\natexlab{a}})}\BibitemShut {NoStop}%
\bibitem [{\citenamefont {{Arun}}\ \emph
  {et~al.}(2005{\natexlab{b}})\citenamefont {{Arun}}, \citenamefont {{Iyer}},
  \citenamefont {{Sathyaprakash}},\ and\ \citenamefont
  {{Sundararajan}}}]{arun-etal-PRD2005-35PNparameterestimation-errata}%
  \BibitemOpen
  \bibfield  {author} {\bibinfo {author} {\bibfnamefont {K.~G.}\ \bibnamefont
  {{Arun}}}, \bibinfo {author} {\bibfnamefont {B.~R.}\ \bibnamefont {{Iyer}}},
  \bibinfo {author} {\bibfnamefont {B.~S.}\ \bibnamefont {{Sathyaprakash}}}, \
  and\ \bibinfo {author} {\bibfnamefont {P.~A.}\ \bibnamefont
  {{Sundararajan}}},\ }\href {\doibase 10.1103/PhysRevD.72.069903} {\bibfield
  {journal} {\bibinfo  {journal} {Phys.~Rev.~D}\ }\textbf {\bibinfo {volume}
  {72}},\ \bibinfo {pages} {069903(E)} (\bibinfo {year}
  {2005}{\natexlab{b}})}\BibitemShut {NoStop}%
\bibitem [{\citenamefont {{Berti}}\ \emph {et~al.}(2005)\citenamefont
  {{Berti}}, \citenamefont {{Buonanno}},\ and\ \citenamefont
  {{Will}}}]{berti-buonanno-will-PRD2005}%
  \BibitemOpen
  \bibfield  {author} {\bibinfo {author} {\bibfnamefont {E.}~\bibnamefont
  {{Berti}}}, \bibinfo {author} {\bibfnamefont {A.}~\bibnamefont {{Buonanno}}},
  \ and\ \bibinfo {author} {\bibfnamefont {C.~M.}\ \bibnamefont {{Will}}},\
  }\href {\doibase 10.1103/PhysRevD.71.084025} {\bibfield  {journal} {\bibinfo
  {journal} {Phys.~Rev.~D}\ }\textbf {\bibinfo {volume} {71}},\ \bibinfo
  {pages} {084025} (\bibinfo {year} {2005})},\ \Eprint
  {http://arxiv.org/abs/arXiv:gr-qc/0411129} {arXiv:gr-qc/0411129} \BibitemShut
  {NoStop}%
\bibitem [{\citenamefont {{Van Den Broeck}}\ and\ \citenamefont
  {{Sengupta}}(2007)}]{vandenbroeck-sengupta-BBHspectro-CQG2007}%
  \BibitemOpen
  \bibfield  {author} {\bibinfo {author} {\bibfnamefont {C.}~\bibnamefont {{Van
  Den Broeck}}}\ and\ \bibinfo {author} {\bibfnamefont {A.~S.}\ \bibnamefont
  {{Sengupta}}},\ }\href {\doibase 10.1088/0264-9381/24/5/005} {\bibfield
  {journal} {\bibinfo  {journal} {Classical Quantum Gravity}\ }\textbf
  {\bibinfo {volume} {24}},\ \bibinfo {pages} {1089} (\bibinfo {year}
  {2007})},\ \Eprint {http://arxiv.org/abs/arXiv:gr-qc/0610126}
  {arXiv:gr-qc/0610126} \BibitemShut {NoStop}%
\bibitem [{\citenamefont {{Ajith}}\ and\ \citenamefont
  {{Bose}}(2009)}]{ajith-bose-PRD2009}%
  \BibitemOpen
  \bibfield  {author} {\bibinfo {author} {\bibfnamefont {P.}~\bibnamefont
  {{Ajith}}}\ and\ \bibinfo {author} {\bibfnamefont {S.}~\bibnamefont
  {{Bose}}},\ }\href {\doibase 10.1103/PhysRevD.79.084032} {\bibfield
  {journal} {\bibinfo  {journal} {Phys.~Rev.~D}\ }\textbf {\bibinfo {volume}
  {79}},\ \bibinfo {pages} {084032} (\bibinfo {year} {2009})},\ \Eprint
  {http://arxiv.org/abs/0901.4936} {arXiv:0901.4936 [gr-qc]} \BibitemShut
  {NoStop}%
\bibitem [{\citenamefont
  {{Vallisneri}}(2008)}]{vallisneri-fisherabuse-PRD2008}%
  \BibitemOpen
  \bibfield  {author} {\bibinfo {author} {\bibfnamefont {M.}~\bibnamefont
  {{Vallisneri}}},\ }\href {\doibase 10.1103/PhysRevD.77.042001} {\bibfield
  {journal} {\bibinfo  {journal} {Phys.~Rev.~D}\ }\textbf {\bibinfo {volume}
  {77}},\ \bibinfo {pages} {042001} (\bibinfo {year} {2008})},\ \Eprint
  {http://arxiv.org/abs/arXiv:gr-qc/0703086} {arXiv:gr-qc/0703086} \BibitemShut
  {NoStop}%
\bibitem [{\citenamefont {{Rodriguez}}\ \emph {et~al.}(2013)\citenamefont
  {{Rodriguez}}, \citenamefont {{Farr}}, \citenamefont {{Farr}},\ and\
  \citenamefont {{Mandel}}}]{rodriguez-farr-farr-mandel-fisherinadequacies}%
  \BibitemOpen
  \bibfield  {author} {\bibinfo {author} {\bibfnamefont {C.~L.}\ \bibnamefont
  {{Rodriguez}}}, \bibinfo {author} {\bibfnamefont {B.}~\bibnamefont {{Farr}}},
  \bibinfo {author} {\bibfnamefont {W.~M.}\ \bibnamefont {{Farr}}}, \ and\
  \bibinfo {author} {\bibfnamefont {I.}~\bibnamefont {{Mandel}}},\ }\href
  {\doibase 10.1103/PhysRevD.88.084013} {\bibfield  {journal} {\bibinfo
  {journal} {\prd}\ }\textbf {\bibinfo {volume} {88}},\ \bibinfo {eid} {084013}
  (\bibinfo {year} {2013})},\ \Eprint {http://arxiv.org/abs/1308.1397}
  {1308.1397 [astro-ph.IM]} \BibitemShut {NoStop}%
\bibitem [{\citenamefont {{Ajith}}(2011)}]{ajith-spin-PRD2011}%
  \BibitemOpen
  \bibfield  {author} {\bibinfo {author} {\bibfnamefont {P.}~\bibnamefont
  {{Ajith}}},\ }\href {\doibase 10.1103/PhysRevD.84.084037} {\bibfield
  {journal} {\bibinfo  {journal} {Phys.~Rev.~D}\ }\textbf {\bibinfo {volume}
  {84}},\ \bibinfo {eid} {084037} (\bibinfo {year} {2011})},\ \Eprint
  {http://arxiv.org/abs/1107.1267} {arXiv:1107.1267 [gr-qc]} \BibitemShut
  {NoStop}%
\bibitem [{\citenamefont {Abbott}\ \emph
  {et~al.}(2016{\natexlab{e}})\citenamefont {Abbott} \emph {et~al.}}]{O1BBH}%
  \BibitemOpen
  \bibfield  {author} {\bibinfo {author} {\bibfnamefont {B.~P.}\ \bibnamefont
  {Abbott}} \emph {et~al.} (\bibinfo {collaboration} {LIGO Scientific and Virgo
  Collaborations}),\ }\href {\doibase 10.1103/PhysRevX.6.041015} {\bibfield
  {journal} {\bibinfo  {journal} {Phys.~Rev.~X}\ }\textbf {\bibinfo {volume}
  {6}},\ \bibinfo {eid} {041015} (\bibinfo {year} {2016}{\natexlab{e}})},\
  \Eprint {http://arxiv.org/abs/1606.04856} {arXiv:1606.04856 [gr-qc]}
  \BibitemShut {NoStop}%
\bibitem [{\citenamefont {Asai}\ \emph {et~al.}(2013)\citenamefont {Asai} \emph
  {et~al.}}]{LIGO-PEpaper2013}%
  \BibitemOpen
  \bibfield  {author} {\bibinfo {author} {\bibfnamefont {J.}~\bibnamefont
  {Asai}} \emph {et~al.} (\bibinfo {collaboration} {LIGO Scientific and Virgo
  Collaborations}),\ }\href {\doibase 10.1103/PhysRevD.88.062001} {\bibfield
  {journal} {\bibinfo  {journal} {\prd}\ }\textbf {\bibinfo {volume} {88}},\
  \bibinfo {eid} {062001} (\bibinfo {year} {2013})},\ \Eprint
  {http://arxiv.org/abs/1304.1775} {arXiv:1304.1775 [gr-qc]} \BibitemShut
  {NoStop}%
\bibitem [{\citenamefont {{Rodriguez}}\ \emph {et~al.}(2014)\citenamefont
  {{Rodriguez}}, \citenamefont {{Farr}}, \citenamefont {{Raymond}},
  \citenamefont {{Farr}}, \citenamefont {{Littenberg}}, \citenamefont
  {{Fazi}},\ and\ \citenamefont
  {{Kalogera}}}]{rodriguez-etal-aLIGO-PE-estimates}%
  \BibitemOpen
  \bibfield  {author} {\bibinfo {author} {\bibfnamefont {C.~L.}\ \bibnamefont
  {{Rodriguez}}}, \bibinfo {author} {\bibfnamefont {B.}~\bibnamefont {{Farr}}},
  \bibinfo {author} {\bibfnamefont {V.}~\bibnamefont {{Raymond}}}, \bibinfo
  {author} {\bibfnamefont {W.~M.}\ \bibnamefont {{Farr}}}, \bibinfo {author}
  {\bibfnamefont {T.~B.}\ \bibnamefont {{Littenberg}}}, \bibinfo {author}
  {\bibfnamefont {D.}~\bibnamefont {{Fazi}}}, \ and\ \bibinfo {author}
  {\bibfnamefont {V.}~\bibnamefont {{Kalogera}}},\ }\href {\doibase
  10.1088/0004-637X/784/2/119} {\bibfield  {journal} {\bibinfo  {journal}
  {Astrophys.~J.}\ }\textbf {\bibinfo {volume} {784}},\ \bibinfo {eid} {119}
  (\bibinfo {year} {2014})},\ \Eprint {http://arxiv.org/abs/1309.3273}
  {arXiv:1309.3273 [astro-ph.HE]} \BibitemShut {NoStop}%
\bibitem [{\citenamefont {{Coe}}(2009)}]{coe-fishermatrix2009}%
  \BibitemOpen
  \bibfield  {author} {\bibinfo {author} {\bibfnamefont {D.}~\bibnamefont
  {{Coe}}},\ }\href@noop {} {\  (\bibinfo {year} {2009})},\ \Eprint
  {http://arxiv.org/abs/0906.4123} {arXiv:0906.4123 [astro-ph.IM]} \BibitemShut
  {NoStop}%
\bibitem [{\citenamefont {{Datta}}\ \emph {et~al.}(2021)\citenamefont
  {{Datta}}, \citenamefont {{Gupta}}, \citenamefont {{Kastha}}, \citenamefont
  {{Arun}},\ and\ \citenamefont {{Sathyaprakash}}}]{schurrefPRD}%
  \BibitemOpen
  \bibfield  {author} {\bibinfo {author} {\bibfnamefont {S.}~\bibnamefont
  {{Datta}}}, \bibinfo {author} {\bibfnamefont {A.}~\bibnamefont {{Gupta}}},
  \bibinfo {author} {\bibfnamefont {S.}~\bibnamefont {{Kastha}}}, \bibinfo
  {author} {\bibfnamefont {K.~G.}\ \bibnamefont {{Arun}}}, \ and\ \bibinfo
  {author} {\bibfnamefont {B.~S.}\ \bibnamefont {{Sathyaprakash}}},\ }\href
  {\doibase 10.1103/PhysRevD.103.024036} {\bibfield  {journal} {\bibinfo
  {journal} {\prd}\ }\textbf {\bibinfo {volume} {103}},\ \bibinfo {eid}
  {024036} (\bibinfo {year} {2021})},\ \Eprint
  {http://arxiv.org/abs/2006.12137} {arXiv:2006.12137 [gr-qc]} \BibitemShut
  {NoStop}%
\bibitem [{\citenamefont {Maggiore}(2008)}]{maggiore-GWvol1}%
  \BibitemOpen
  \bibfield  {author} {\bibinfo {author} {\bibfnamefont {M.}~\bibnamefont
  {Maggiore}},\ }\href@noop {} {\emph {\bibinfo {title} {Gravitational Waves:
  Volume 1}}}\ (\bibinfo  {publisher} {Oxford University Press},\ \bibinfo
  {address} {Oxford},\ \bibinfo {year} {2008})\BibitemShut {NoStop}%
\bibitem [{\citenamefont {{Hogg}}(1999)}]{hogg}%
  \BibitemOpen
  \bibfield  {author} {\bibinfo {author} {\bibfnamefont {D.~W.}\ \bibnamefont
  {{Hogg}}},\ }\href@noop {} {\  (\bibinfo {year} {1999})},\ \Eprint
  {http://arxiv.org/abs/astro-ph/9905116} {arXiv:astro-ph/9905116 [astro-ph]}
  \BibitemShut {NoStop}%
\bibitem [{\citenamefont {Ade}\ \emph {et~al.}(2016)\citenamefont {Ade} \emph
  {et~al.}}]{planck2015-cosmoparam}%
  \BibitemOpen
  \bibfield  {author} {\bibinfo {author} {\bibfnamefont {P.~A.~R.}\
  \bibnamefont {Ade}} \emph {et~al.} (\bibinfo {collaboration} {Planck
  Collaboration}),\ }\href {\doibase 10.1051/0004-6361/201525830} {\bibfield
  {journal} {\bibinfo  {journal} {Astron.~Astrophys.}\ }\textbf {\bibinfo
  {volume} {594}},\ \bibinfo {eid} {A13} (\bibinfo {year} {2016})},\ \Eprint
  {http://arxiv.org/abs/1502.01589} {arXiv:1502.01589 [astro-ph.CO]}
  \BibitemShut {NoStop}%
\bibitem [{\citenamefont {\mbox{2012 IAU Resolution B2}}(2012)}]{IAU2012}%
  \BibitemOpen
  \bibfield  {author} {\bibinfo {author} {\bibnamefont {\mbox{2012 IAU
  Resolution B2}}},\ }\href@noop {} {}\bibinfo {howpublished}
  {\url{https://www.iau.org/static/resolutions/IAU2012_English.pdf}} (\bibinfo
  {year} {2012})\BibitemShut {NoStop}%
\bibitem [{\citenamefont {\mbox{2015 IAU Resolution B2}}(2015)}]{IAU2015}%
  \BibitemOpen
  \bibfield  {author} {\bibinfo {author} {\bibnamefont {\mbox{2015 IAU
  Resolution B2}}},\ }\href@noop {} {}\bibinfo {howpublished}
  {\url{https://www.iau.org/static/resolutions/IAU2015_English.pdf}} (\bibinfo
  {year} {2015})\BibitemShut {NoStop}%
\bibitem [{\citenamefont {{Husa}}\ \emph {et~al.}(2016)\citenamefont {{Husa}},
  \citenamefont {{Khan}}, \citenamefont {{Hannam}}, \citenamefont
  {{P{\"u}rrer}}, \citenamefont {{Ohme}}, \citenamefont {{Forteza}},\ and\
  \citenamefont {{Boh{\'e}}}}]{husa-khan-etalPRD2016}%
  \BibitemOpen
  \bibfield  {author} {\bibinfo {author} {\bibfnamefont {S.}~\bibnamefont
  {{Husa}}}, \bibinfo {author} {\bibfnamefont {S.}~\bibnamefont {{Khan}}},
  \bibinfo {author} {\bibfnamefont {M.}~\bibnamefont {{Hannam}}}, \bibinfo
  {author} {\bibfnamefont {M.}~\bibnamefont {{P{\"u}rrer}}}, \bibinfo {author}
  {\bibfnamefont {F.}~\bibnamefont {{Ohme}}}, \bibinfo {author} {\bibfnamefont
  {X.~J.}\ \bibnamefont {{Forteza}}}, \ and\ \bibinfo {author} {\bibfnamefont
  {A.}~\bibnamefont {{Boh{\'e}}}},\ }\href {\doibase
  10.1103/PhysRevD.93.044006} {\bibfield  {journal} {\bibinfo  {journal}
  {\prd}\ }\textbf {\bibinfo {volume} {93}},\ \bibinfo {eid} {044006} (\bibinfo
  {year} {2016})},\ \Eprint {http://arxiv.org/abs/1508.07250} {arXiv:1508.07250
  [gr-qc]} \BibitemShut {NoStop}%
\bibitem [{\citenamefont {{Hofmann}}\ \emph {et~al.}(2016)\citenamefont
  {{Hofmann}}, \citenamefont {{Barausse}},\ and\ \citenamefont
  {{Rezzolla}}}]{hofmann-barausse-rezzola-finalspinApJL2016}%
  \BibitemOpen
  \bibfield  {author} {\bibinfo {author} {\bibfnamefont {F.}~\bibnamefont
  {{Hofmann}}}, \bibinfo {author} {\bibfnamefont {E.}~\bibnamefont
  {{Barausse}}}, \ and\ \bibinfo {author} {\bibfnamefont {L.}~\bibnamefont
  {{Rezzolla}}},\ }\href {\doibase 10.3847/2041-8205/825/2/L19} {\bibfield
  {journal} {\bibinfo  {journal} {Astrophys.~J.~Lett.}\ }\textbf {\bibinfo
  {volume} {825}},\ \bibinfo {eid} {L19} (\bibinfo {year} {2016})},\ \Eprint
  {http://arxiv.org/abs/1605.01938} {arXiv:1605.01938 [gr-qc]} \BibitemShut
  {NoStop}%
\end{thebibliography}%
\end{document}